\documentclass[12pt]{iopart}
\expandafter\let\csname equation*\endcsname\relax
\expandafter\let\csname endequation*\endcsname\relax
\usepackage{amsmath}
\usepackage{iopams}          %
\usepackage{mathtools}
\usepackage{bm}
\usepackage{booktabs}
\usepackage{multirow}
\usepackage{array}
\usepackage{enumitem}
\usepackage{graphicx}
\usepackage[round,authoryear]{natbib}

\usepackage[ruled,vlined,linesnumbered]{algorithm2e}
\usepackage{longtable}   %
\usepackage[hidelinks]{hyperref}

\newcommand{\sigfair}{\sigma_{\mathrm{fair}}}
\newcommand{\sigCRB}{\sigma^{\mathrm{CRB}}}
\newcommand{\WED}{\mathrm{WED}}
\newcommand{\DIF}{\mathrm{DIF}}
\newcommand{\NoFC}{N_0\mathrm{FC}}
\newcommand{\Nzero}{N_{0}}
\newcommand{\Nzeroray}[1]{N_{0,#1}}

\newcommand{\svec}[1]{\bm{#1}}
\newcommand{\E}{\mathbb{E}}
\newcommand{\Var}{\mathrm{Var}}
\newcommand{\Xhat}{\hat{X}}
\newcommand{\suppsec}[1]{supplementary section~\ref{#1}}

\begin{document}

\title[Flux compensation for Temporal CT]{Flux Compensation as a Prerequisite for Temporal~CT:
Per-Bundle Cram\'er--Rao Floors and Estimators That Reach Them}
\author{G M Besson}
\address{ForeVision XRCT, Boulder, WY 82923, USA}
\ead{ForeVision\_XRCT@protonmail.com}

\begin{abstract}
\noindent\textbf{Objective.}
In Temporal computed tomography (TCT) each detector reading sums the
photons of up to three rays.  Earlier work measured the resulting
per-bundle cost as a Dose Inflation Factor (DIF).  It used one incident
count $\Nzero$ for all rays and trained its learned prior on one
patient; this paper removes both restrictions.

\noindent\textbf{Approach.}
\textbf{$\NoFC$} ($\Nzero$ flux compensation), a bow-tie and
tube-current-modulation co-design, approximately equalizes each bundle's detected
flux. \textbf{SNN2} is a bias-corrected, box-projected Gauss--Newton
Poisson likelihood iteration. \textbf{V8c}, a neural network trained on
19 patients, corrects SNN2. \textbf{T2bp}, an empirical-Bayes
post-processor without trained weights, combines per-bundle
Poisson likelihoods with a prior from the patient's own bundles.

\noindent\textbf{Main results.}

On 23 non-test volumes, $\NoFC$ lowers the
per-volume median middle-path Cram\'er--Rao floor, as a DIF, from
$23$--$48$ under a flat-fluence baseline to $3.3$--$6.3$ at
$35$--$62\%$ of the baseline's incident photons, near the
equal-attenuation $13/3$ and $1.4$--$1.7$ times below GE-type
bow-ties.
For bundles with $X_2<6$ ($83\%$), SNN2 attains this floor (median
mean-squared-error ratio $1.02$); holding air paths at zero, a
box-constrained maximum likelihood estimate lowers the ratio to $0.88$,
and T2bp$\,\circ\,$SNN2 to $0.89$ ($0.74$--$1.26$);
T2bp's per-patient prior adds measurably only in the deeper bins (a
median $12$--$29\%$ over all bins).
T2bp$\,\circ\,$SNN2 beats T2bp$\,\circ\,$V8c on three of four $\NoFC$ test
cells, but trails or ties it on the four $\NoFC$ validation
volumes.
V8c beats SNN2 in distribution under baseline
acquisition, gains at most $9\%$ under $\NoFC$ (losing $14\%$ on the
largest in-distribution test patient), and fails on a patient acquired
outside its corpus's exposure rule.

\noindent\textbf{Significance.}
Flux compensation, chiefly the bow-tie, is a prerequisite;
classical estimation attains the resulting limit. The
remaining three- to six-fold gap to single-source CT at equal fluence,
and the high-attenuation tail, are targets for inter-bundle priors.
\end{abstract}

\noindent{\it Keywords}: Temporal CT, multi-source CT, bow-tie filter, tube-current modulation, Cram\'er--Rao bound, maximum-likelihood estimation, empirical Bayes

\maketitle

\section{Introduction}
\label{sec:intro}

\subsection{Background and Paper~1 results}
\label{sec:intro:background}

In Temporal CT (TCT), multiple x-ray sources and an extended detector
arc rotate on two independent coaxial drums, exposing the patient to
simultaneous projections, for high temporal resolution, notably in
cardiac imaging
\citep{besson2015medphys,besson2016spie}.  At a specific angular
sampling condition a sliding-window source arrangement groups detector
readings into \emph{bundles}: $M=5$ readings of one nominal
detector cell sampling three line integrals $(X_1, X_2, X_3)$ through a
$5\!\times\!3$ binary system matrix, a deliberately simplified
\emph{model problem} rather than a specific scanner's geometry.
Recovering $(X_1, X_2, X_3)$ from the five summed readings, a
demultiplexing step before any reconstruction, is the subject of
Paper~1 \citep{besson2026pmb} and of this work.

Paper~1 measured the cost of summing through the per-bundle
Cram\'er--Rao bound (CRB) on the line integrals' variance.  At equal
attenuation, matching the noise of an equal-dose single-source scan
takes $7/3$ times the photons for the endpoints and $13/3$ for the middle
path; in real anatomy, where a heavily attenuated middle ray is summed
with brighter neighbors, the middle-path cost rises several-fold.  The
floor informs two decisions: how many photons a TCT scan needs to match
single-source noise, and whether to seek gains in the estimator or the
acquisition, since no unbiased estimator can go below the floor but the
acquisition can lower it (Section~\ref{sec:n0fc}).  Paper~1 left two
limitations,
both addressed here: its learned prior was trained on a single patient,
and all rays shared one incident count $\Nzero$.

\subsection{Contributions}
\label{sec:intro:contributions}

This paper supplies two tools and makes three findings.  The tools are
SNN2, a prior-free, bias-corrected Poisson likelihood iteration
(Section~\ref{sec:snn2}), and the per-ray dose inflation factor (DIF;
Section~\ref{sec:model:dif}); the estimator names are development labels.

\begin{enumerate}[nosep, leftmargin=*]
  \item \textbf{Flux compensation is a prerequisite in TCT.}  $\NoFC$
        ($\Nzero$ flux compensation), a fitted bow-tie with per-slice
        tube-current modulation (TCM), lowers the per-volume median
        middle-path CRB floor (23 non-test patients) from a DIF of
        $23$--$48$ under a flat-fluence baseline to $3.3$--$6.3$, near the
        equal-attenuation $13/3$, with $35$--$62\%$ of the baseline's
        incident photons.
  \item \textbf{Estimators without trained weights reach the floor.}
        Under $\NoFC$, SNN2 attains it for the $83\%$ of bundles with
        $X_2<6$, and T2bp, a per-patient empirical-Bayes post-processor
        tuned on data including the test patients
        (Section~\ref{sec:discussion:limitations}), goes below it only
        through bias; T2bp on SNN2 and on the box-constrained estimate are
        the two most accurate estimators on the two test volumes of largest
        water-equivalent diameter (WED).
  \item \textbf{Our corpus network does not carry over.}  V8c
        improves on SNN2 under the baseline but gains
        little or loses under $\NoFC$, and fails on a patient acquired
        outside its corpus's exposure rule.
\end{enumerate}

\subsection{Scope and relation to companion work}
\label{sec:intro:scope}

This work is per-bundle: each bundle is an independent inversion, and
cross-bundle structure enters only through population-level priors
(V8c's corpus prior, T2bp's per-patient prior).  Inter-bundle
correlations, characterized in Paper~1, are left to a third paper in this
series (Paper~3), motivated by the residual bias of Section~\ref{sec:bridge}.
Paper~3, the decisive test for TCT, asks whether they close the remaining
gap to single-source CT, and builds on T2bp for practical reasons
(Section~\ref{sec:bridge}).  Image reconstruction is beyond this paper and
Paper~1.  The per-bundle floor still
bears on image noise: a linear reconstruction's pixel variance is a
weighted sum of nearby rays' variances and within-bundle covariances.

\section{Forward model and notation}
\label{sec:model}

\subsection{TCT bundle measurement model with per-ray $\Nzero$}
\label{sec:model:forward}
\begin{figure}[!t]
\centering
\includegraphics[width=\linewidth]{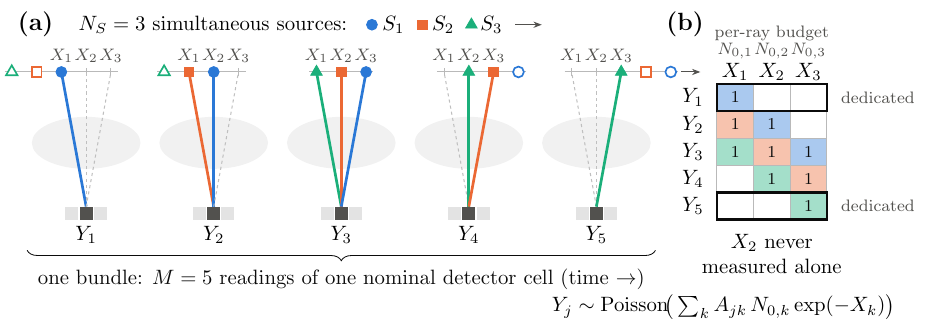}
\caption{The TCT bundle.  (a)~One nominal detector cell over its $M=5$
readings $Y_1,\dots,Y_5$.  The three rays $X_1,X_2,X_3$ are fixed, their
source ends one step apart along the source drum; the $N_S=3$ active
sources, also one step apart, advance one step from $Y_j$ to $Y_{j+1}$
(arrow).  The cell's readings in between belong to other bundles.  A ray
contributes to a reading only while a source sits at its end (solid,
colored by source; dashed: not lit), so the readings sum one, two,
three, two, then one ray, and each ray is lit once by each source.
Schematic; geometry in \citet{besson2015medphys}.  (b)~The $5\times3$
binary matrix $A$ relating readings $Y_j$ to line integrals $X_k$
(Section~\ref{sec:model:forward}), adapted from Paper~1
\citep{besson2026pmb}; its nine nonzero entries, tinted by source, are
the Poisson deviates $d[j,k]$.  Each ray has its own incident count
$\Nzeroray{k}$ (Section~\ref{sec:n0fc}).  Only the middle path $X_2$
lacks a dedicated reading.}
\label{fig:geometry}
\end{figure}

A \emph{bundle} is the elementary TCT measurement unit
(Figure~\ref{fig:geometry}): $M=5$ readings (rows $j=1,\dots,5$) of a
single nominal detector cell, one source step apart. It samples three line
integrals $\svec{X}=(X_1,X_2,X_3)$, the three rays of the bundle
($k=1,2,3$), each illuminated in turn by all three sources.  A fixed
binary $5{\times}3$ matrix $A$ records which rays reach each row,
with nonzero pattern (row\,$\to$\,rays) $1\!\to\!\{1\}$,
$2\!\to\!\{1,2\}$, $3\!\to\!\{1,2,3\}$, $4\!\to\!\{2,3\}$,
$5\!\to\!\{3\}$: the endpoints $X_1,X_3$ are each measured in isolation
(rows~1 and~5), while the middle path $X_2$ appears only superposed with
a neighbor (rows~2--4).  Each ray carries a nominal air-scan count
$\Nzeroray{k}$, set by the $\NoFC$ design (Section~\ref{sec:n0fc}) and
generalizing Paper~1's single $\Nzero$; it is shared by the three
sources that illuminate the ray, identical in every row it reaches, and
known exactly to the estimators and bounds, as in Paper~1's known-flux
reference \citep{besson2026pmb}.  Ray~$k$ contributes mean intensity
$p_{jk}=\Nzeroray{k}\,A_{jk}\,e^{-X_k}$ to row~$j$, whose mean is
$y_j=\sum_k p_{jk}$.  The row counts $Y_j$ are $\mathrm{Poisson}(y_j)$;
the simulation draws each as the sum of independent per-source deviates
$d[j,k]$, one per nonzero $(j,k)$ (\suppsec{app:snn2:deviates}).

\subsection{Generalized Fisher information matrix and CRB}
\label{sec:model:fim}
The per-bundle Fisher information matrix (FIM) for $\svec{X}$ carried by the five
row counts is
\begin{equation}
  F_{km}(\svec{X}) \;=\; \sum_{j=1}^{5} \frac{p_{jk}\,p_{jm}}{y_j},
  \label{eq:fim}
\end{equation}
a symmetric $3{\times}3$ matrix whose Cram\'er--Rao bound
$\sigma_k^2=[F^{-1}]_{kk}$ is the per-path variance floor for any
unbiased estimator; with equal $\Nzeroray{k}$ it reduces to Paper~1's
FIM.  Each reading informs a ray in proportion to the ray's detected
photons $p_{jk}$ times its share $p_{jk}/y_j$ of the reading, which is
why a dim ray in a bright sum is poorly measured
(Section~\ref{sec:n0fc:problem}).

\subsection{Dose inflation factor}
\label{sec:model:dif}

Paper~1 measures TCT's per-bundle cost by the \emph{Dose Inflation
Factor} (DIF), the squared ratio of an estimator's root-mean-square
error to $\sigfair(x)=e^{x/2}/\sqrt{3\Nzero}$, the equal-dose
single-source floor at $3\Nzero$ incident photons per ray
\citep{besson2026pmb}.  For an unbiased estimator it is the variance
ratio
\begin{equation}
  \DIF_k(X_1, X_2, X_3) \;=\;
  \frac{\sigma^2_{\mathrm{algo},k}(X_k)}{\sigma^2_{\mathrm{fair}}(X_k)}.
  \label{eq:dif_def}
\end{equation}
At the CRB both variances scale as $1/\Nzero$, so a DIF of $d$ means
that $d$ times the photons restore $\sigfair$.  Dose here is the
incident photon budget per ray, not absorbed dose; at equal attenuation
the CRB gives Paper~1's $7/3$ (endpoints) and $13/3$ (middle).  DIF is
computed per bundle from the per-ray FIM~\eqref{eq:fim} and pooled per
bin and path as in Paper~1, averaging squares before taking any ratio.
This gives an estimator's RMSE ratio
$\mathrm{RMSE}_b/\bar\sigma_{\mathrm{fair},b}$ and the realized floor
$\bar\sigma^{\mathrm{CRB}}_b/\bar\sigma_{\mathrm{fair},b}$; the square of
their ratio is the estimator's mean squared error over the mean CRB
variance of the same bundles.  Volumes are summarized by their median
middle-path floor (Section~\ref{sec:results:floor}).

Two extensions are needed.  Under the per-ray counts
$\Nzeroray{k}$ that $\NoFC$ introduces,
$\sigma^2_{\mathrm{fair}}(X_k)=e^{X_k}/(3\,\Nzeroray{k})$
(\suppsec{app:dif:crb}).  And the DIF factorizes as
\begin{equation}
  \DIF_k \;=\; \DIF_k^{\mathrm{struct}}\,\cdot\,
               \DIF_k^{\mathrm{contrast}}(X_1, X_2, X_3),
  \label{eq:dif_factor_main}
\end{equation}
where the structural factor ($7/3$, $13/3$, $7/3$ for $k=1,2,3$) is the
CRB value at equal detected fluences, fixed by the system matrix~$A$,
and the contrast factor is the remainder, set by the spread of the three
detected fluences.  Acquisition can change only the contrast factor
(Section~\ref{sec:n0fc}).

Because an estimator can go below the CRB by trading bias for variance,
bias is charged at the same rate, as in Paper~1:
\begin{equation}
  \DIF_k^{\mathrm{MSE}} \;=\;
  \frac{\sigma^2_{\mathrm{algo},k} + \mathrm{bias}^2_{\mathrm{algo},k}}
       {\sigma^2_{\mathrm{fair}}}.
  \label{eq:dif_mse}
\end{equation}
Every estimator DIF in this paper uses~\eqref{eq:dif_mse}, estimated by
averaging squared errors over many bundles, per bin or, for per-volume
summaries, within cells of near-constant floor (\suppsec{app:dif:single});
the floor is~\eqref{eq:dif_def} at the CRB.  Bins are unit-width in the
true middle-path line integral,
\begin{equation}
  \text{bin } b:\ X_2^{\mathrm{true}}\in[b-1,b),\ b=1,\dots,9;
  \qquad \text{bin }10:\ X_2^{\mathrm{true}}\ge9,
  \label{eq:x2bins}
\end{equation}
where $X_2=6$ is $0.25\%$ transmission, about $32$~cm of water; true
$X_2$ exceeds $9.2$ in four volumes (at most $0.08\%$ of a volume's
bundles; up to $11.5$ in LIDC-0766).  The pool RMSE of a set of bundles
is the count-weighted root-mean-square of its per-bin RMSEs.

\section{Data and simulation}
\label{sec:library}

\subsection{Patient corpus and TCT signature}
\label{sec:library:corpus}

The corpus comprises 26 chest CT volumes, 25 from LIDC-IDRI
\citep{armato2011} and PIS, a de-identified clinical volume whose scaled
form PIS\_NFLTD (``inflated'') is Paper~1's reference patient
\citep{besson2026pmb}.  They were selected to span, as measured by a
six-component signature ($\WED$, ellipticity, line-integral bin entropy
and 95th percentile, and mean and 95th-percentile intra-bundle
variance), the chest-anatomy variation that drives bundle statistics;
their $\WED$ \citep{mccollough2014} spans $[206.9, 381.2]$~mm
(\suppsec{app:split}).  The assortment suits the paper's per-bundle scope:
the $23$ non-test volumes hold $1.90\times10^9$ bundles per acquisition,
and $\NoFC$ brings the floor near $13/3$ in every WED stratum
(Table~\ref{tab:corpus-dif-stratum}).

\subsection{WED-stratified train/val/test split}
\label{sec:library:split}

The 26 volumes are split 19/4/3 patient-wise into train/val/test,
stratified by $\WED$.  Test holds PIS ($244.4$~mm), LIDC-0951
($206.9$~mm, below the training range) and LIDC-0116 ($334.5$~mm, the
only one with appreciable bin~9--10 population); validation holds
LIDC-0052, -0628, -0709 and -0766.  Supplementary section~\ref{app:split}
explains each placement; Section~\ref{sec:discussion:limitations} gives
the development choices that used validation or test patients.

\paragraph{Test cells.} The flat-fluence baseline has no bow-tie and a
constant tube current, with uniform per-ray $\Nzero=N_{\mathrm{base}}$
set per patient, as by automatic exposure control, for a target mean
detected count of $10^5$ per source ray (\suppsec{app:n0fc:nbase}).
The test volumes are the three test patients and PIS\_NFLTD, Paper~1's
patient (Section~\ref{subsec:paper1-floor}): PIS with its in-plane voxel
size scaled by $1.40$ (WED~$\approx342$~mm).  Acquired with PIS's
design (bow-tie, tube-current knots and $N_{\mathrm{base}}$), it detects
under the baseline $38\%$ of the target count that every other volume
meets: it is the out-of-distribution (OOD) probe, inside the training
range in anatomy (training WED reaches $381.2$~mm) but not in acquisition
(Section~\ref{sec:results:v8c-ood} also tests it at the target count);
the other volumes are in
distribution.  Each test volume is evaluated on every slice under the
baseline and $\NoFC$: eight test cells.

\subsection{Simulation, inversion, and training pipeline}
\label{sec:library:pipeline}

Stage~1 converts CT numbers to monoenergetic attenuation, with water at
the effective energy of a $120$~kVp beam (about $73$~keV):
$\mu=0.019\,(1+\mathrm{HU}/1000)$~mm$^{-1}$, clipped at zero.  It then
forward-projects the clinical reconstruction into line integrals with
ASTRA's \texttt{cyl\_cone\_vec} projector \citep{vanaarle2016}, adding
no noise.  The reconstruction's own noise, blur and artifacts remain in
these line integrals, which are the exact reference for every error
reported here.  The geometry is third-generation equiangular fan-beam,
used only to generate the line integrals, not to model the TCT gantry:
source-to-isocenter $500$~mm, source-to-detector $1000$~mm (curved
detector), $999$ channels over a $60^\circ$ fan, $500$~mm field of view
and $1024$ views per slice.  Bundle~$c$ ($c=1,\dots,333$) takes
channels $c$, $c+333$ and $c+666$, three rays $20^\circ$ apart that in
TCT reach one nominal detector cell from three source positions
(Figure~\ref{fig:geometry}); the middle ray passes within $87$~mm of the
isocenter and the endpoints $87$--$250$~mm from it, so $X_2$ is
usually the longest path and a bow-tie can dim the endpoints relative to
it.  Stage~2 applies the $\NoFC$ parameters
(Section~\ref{sec:n0fc:param}) and draws the counts
(\suppsec{app:snn2:deviates}); Stage~3 inverts every bundle with SNN2
(Section~\ref{sec:snn2}; supplementary Algorithm~\ref{alg:snn2}), whose
estimate V8c corrects (Stage~4) and T2bp post-processes
(Section~\ref{sec:t2bp}).  \emph{Deployable} means usable on a new
patient's $\NoFC$ scan without ground truth; the scan's tube-current
schedule is itself an oracle, fitted to the true attenuation
(Section~\ref{sec:n0fc:param}).

\section{SNN2: Gauss--Newton per-bundle inversion}
\label{sec:snn2}

SNN1, Paper~1's iteration \citep[Section~S2.3]{besson2026pmb}, updates
the rays one reading at a time and stalls where the middle ray supplies a
few percent of each reading (\suppsec{app:snn2}).  SNN2 updates the three
rays jointly, with a Gauss--Newton (Fisher-scoring) step on the Poisson
log-likelihood, and converges in about four iterations.

\subsection{Gauss--Newton update}
\label{sec:snn2:gn}

With $\Delta Y_j = Y_j - y_j$ the residual of reading~$j$, the score is
\begin{equation}
  s_k(\svec{X}) = -\sum_{j=1}^{5} p_{jk}\,\Delta Y_j / y_j,
  \label{eq:score}
\end{equation}
each residual weighted by the ray's share of the reading.  Each
iteration solves the $3{\times}3$ system $F\,\svec{\delta}=\svec{s}$ in
closed form, with $F$ evaluated at the current estimate, and keeps the
estimate in the box $0\le X_k\le X_{\max}=9.5$; the warm starts, the
step limit and the stopping rule are in supplementary
Algorithm~\ref{alg:snn2} and \suppsec{app:snn2:clamp}.  Under the
baseline, the bundles that reach the $20$-iteration ceiling (per-volume
median $5.3\times10^{-3}$) carry $3$--$39\%$ of SNN2's squared error;
under $\NoFC$ ($1.3\times10^{-7}$) they are a negligible tail.

\subsection{Converged estimate and its bias}
\label{sec:snn2:mle}

SNN2 converges to the per-bundle Poisson maximum-likelihood estimate
(MLE) where no box bound is active: its update is iteratively reweighted
least squares in the intensities (\suppsec{app:snn2:mle}).  Where a bound
is active, as for a coordinate in $14$--$59\%$ of the $X_2<6$ bundles,
the projected fixed point is not a Karush--Kuhn--Tucker point of the
box-constrained problem.  The active coordinate is almost always an
endpoint ray through air (\suppsec{app:snn2:mle}), and SNN2 holds it at
zero without re-fitting the middle path; Section~\ref{sec:results:dif}
compares an exactly solved box-constrained estimate, which re-fits it.
At high attenuation the logarithm of a small
Poisson count overestimates the line integral, for a single ray with $n$
detected photons on average by about $1/(2n)$.  The MLE's upward bias has
this origin and, to first order, falls as $1/\Nzero$ when the incident
counts are scaled together (\suppsec{app:biasstruct:shift}); it turns
negative only near the box limit, which caps the estimates
(\suppsec{app:snn2_residual:structure}).  Near an active bound the
unbiased CRB is not a valid floor, so \suppsec{app:t2bp:sensitivity}
reports the near-boundary and truth-interior bundles separately.

\subsection{Second-order bias correction}
\label{sec:snn2:bias}

The upward bias is second order in the noise (Jensen's inequality), so
where the noise is small a one-shot second-order correction removes part
of it.  A smooth taper confines it to where the largest per-path CRB
standard deviation is below about $0.15$, at low and moderate attenuation
and, under $\NoFC$, higher bins; beyond, it would overcorrect
(\suppsec{app:snn2:bias}).  All SNN2 results include it; beyond the
taper, estimates are uncorrected.  Paper~1's reference runs without it
\citep{besson2026pmb}; Section~\ref{sec:results:dif} reports the
box-constrained estimate both ways.

\section{$\NoFC$: bow-tie and TCM co-design}
\label{sec:n0fc}

SNN2 approaches the CRB, but no estimator lowers the CRB itself; only
the acquisition can.  In single-source CT a bow-tie and tube-current
modulation improve a scanner that would work without them: they serve
dose and noise objectives and limit the detector's dynamic range
(Section~\ref{sec:disc:prior}).  In TCT the same hardware, designed for
TCT, is a prerequisite.  Each reading sums up to three rays, so under
uniform flux the bright rays' photons
swamp the dim ray's information (Section~\ref{sec:n0fc:problem}), and
the per-volume median middle-path floor is a DIF of $23$--$48$; $\NoFC$
brings it to $3.3$--$6.3$ (Section~\ref{sec:results:floor}).  Even a
generous ceiling of three times a single-source scan's photons (a DIF of
$3$, Section~\ref{sec:model:dif}), justifiable only by a substantial
diagnostic gain, is exceeded $8$--$16$-fold under flat fluence and
$1.1$--$2.1$-fold under $\NoFC$: flux compensation is necessary, not
sufficient, and whether the rest can be closed is Paper~3's question
(Section~\ref{sec:intro:scope}).  It
redistributes the per-ray counts $\Nzeroray{k}$ to approximately equalize each
bundle's detected fluence (details in \suppsec{app:n0fc}).

\subsection{The intra-bundle dynamic-range problem}
\label{sec:n0fc:problem}

Each bundle samples five row sums of up to three exponentially attenuated
fluences.  Where the line integrals differ, as in real anatomy, the
detected fluences $\Phi_k = \Nzeroray{k}\,e^{-X_k}$ span orders of
magnitude under uniform $\Nzero$, and the dim ray is measured like a
primary beam under heavy scatter.  A reading carries
$\Phi_{\mathrm{dim}}^2/y_j=\Phi_{\mathrm{dim}}/(1+\Phi_{\mathrm{bright}}/\Phi_{\mathrm{dim}})$
of Fisher information about it (two rays, the bright one known): its
own photons' worth divided by one plus the bright-to-dim ratio, exactly
as known scatter divides it by one plus the scatter-to-primary ratio.
At
$(X_1,X_2,X_3)=(3,6,3)$ under uniform $\Nzero$ the middle ray supplies
under $5\%$ of each reading it enters (\suppsec{sec:snn2:snn1}).  The
factorization~\eqref{eq:dif_factor_main} isolates the effect: under
uniform $\Nzero$ the contrast penalty $\DIF_2^{\mathrm{contrast}}$, which
multiplies the structural $13/3$, has a per-volume median of
$5.4$--$11.1$ (the baseline floors of Section~\ref{sec:results:floor}
divided by $13/3$).  Equalizing $\Phi_k$ within each bundle through
$\Nzeroray{k}$ designs it away.  Equal detected fluence is also a
single-source objective, realized by fluence-field modulation
\citep{szczykutowicz2013,szczykutowicz2014}, though not the
variance-optimal one there, which for independent rays follows the
square-root law $N_i\propto e^{X_i/2}$ \citep{gies1999}; equalization
needs $\Nzeroray{k}\propto e^{X_k}$, twice the exponent
(\suppsec{app:n0fc:profiles}).

The ideal prescription $\Nzeroray{k}\propto e^{+X_k}$ would make $\Phi_k$
uniform and $\DIF_k^{\mathrm{contrast}}$ unity, but it needs per-bundle
tube-current ratios up to about $665{\times}$, beyond the per-rotation
range of our schedules (at most $89{\times}$; \suppsec{app:n0fc:tcm}).
$\NoFC$ therefore multiplies two mechanisms: a global hardware bow-tie
$B(\gamma)$ that passively pre-attenuates bright peripheral rays, and a
per-(patient, slice) tube-current schedule $\tau(\theta)$
(\suppsec{app:n0fc:theory}).

Equal detected fluence is a tractable surrogate for, not a minimizer of, the per-bundle variance. Over the corpus it comes
within a median
$2.2\%$ (mean $7.0\%$) of the allocation that minimizes the summed CRB
of the three paths, and changing the design target does not pay
(\suppsec{app:n0fc:opt}).

\subsection{Parameterization and fitting}
\label{sec:n0fc:param}

The $\NoFC$ per-ray fluence factorizes as
\begin{equation}
  \Nzeroray{k} \;=\; N_{\mathrm{base}}\,
  B(\gamma_k)\,\exp\!\bigl(\tau(\theta_k)\bigr),
  \label{eq:n0fc}
\end{equation}
with $N_{\mathrm{base}}$ a scalar normalization, $B(\gamma)$ a global
hardware bow-tie acting on ray~$k$'s fan angle $\gamma_k$, and
$\tau(\theta)$ a per-(patient, slice) log-modulation of the tube current
at the angle $\theta_k$ of the ray's source position: a bow-tie is one
physical filter, while tube current is controllable per view.  The
bow-tie $B^*$ is fitted once, on reference slices of three non-test
patients, so no test-pool data enters it.  The TCM $\tau^*$ is then
fitted to each (patient, slice), test patients included, from that
patient's own attenuation: a closed-loop oracle design
\citep{gies1999,hsieh2015}, optimal for equalization, not for the CRB.
Its oracle status matters little: the median floor comes from the
bow-tie, and schedules not fitted to the slice move it by $-4$ to $+3\%$
(Section~\ref{sec:results:floor}).
Both fits minimize the within-row spread of $\ln(\Nzeroray{k}e^{-X_k})$
over the multi-ray rows, with incident counts clipped to $[10^3,10^6]$
photons per ray (\suppsec{app:n0fc}).  The baseline and $\NoFC$ share
$N_{\mathrm{base}}$, so $\NoFC$ delivers fewer photons: $35$--$62\%$ of
the baseline's incident photons on rays crossing the patient, per corpus
volume (median $43\%$), and $38$--$69\%$ of the photons that interact
there, a closer proxy for absorbed dose (\suppsec{app:n0fc:nbase}).  DIF,
normalized per ray, does not depend on this total, but the RMSE
comparisons between the regimes give the baseline about twice $\NoFC$'s
incident photons.

\subsection{Effect on the Cram\'er--Rao floor}
\label{sec:n0fc:fisher}

\begin{figure}[!t]
\centering
\includegraphics[width=\linewidth]{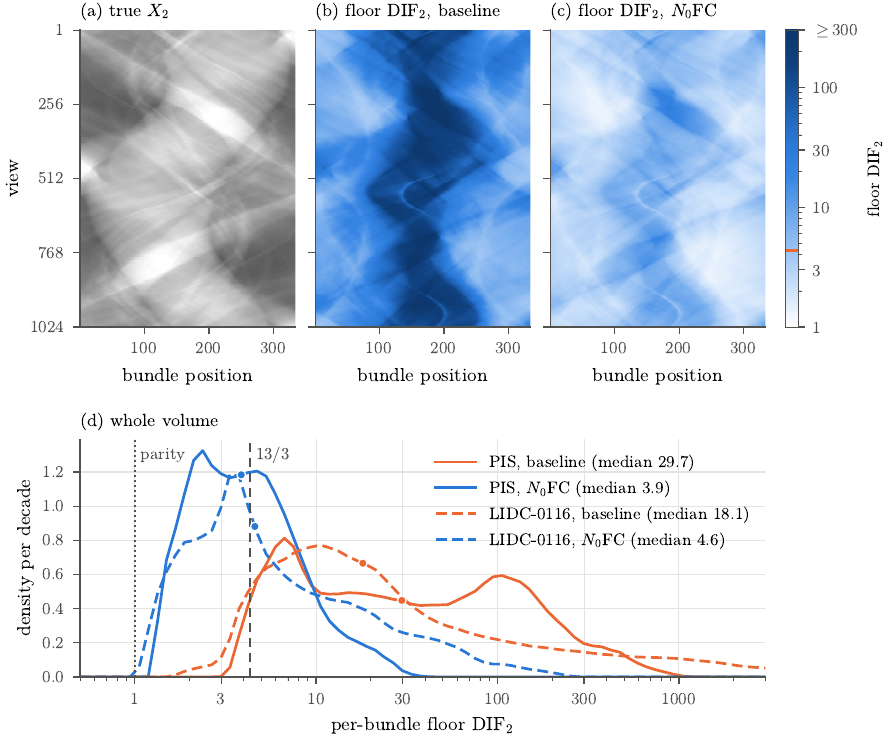}
\caption{What $\NoFC$ does to the per-bundle middle-path floor.  Top: one
PIS slice (slice $100$ of $200$; $1024$ views $\times$ $333$ bundles).
(a)~True $X_2$.  (b,~c)~Floor
$\DIF_2=[F^{-1}]_{22}/\sigma^2_{\mathrm{fair}}(X_2)$ ($F$ at the true line
integrals) under the baseline and under $\NoFC$, on a common log scale
(orange tick: $13/3$).  (d)~The per-bundle floor over the whole volume
for PIS and LIDC-0116, as density per decade of $\DIF_2$; dots: medians.
No estimator is involved.}
\label{fig:dif-maps}
\end{figure}

With~\eqref{eq:n0fc}, the per-ray FIM~\eqref{eq:fim} gives the CRB
exactly for any fluence pattern.  With $q_j=\min_k\Phi_k/\max_k\Phi_k$
the detected-fluence ratio of a multi-ray row, $\DIF_2$ is $13/3$ at
equalization ($q_j=1$), rises somewhat more slowly than $1/q$ when the
middle ray is the dimmest, as it typically is at high attenuation under
uniform $\Nzero$, and falls when it is the brightest (supplementary
Table~\ref{tab:dif_pattern}).  The endpoint floors rise modestly in
exchange (per-patient median $\DIF$ from $1.95$--$2.09$ to
$2.24$--$2.51$), though their standard deviations rise
$1.3$--$16.5$-fold (\suppsec{app:n0fc:emp:test}): DIF is relative to each
ray's own incident count $\Nzeroray{k}$ (Section~\ref{sec:model:dif}), and
the bow-tie removes most of the endpoints' photons.
Figure~\ref{fig:dif-maps} shows $\NoFC$ lowering the floor almost
everywhere on one test patient, most where a bright endpoint dilutes the
middle path.  Over the whole PIS volume the median $\DIF_2$ falls from
$29.7$ to $3.9$, and the share of bundles below $13/3$ rises from $3\%$
to $56\%$ (LIDC-0116: $18.1\to4.6$; $7\%\to48\%$).

\section{V8c: multi-patient neural-network residual corrector}
\label{sec:v8c}

\label{sec:library:v8c}%
\label{sec:library:train}%
V8c is a fully connected residual network ($\sim3.0$M parameters) that
corrects SNN2's three estimates from $26$ inputs describing only the
bundle being corrected, unlike T2bp's prior: its five normalized
readings and their logarithms, a count level ($\log_{10}$ of the
bundle-mean incident count), SNN2's estimates, and Fisher-information
features at them (\suppsec{app:v8c:arch}).  One network is trained per
acquisition, once and with one seed, on SNN2's residuals over the
$1.58\times10^{9}$ bundles of the $19$ training patients, with a robust
loss weighting the middle path four times each endpoint
(supplementary sections~\ref{app:v8c:loss} and~\ref{app:v8c:train}).
The checkpoint, chosen on the four validation patients by the
middle-path standard deviation at $X_2\in[5,8)$, is epoch~3 (baseline)
and epoch~1 ($\NoFC$), the ``val-best'' of
Section~\ref{sec:results}.  Under $\NoFC$ the validation score worsens
after epoch~1 while the training loss keeps falling, consistent with the
small residual SNN2 leaves there (Section~\ref{sec:results:snn2}),
though the robust loss and the sparse high-attenuation tail, untested,
may contribute.

\section{T2bp: per-patient empirical-Bayes post-processor}
\label{sec:t2bp}

Unlike V8c, T2bp builds its prior \emph{from the test patient's own scan}.
Where few photons survive, a bundle's counts fit a wide range of
middle-path values; T2bp replaces SNN2's estimate by their average, each
weighted by its Poisson likelihood and by how common it is among this
patient's bundles with a similar endpoint path.  This plug-in,
early-stopped empirical Bayes \citep{robbins1956,efron2011} uses no
trained weights and reads no ground truth when applied, so it cannot
inherit a mismatched training distribution; it depends instead on its
\emph{base estimator}, the one whose estimates it refines (SNN2, V8c or
the box-constrained estimate; Section~\ref{sec:results:composition};
\suppsec{app:t2bp}).

\subsection{Per-patient prior}
\label{sec:t2bp:prior}

The prior is the smoothed histogram of the SNN2 estimates of the
patient's $N_b \sim 10^7\text{--}10^8$ bundles, conditioned on the
endpoint estimate:
\begin{equation}
  p(X_2 \mid \Xhat_3)
  \;\propto\; \sum_{n=1}^{N_b}
      K_{h_3}\!(\Xhat_3 - \Xhat_{n,3}^{\mathrm{SNN2}})
      K_{h_2(X_2)}\!(X_2 - \Xhat_{n,2}^{\mathrm{SNN2}})
\label{eq:t2bp-prior}
\end{equation}
with $K_h$ a Gaussian kernel, normalized over $X_2$ for each $\Xhat_3$.
The endpoint $X_3$, measured alone in reading $Y_5$, has a precise SNN2
estimate (standard deviation at most $1.2$ times its CRB on the test
patients, \suppsec{app:n0fc:emp:test}); conditioning on it
recovers from the patient's own bundles what V8c learns from the corpus:
how deep the middle ray runs given its neighbor.  The widths are
$h_3=0.15$ and $h_2=\max\bigl(0.02,\,0.5\,\sigma_L(X_2)\bigr)$, with
$\sigma_L(X_2)$ the median CRB width of bundles whose SNN2 estimate is
near $X_2$, computed without ground truth
(\suppsec{app:t2bp:deploy}).  Fixed at design, the three constants
matter little: halving or doubling each moves the $X_2<6$ error by under
$1\%$, except the $0.5$ on two baseline cells, where halving it lowers the
error by $13$--$19\%$ and doubling it raises it by $28$--$38\%$ (\suppsec{app:t2bp:sensitivity}, with the prior's
self-influence).
\label{sec:t2bp:bandwidth}

\subsection{Posterior mean and iteration}
\label{sec:t2bp:posterior}

\begin{figure}[!t]
\centering
\includegraphics[width=\linewidth]{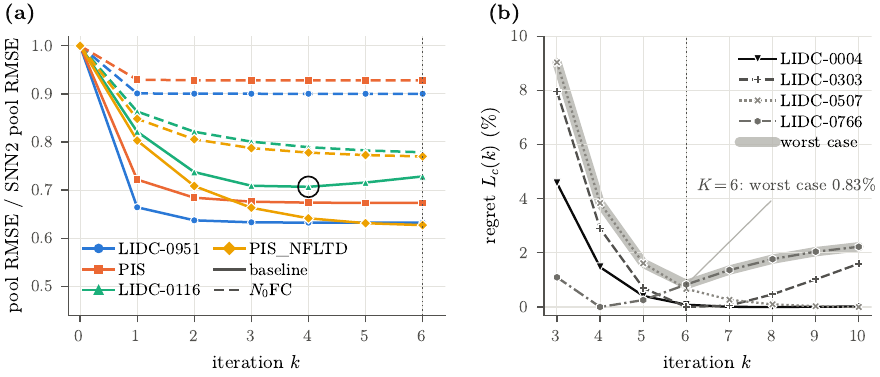}
\caption{The T2bp iteration and the choice of $K$.  (a)~Pool $X_2$ RMSE of
T2bp$\,\circ\,$SNN2 after $k$ iterations, relative to SNN2-alone ($k=0$),
for the eight test cells (solid: baseline; dashed: $\NoFC$).  All
decrease monotonically except LIDC-0116 under the baseline, lowest at
$k=4$ (ring; \suppsec{app:t2bp:deploy}).  (b)~Choosing $K$ on four corpus
patients, three from training and LIDC-0766 from validation (baseline,
every 16th slice): each one's regret
$L_c(k)=100\,[R_c(k)-\min_j R_c(j)]/\min_j R_c(j)$, with $R_c(k)$ its pool
RMSE after $k$ iterations, and the worst case $\max_c L_c(k)$ (gray
band), which $K=6$ minimizes ($0.83\%$; rule in \suppsec{app:t2bp:deploy}).
Regrets at $k=1,2$ ($6$--$51\%$) are off scale.  The rule
uses only these patients but was written after the per-iteration
test-cell errors had been inspected (Section~\ref{sec:t2bp:posterior}).}
\label{fig:t2bp-iter}
\end{figure}

Each bundle's T2bp estimate is the posterior mean
\begin{equation}
  \Xhat_2^{\mathrm{T2bp}} =
    \frac{\sum_i x_i\, p_i\, w_i}{\sum_i p_i\, w_i},
  \qquad
  p_i = L(Y \mid x_i)\, p(x_i \mid \Xhat_3),
\label{eq:t2bp-posterior}
\end{equation}
where $L(Y\mid x_i)$ is the bundle's Poisson likelihood with the
endpoints at their SNN2 values, on a likelihood-adapted grid with
trapezoidal weights $w_i$ (\suppsec{app:t2bp:grid}).  The refined
estimates rebuild the prior, and the posterior is recomputed.  Posterior
means are less spread than the true values \citep{louis1984}, so each
pass narrows the prior; where the likelihood width is comparable to the
anatomical spread, the error passes through a minimum and then rises
through bias (Figure~\ref{fig:t2bp-iter}), so T2bp is stopped early, as
maximum-likelihood expectation-maximization (MLEM) reconstruction is
\citep{barrett1994}.  We run a fixed $K=6$ iterations, chosen
on four corpus patients selected for high error or slow convergence; the
rule, written after the per-iteration test-cell errors had been
inspected, did not select the count they favor
(\suppsec{app:t2bp:deploy}).  On the eight test cells, these truth-free
settings change pool RMSE by $-1.7$ to $+2.5\%$
against an earlier implementation whose width and stop used the ground
truth.

\section{Results}
\label{sec:results}

\subsection{Estimators and metric}
\label{sec:results:deployable}

The estimators are SNN2-alone (Section~\ref{sec:snn2}), V8c-alone
(Section~\ref{sec:v8c}), T2bp$\,\circ\,$SNN2 and T2bp$\,\circ\,$V8c
(Section~\ref{sec:t2bp}; $K$ was chosen on the SNN2 base). On the test
cells the metric is the pool $X_2$ RMSE, with bias and variance per
unit-width bin~\eqref{eq:x2bins}; Section~\ref{sec:results:dif}
computes per-volume DIFs from the T2bp$\,\circ\,$SNN2 posteriors.

\subsection{The floor across the corpus}
\label{sec:results:floor}

\begin{figure}[!t]
\centering
\includegraphics[width=\linewidth]{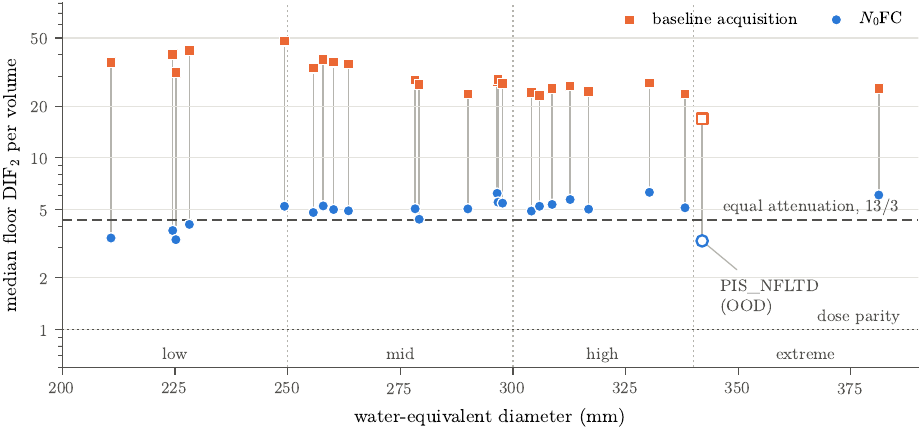}
\caption{Per-volume median middle-path floor $\DIF_2$ against WED for the
$23$ non-test volumes, under the baseline (squares) and
$\NoFC$ (circles); segments join each patient's two values.  Open
markers: PIS\_NFLTD.  Thin vertical lines: the WED strata of
Table~\ref{tab:corpus-dif-stratum}.  Dashed: $13/3$; dotted: dose parity.
Ranges $23.2$--$48.0$ (baseline) and $3.33$--$6.30$ ($\NoFC$), a
$4.2$--$10.6$-fold gain (median $5.2$); PIS\_NFLTD $16.9\to3.28$.
Analytic floor at the true line integrals; no estimator is involved.
Per-patient gains: supplementary Fig.~\ref{fig:n0fc_floor_gain}.}
\label{fig:floor-wed}
\end{figure}

The acquisition alone sets the floor. For the $23$ non-test volumes
(Section~\ref{sec:library:split}) and PIS\_NFLTD, it is each
volume's median middle-path $\DIF_2$ over its bundles, exact from the
FIM at the true line integrals (Table~\ref{tab:corpus-dif-stratum},
Figure~\ref{fig:floor-wed}).

\begin{table}[!t]
\renewcommand{\arraystretch}{1.05}
\caption{Middle-path floor $\DIF_2$ by WED stratum ($23$ non-test
volumes; PIS\_NFLTD separately): per-volume medians, baseline and $\NoFC$,
and the per-volume gain, each a median over the stratum (so the gain
need not equal the ratio of the two floors); the MSE of SNN2 and
T2bp$\,\circ\,$SNN2 over the floor under $\NoFC$ for bundles with $X_2<6$ (about $83\%$ of bundles, $25\%$ to
nearly $100\%$ per volume; pooled over the stratum,
Section~\ref{sec:model:dif}); and T2bp$\,\circ\,$SNN2's RMSE reduction
from baseline to $\NoFC$ (stratum median).  The high stratum's SNN2
value includes LIDC-0766, at $1.69$ alone (Section~\ref{sec:results:dif}).
The baseline uses about twice
$\NoFC$'s incident photons (Section~\ref{sec:n0fc:param}).}
\label{tab:corpus-dif-stratum}
\centering
\footnotesize
\setlength{\tabcolsep}{4pt}
\begin{tabular}{lccccccc}
\toprule
 & & \multicolumn{3}{c}{CRB floor, median $\DIF_2$} & \multicolumn{2}{c}{MSE / floor} & RMSE \\
\cmidrule(lr){3-5}\cmidrule(lr){6-7}
WED stratum & $n$ & baseline & $\NoFC$ & gain & SNN2 & T2bp & reduction \\
\midrule
low ($<250$\,mm)    & 5  & 40.1 & 3.77 & 10.3 & 1.01 & 0.77 & $76\%$ \\
mid ($250$--$300$)  & 10 & 28.5 & 5.04 & 5.9  & 1.05 & 0.92 & $68\%$ \\
high ($300$--$340$) & 7  & 24.4 & 5.22 & 4.6  & 1.28 & 1.03 & $60\%$ \\
extreme ($\geq340$) & 1  & 25.5 & 6.07 & 4.2  & 1.05 & 0.97 & $45\%$ \\
\midrule
PIS\_NFLTD (OOD)    & 1  & 16.9 & 3.28 & 5.1  & 1.03 & 1.03 & $46\%$ \\
\bottomrule
\end{tabular}
\end{table}

\paragraph{The $\NoFC$ headline.}
Under the flat-fluence baseline the per-volume median floor is
$\DIF_2=23$--$48$ across the corpus patients. The $\NoFC$ co-design
(Section~\ref{sec:n0fc}) lowers it to $3.3$--$6.3$ for all $23$
non-test volumes (median $5.0$), a $4.2$--$10.6$-fold reduction
(median $5.2$), largest for the smallest patients (\suppsec{app:dif},
Fig.~\ref{fig:n0fc_floor_gain}). The range brackets the
equal-attenuation value $13/3\approx4.33$: equalizing the detected
fluence within each bundle removes most of the contrast factor of
Section~\ref{sec:model:dif}, leaving about the structural factor of the
five-row measurement. These floors use the oracle TCM of
Section~\ref{sec:n0fc}, fitted to each slice's own attenuation (PIS's
knots for PIS\_NFLTD), but the median comes from the bow-tie. Taking the
TCM from up to ten slices away or using one angular profile per patient
moves the per-volume median floor by $-4$ to $+3\%$ over all $24$
volumes, and dropping it lowers the median by $0.5$--$16\%$, while the
mean floor rises by up to $33\%$ (\suppsec{app:n0fc:tcm}): the TCM acts
on the tail.

\begin{figure}[!t]
\centering
\includegraphics[width=\linewidth]{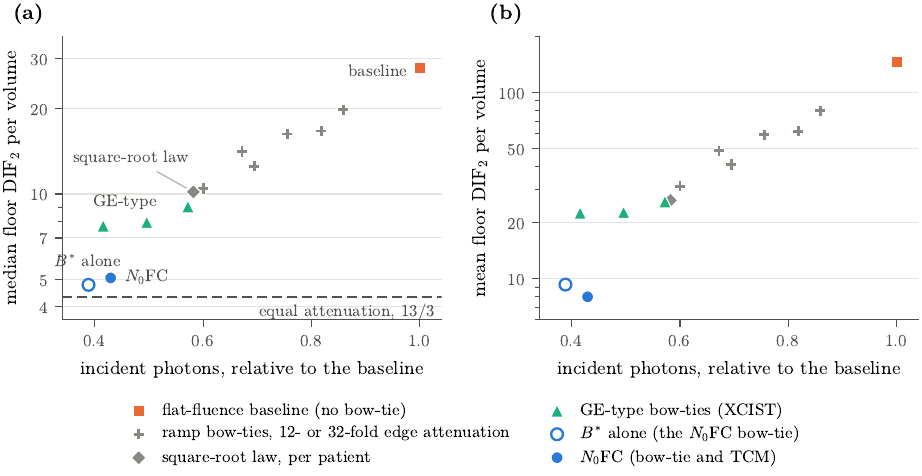}
\caption{Middle-path floor $\DIF_2$ against incident photons for the
designs of supplementary Table~\ref{tab:bowtie_compare}, on the $23$
non-test volumes ($24$ sampled slices each): (a)~each
volume's median over its bundles, (b)~its mean; both coordinates are
medians over the volumes.  Photons: on rays through the patient,
relative to the flat-fluence baseline (square).
Plus signs: ramp bow-ties with $12$- or $32$-fold edge attenuation;
diamond: a per-patient square-root-law bow-tie; triangles: XCIST's three
GE-type profiles at $120$~kVp \citep{wu2022}; circles: $\NoFC$ (filled)
and its bow-tie $B^*$ alone (open).  All use the same reference tube
current; only $\NoFC$ modulates it.  GE-type over $\NoFC$ median floor:
$1.44$--$1.72$ (median over volumes; $1.20$--$2.29$ per volume).  Dashed:
$13/3$.}
\label{fig:bowtie-designs}
\end{figure}

\paragraph{Against a conventional bow-tie.}
Clinical scanners, unlike the baseline, use a bow-tie. The XCIST
simulator's three GE-type profiles \citep{wu2022}, converted to
transmission at $120$~kVp, lower the corpus median floor from $27.9$ to
$7.7$--$9.0$ at the same reference tube current and a similar photon
count: two-thirds to three-quarters of $\NoFC$'s reduction on a log
scale. $\NoFC$'s floor is a further $1.4$--$1.7$-fold lower (median over volumes),
almost all from its deeper bow-tie (Figure~\ref{fig:bowtie-designs};
supplementary Table~\ref{tab:bowtie_compare}). For equal median
precision the GE-type bow-ties need $0.16$--$0.19$ of the baseline's
incident photons, and $\NoFC$ $0.11$.

\paragraph{The high-attenuation tail.}
The per-volume \emph{median} sits near $13/3$, but the distribution is
right-skewed: under $\NoFC$ the per-volume mean floor is $5.2$--$16.0$
(\suppsec{app:dif:empirical}). The corpus median of the per-volume mean
falls $18$-fold, from $145$ to $8.0$, against $5.5$-fold for the median
($27.9$ to $5.1$; supplementary Table~\ref{tab:bowtie_compare}).
The tail matters more for images: each ray's variance is its floor
times $\sigfair^2\propto e^{X}$ (Section~\ref{sec:model:dif}), and for
equal mean CRB variance over the rays through the patient $\NoFC$ needs
$0.02$ of the baseline's incident photons, against $0.10$ for a
per-patient square-root-law bow-tie (\suppsec{app:n0fc:profiles}).

\subsection{SNN2 against the CRB and the effect of $\NoFC$}
\label{sec:results:snn2}

\label{sec:n0fc:empirical}
SNN2 reads each ray's $\Nzeroray{k}$ (Section~\ref{sec:model:forward}),
so its changes between acquisitions are the acquisition's alone.

At the true line integrals of the four validation patients, $\sigCRB_2$ falls
$1.7$--$1.9$-fold at bins~4--5, where the middle ray gets $0.76$--$0.88$
of the baseline's photons, and $12$--$14$-fold at bins~9--10, where it
gets $2.8$--$3.0$ times them, so extra photons explain part of that
fall. Across bins~6--10, SNN2's $\mathrm{SD}_2$ falls $3$--$3.7$-fold and
$|\mathrm{Bias}_2|$ $5$--$19$-fold (\suppsec{app:n0fc:emp:val}). Its
ratio of standard deviation to the CRB at its own estimates (near $1$ if
efficient and unbiased) is descriptive only, as near-singular Fisher
matrices inflate that CRB in photon-starved bins (supplementary
Table~\ref{tab:t2bp:sigmaL}). At $X_2\in[7,8)$ it moves from $0.31$ to
$1.12$; a dose-reading check shows independently that the baseline's
box $0\le X_k\le9.5$ acts there as a prior and biases the estimate
(\suppsec{app:dif:dose}); under $\NoFC$ the estimate there is nearly
efficient.

\paragraph{Test patients.}
Pooled over the three test patients ($\WED$ 207--335\,mm), the
middle-path RMSE falls from $0.240$ to $0.079$ (supplementary
Table~\ref{tab:2x2_pool} and \suppsec{app:n0fc:emp:test}). On the hardest bin, LIDC-0116's bin~10
($77{,}000$ bundles, $X_2\!\ge\!9$), where $\NoFC$ gives the middle ray
$2.2$ times the baseline's incident photons, $\mathrm{RMSE}_2$ falls
$73\%$ ($1.46\!\to\!0.40$), with a $12.6\times$ bias collapse
($0.76\!\to\!0.060$).

Under $\NoFC$ the SNN2 residual that V8c is trained to correct is
$3$--$5$ times smaller at high attenuation
(\suppsec{app:snn2_residual:structure}): V8c has less
to correct.

\subsection{V8c in and out of distribution}
\label{sec:results:v8c-all}

\subsubsection{In distribution.}
\label{sec:results:v8c}

At its val-best checkpoint on the three test patients, V8c's pool $X_2$
RMSE falls from baseline to $\NoFC$ by $4.8\times$ (LIDC-0951),
$3.9\times$ (PIS) and $1.7\times$ (LIDC-0116), contracting with WED as
SNN2's residual does (Section~\ref{sec:results:snn2}). Under $\NoFC$ it
gains at most $9\%$ over SNN2 (LIDC-0951) and loses $14\%$ on LIDC-0116
(Table~\ref{tab:composition-10cell}), where it is biased low at
$X_2\ge8$ ($-0.39$ and $-0.64$; SNN2 within $0.06$), in two bins holding
$7\%$ of the bundles and most of the excess error. A V8c trained under
the baseline does not transfer to $\NoFC$ data (\suppsec{app:v8c:cross}).

\subsubsection{Out of distribution.}
\label{sec:results:v8c-ood}

\begin{figure}[!t]
\centering
\includegraphics[width=\linewidth]{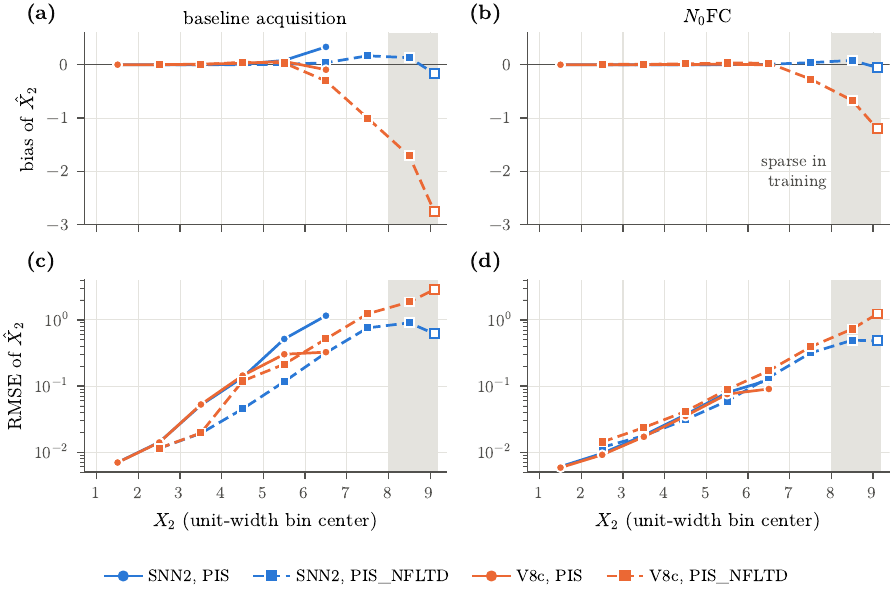}
\caption{V8c's out-of-distribution failure.  Bias (a,~b) and RMSE (c,~d)
per $X_2$ bin for SNN2-alone (blue) and V8c-alone (orange) on
PIS (circles, solid) and PIS\_NFLTD (squares, dashed: the same anatomy at
$1.40\times$ in-plane spacing, acquired at PIS's photon level), under
baseline (a,~c) and $\NoFC$ (b,~d) acquisition.  Shaded: $X_2\ge8$
($0.26\%$ of the training pool, $0.78\%$ of PIS\_NFLTD).  Open markers:
fewer than $10^4$ bundles.  Values: \suppsec{app:v8c:ood}.}
\label{fig:v8c-ood}
\end{figure}

On the OOD probe PIS\_NFLTD under the baseline, V8c-alone's pool $X_2$
RMSE is $0.471$ against SNN2-alone's $0.279$ on the same bundles: the
corpus-trained network is \emph{worse} than the estimator it was built
to refine. Figure~\ref{fig:v8c-ood} decomposes the failure per bin, and
\suppsec{app:v8c:ood} shows that unseen attenuation alone does not
explain it.

Under the baseline, the acquisition explains most of the failure.
PIS\_NFLTD's $N_{\mathrm{base}}$ was copied from PIS ($229{,}266$), not
set by the corpus rule ($602{,}461$). Regenerating its test bundles at
the rule's value (recalibration) lowers the V8c-alone/SNN2-alone RMSE
ratio from $1.69$ to $0.85$, against $0.70$--$0.78$ on the
in-distribution test patients, and removes V8c's deep bias
(\suppsec{app:biasstruct:e2}). Changing only V8c's count-level input
reproduces most of that change at the lower counts and part of it at the
higher; how the input produces the failure is not resolved
(\suppsec{app:biasstruct:e3}).  The failure
is not specific to low counts: on PIS acquired at $602{,}461$, $2.6$
times its own rule value, V8c-alone is also worse than SNN2-alone (ratio
$1.46$, against $0.73$ at PIS's own value; supplementary
Table~\ref{tab:tests}).

Under $\NoFC$ V8c's RMSE on PIS\_NFLTD falls to $0.159$, but its bias
keeps its signs and structure; recalibration moves the ratio only from
$1.30$ to $1.22$, against $1.14$ on LIDC-0116, nearest in size.

\subsection{T2bp composition: T2bp$\,\circ\,$SNN2 vs T2bp$\,\circ\,$V8c}
\label{sec:results:composition}

\begin{figure}[!t]
\centering
\includegraphics[width=\linewidth]{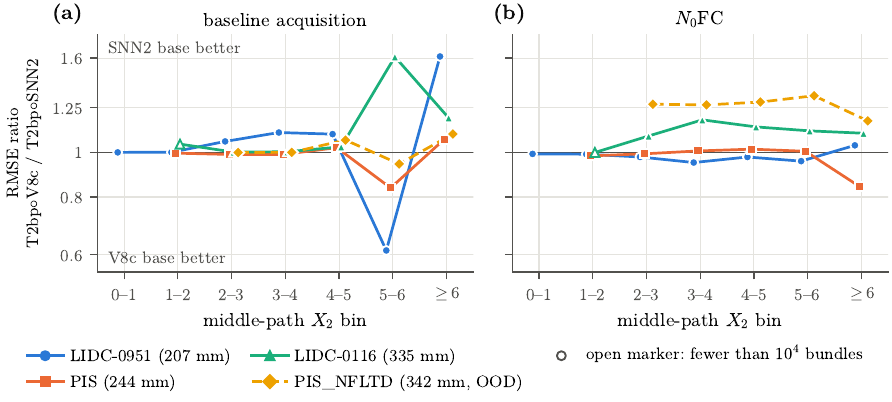}
\caption{Middle-path RMSE ratio, T2bp$\,\circ\,$V8c over T2bp$\,\circ\,$SNN2,
per unit-width $X_2$ bin on the four test volumes of
Table~\ref{tab:composition-10cell}, after the fixed $K=6$ iterations, under
(a) baseline and (b) $\NoFC$ acquisition; above~$1$ T2bp$\,\circ\,$SNN2 is
better, and $X_2\geq6$ is one bin.  Under $\NoFC$ T2bp$\,\circ\,$V8c
is $8$--$33\%$ worse in every populated bin of LIDC-0116 and PIS\_NFLTD,
and within $5\%$ on LIDC-0951 and PIS below $X_2=6$; under the baseline
its leads on PIS and LIDC-0951 come from $X_2\in[5,6)$ (ratios $0.84$
and $0.61$).}
\label{fig:composition-bins}
\end{figure}

Table~\ref{tab:composition-10cell} and Figure~\ref{fig:composition-bins}
compare T2bp$\,\circ\,$SNN2 with T2bp$\,\circ\,$V8c on the four test
volumes, isolating what V8c's learned correction of SNN2 adds.  A cell
is \emph{tied} when a paired bootstrap's $95\%$ interval over contiguous
$8$-slice blocks includes zero (\suppsec{app:valcomp}).

\begin{table}[t]
\centering
\caption{Pool $X_2$ RMSE of the six estimators on the eight test cells,
whose volumes were also used in development (Table~\ref{tab:dev-provenance}).
Box MLE: the exactly solved box-constrained estimate of
Section~\ref{sec:results:dif}, uncorrected.
All six are evaluated on all slices of each test volume;
T2bp values are taken after the fixed $K=6$ iterations of
Section~\ref{sec:t2bp:posterior}, which use no ground truth.  The baseline
acquisition uses about twice $\NoFC$'s incident photons
(Section~\ref{sec:n0fc:param}).  Winner: $\star$ (T2bp$\,\circ\,$SNN2) or
$\bullet$ (T2bp$\,\circ\,$V8c) when the paired bootstrap's $95\%$ interval
excludes zero, otherwise tied (supplementary Table~\ref{tab:tccomp}); the
box-constrained base estimator in place of SNN2 gives the same verdicts
(supplementary Table~\ref{tab:kktcomp}).}
\label{tab:composition-10cell}
\small
\setlength{\tabcolsep}{4pt}
\begin{tabular}{llrrrrrrl}
\toprule
 & & \multicolumn{3}{c}{Alone} & \multicolumn{3}{c}{T2bp$\,\circ\,$base estimator} & \\
\cmidrule(lr){3-5}\cmidrule(lr){6-8}
Patient & Regime & SNN2 & Box MLE & V8c & SNN2 & Box MLE & V8c & Winner \\
\midrule
\multicolumn{9}{l}{\textit{Flat-fluence baseline (reference)}} \\
PIS         & baseline & $0.158$ & $0.144$ & $0.115$ & $0.1065$ & $0.1052$ & $0.0992$ & $\bullet$ \\
LIDC-0951   & baseline & $0.168$ & $0.143$ & $0.118$ & $0.1064$ & $0.1035$ & $0.1016$ & tied \\
LIDC-0116   & baseline & $0.415$ & $0.415$ & $0.325$ & $0.302$ & $0.311$ & $0.364$ & $\star$ \\
PIS\_NFLTD  & baseline & $0.279$ & $0.279$ & $0.471$ & $0.175$ & $0.175$ & $0.188$ & tied \\
\midrule
\multicolumn{9}{l}{\textit{$\NoFC$ regime (oracle TCM)}} \\
PIS         & $\NoFC$     & $0.0312$ & $0.0292$ & $0.0296$ & $0.0290$ & $0.0289$ & $0.0292$ & $\star$ \\
LIDC-0951   & $\NoFC$     & $0.0268$ & $0.0243$ & $0.0244$ & $0.0241$ & $0.0241$ & $0.0234$ & $\bullet$ \\
LIDC-0116   & $\NoFC$     & $0.1637$ & $0.1591$ & $0.1865$ & $0.127$ & $0.127$ & $0.140$ & $\star$ \\
PIS\_NFLTD  & $\NoFC$     & $0.123$ & $0.121$ & $0.159$ & $0.0945$ & $0.0943$ & $0.113$ & $\star$ \\
\bottomrule
\end{tabular}
\end{table}

\paragraph{On the $\NoFC$ test cells the SNN2 base is the stronger partner.}
On the four $\NoFC$ cells, T2bp$\,\circ\,$SNN2 wins three: the two largest-WED cells, LIDC-0116 and the
OOD PIS\_NFLTD, by $9\%$ and $16\%$ lower pool RMSE, and PIS, narrowly, by $1\%$.
T2bp$\,\circ\,$V8c's only win is LIDC-0951, the lowest-WED patient, by $3\%$
($0.0007$).  Across the four cells the SNN2-base margin rises
with WED ($-3$, $+1$, $+9$ and $+16\%$ from LIDC-0951 to PIS\_NFLTD): for
the larger volumes here, the OOD probe included, the SNN2 base is the stronger
composition partner.

\paragraph{The validation pool reverses the ordering.}
On the four validation volumes, whose water-equivalent diameters
($225$--$317$~mm) fill the band the test cells leave open, the V8c base
leads under $\NoFC$ on three and ties the fourth, by $1.6$--$12.6\%$.
Interpolating the test-cell margins linearly in WED predicts the
opposite on the three larger volumes, an SNN2-base lead of $3$--$8\%$,
and only a $1\%$ V8c-base lead on the smallest; as in
Section~\ref{sec:results:v8c}, V8c-alone's advantage shrinks with WED
(\suppsec{app:valcomp}).  We give the ordering as a property of the
cells evaluated, not a recommendation.

\paragraph{Baseline regime.}
Under baseline acquisition T2bp$\,\circ\,$V8c wins PIS by $7\%$ and
leads LIDC-0951 by $5\%$, a tie; $\NoFC$ reverses the advantage on PIS
and cuts it to $3\%$ on LIDC-0951, where it becomes a win.  T2bp lowers
pool RMSE on all eight cells for the SNN2 base ($7$--$37\%$), and on
seven for the V8c base (\suppsec{app:bsregimes}).

\paragraph{The endpoints.}  T2bp leaves them unchanged, so under $\NoFC$
its gain on all three paths is $58$--$89\%$ of its middle-path gain
(\suppsec{app:t2bp:threepath}).

\subsection{T2bp across the corpus}
\label{sec:results:dif}

T2bp$\,\circ\,$SNN2, without trained weights, was run with the truth-free
settings on Section~\ref{sec:results:floor}'s corpus sweep ($24$
patients, both acquisitions); $K$ was chosen on four of them
(Section~\ref{sec:t2bp:posterior}; summaries without them:
\suppsec{app:t2bp:deploy}).  Table~\ref{tab:corpus-dif-stratum} gives
SNN2's and T2bp$\,\circ\,$SNN2's mean squared errors relative to the floor.

Under $\NoFC$, SNN2 attains the floor for bundles with $X_2<6$ ($83\%$
of bundles): its mean squared error is $1.01$--$1.07$ times their mean
CRB variance on $21$ of the $23$ non-test volumes (median $1.02$; LIDC-0510 $1.14$;
LIDC-0766 $1.69$, from rare unequalized bundles,
\suppsec{app:dif:snn2tail}).  T2bp$\,\circ\,$SNN2 lowers the ratio on
every patient, to $0.74$--$1.26$ (median $0.89$).  That is not the
prior's doing: an exactly solved box-constrained maximum-likelihood
estimate, with neither prior nor training, reaches a median $0.88$, and
$0.94$ with the bias correction of \suppsec{app:snn2:bias}, which T2bp
improves on for $20$ of $22$ patients (the 23rd failed a consistency
check fixed in advance; with it, $20$ of $23$).  A reading written down
with its thresholds before its run asked what share of T2bp's gain at
$X_2<6$ this corrected estimate reproduces: above $0.70$ the gain would
be the box's, below $0.30$ the prior's.  It came out mixed, $0.63$.  Two
later readings fixed the same way found the prior's contribution there
unmeasurable: T2bp changes the uncorrected estimate's mean squared
error by $-2\%$, and by $+2\%$ against a box-constrained estimate when
both are given the true endpoint values.
The uncorrected estimate's gain comes almost entirely from bundles with
an endpoint ray through air at the bound, where it re-fits the middle
path (Section~\ref{sec:snn2:mle}); elsewhere it matches SNN2
(\suppsec{app:snn2:mle}).  Like T2bp's, its sub-floor error comes with
bias (\emph{post hoc}: chosen after the results were seen;
\suppsec{app:t2bp:sensitivity}).  The prior earns its
keep deeper: over all bins, on the $14$ cells of the later readings (the
mechanism corpus: the eight test cells and six non-test volumes under
$\NoFC$), T2bp lowers the uncorrected estimate's error by a median $12\%$
over the eleven cells that pass every check fixed in advance and $29\%$
over all fourteen, significantly on $13$ cells and on none the other way;
at $X_2\ge6$, with the true endpoints given to both, by $15\%$ and
$21\%$.  At full volume on the eight test cells
(Table~\ref{tab:composition-10cell}; readings fixed before the run,
\suppsec{app:valcomp}), T2bp lowers the uncorrected estimate's mean
squared error by a median $19\%$ under $\NoFC$ ($2\%$ on PIS and
LIDC-0951, $37$--$40\%$ on LIDC-0116 and PIS\_NFLTD) and $47\%$ under
the baseline: the per-patient prior's gain does not depend on a weak
base estimator.  Under $\NoFC$ T2bp's error is within $2\%$ whichever
classical base estimator it starts from; under the baseline acquisition,
starting from the uncorrected box-constrained estimate lowers it by
$2.7\%$ on LIDC-0951 and raises it by $2.9\%$ on LIDC-0116.  The
acquisition, not the estimator, sets the typical bundle's normalized
MSE: T2bp's per-volume median MSE-DIF is $2.9$--$5.9$ (median $4.6$;
PIS\_NFLTD $3.3$), against $23$--$43$ under flat fluence (floor
$23$--$48$).  The sub-floor gain costs no small-lesion contrast at
$X_2<6$ but does above (Section~\ref{sec:discussion:limitations}).  At
$X_2<6$ it comes from bundles with a path near a box bound: under
$\NoFC$, SNN2, T2bp$\,\circ\,$SNN2 and the box-constrained estimate, with
or without the bias correction, all sit at $1.02$--$1.04$ times the floor
on the other bundles (post hoc; \suppsec{app:t2bp:sensitivity}).  T2bp
goes far below the floor only in the sparse high-attenuation tail,
paying the shrinkage bias that \eqref{eq:dif_mse} charges
(\suppsec{app:paper1floor}).

\subsection{The Paper~1 regime under the unbiased floor}
\label{subsec:paper1-floor}

On PIS\_NFLTD, where Paper~1 evaluates its network
\citep[Section~5.3]{besson2026pmb}, we compare T2bp$\,\circ\,$SNN2 with
the unbiased floor (\suppsec{app:paper1floor}).  Under baseline
acquisition T2bp \emph{rides} the floor, within $0.98$--$1.07$ of it through $[6,7)$,
falling below it only in the sparse tail, where the box truncates the
error: \emph{not an efficiency comparison}.  A prior buys variance below
the floor only where the measurement is weak, and pays in bias
\citep{hero1996}.  Under $\NoFC$ it rides a floor that $\NoFC$ itself
lowers, from an RMSE ratio of $8.24$ to $3.26$ at $[6,7)$.

\section{Discussion}
\label{sec:discussion}

\subsection{Relation to prior work}
\label{sec:disc:prior}

Fluence-field modulation and dynamic bow-ties shape the per-ray fluence
in single-source CT for noise, dose or detector dynamic range
\citep{bartolac2011,hsiehpelc2013,szczykutowicz2013,szczykutowicz2014,gang2017}.
$\NoFC$ uses the conventional part of that hardware (a static bow-tie
and tube-current modulation) with the dynamic-range target, equal detected fluence,
applied to the three rays of a TCT bundle, where it nears the
variance optimum (Section~\ref{sec:n0fc:problem}). Dual-source CT
records each source on its own detector \citep{flohr2006}, and
inverse-geometry CT fires its source spots in sequence
\citep{schmidt2004} and can shape the fluence spot by spot, a virtual
bow-tie \citep{deman2016}; TCT sums sources in one reading, coupling
the three rays' information. Single-photon emission
CT transmission scans with overlapping line-source beams also sum sources, and
maximum-likelihood reconstruction of the attenuation map models that
overlap \citep{yu2000}. Per-bundle inversion resembles low-dose sinogram restoration
\citep{lariviere2005,wang2006}, but must first separate the rays each
reading sums.

\subsection{What the design assumptions imply}
\label{sec:disc:design}

Equalization's doubled exponent (Section~\ref{sec:n0fc:problem}) means
$150$--$1{,}100$-fold edge attenuation for a center-to-edge $\Delta X$ of
$5$--$7$, against $12$--$33$-fold, the range of current clinical
filters, under the square-root law. $\NoFC$'s fitted bow-tie
($155$-fold) sits at the low end of that range, and its fit wanted a
steeper ramp still (\suppsec{app:n0fc:objective}). A filter
that deep may not be practical; in a polychromatic beam it would also
harden the spectrum and add scatter from the filter itself, neither of
which is modeled. Conventional bow-ties capture most of the gain
(Section~\ref{sec:results:floor}). $\NoFC$ itself only approximates
equalization: under the realized allocation the median bundle's summed
CRB is $1.31$ times its value at exact equalization
(\suppsec{app:n0fc:opt}). The tube-current schedules are an
oracle acting on the tail; a realizable one, smooth along $z$ and
limited in range and slew rate, is left to future work. Each schedule is
set from the slice's true attenuation, so the per-ray counts every
estimator reads as known carry information about it: no step reads the
ground truth, but the acquisition was designed from it.

\subsection{Residual bias structure: the bridge to inter-bundle priors}
\label{sec:bridge}

At $X_2\geq6$ under baseline acquisition, T2bp turns SNN2's mean bias
(estimate minus truth) of $+0.33$ (PIS) and $+0.57$ (LIDC-0951) into
$-0.18$ and $-0.23$, and overcorrects LIDC-0116's $+0.06$ to $-0.12$.
Under $\NoFC$ that bias is already small ($+0.005$ to $+0.025$), and
T2bp reduces variance. What remains is small at most bins
($|\mathrm{bias}|\leq0.005$) but concentrated at the deepest bin,
where $|\mathrm{bias}|/\sigma$ reaches $0.25$--$0.49$ (baseline) and
$0.05$--$0.82$ ($\NoFC$).

SNN2 sees a bundle only through its counts and per-ray $\Nzero$, so its
bias follows the photon level, not body size: comparing at equal
detected fluence $\Nzeroray{k}e^{-X_k}$ rather than equal $X$ removes
$92$--$100\%$ of the between-patient differences
(supplementary sections~\ref{app:biasstruct:shift} and~\ref{app:biasstruct:cells}), and
since $N_{\mathrm{base}}$ rises with patient size (Spearman $\rho=0.97$ with
WED; \suppsec{app:n0fc:nbase}), differences at matched $X$ reflect that
calibration. T2bp also depends on the patient through its prior: a
truth-free split of its bias into SNN2's error carried through and
shrinkage toward the prior attributes $84\%$ (baseline) and $87\%$
($\NoFC$) to shrinkage at $X_2\in[6,9)$ (\suppsec{app:biasstruct:t2bp}).
This analysis does not separate prior and photon level.

We expect what remains to need information from neighboring bundles:
Paper~1 found the line-integral field strongly correlated between them
\citep{besson2026pmb}.

The two estimators reach that information at very different cost. T2bp's
conditioning is a modeling choice: widening Equation~(\ref{eq:t2bp-prior})
from the bundle's own endpoint estimate $\Xhat_3$ to its neighbors' estimates changes the
conditioning set, not the machinery. T2bp stays per-patient and without trained weights, so
the question is computational (neighborhood size, grid cost) and
answerable on this corpus. V8c maps
a fixed per-bundle feature vector, so inter-bundle information would
need a different architecture, a corpus organized by neighborhoods, and
the generalization question reopened; model size was not the limit on
an earlier $13$-patient corpus (\suppsec{app:v8c:capacity}). We
therefore take the inter-bundle step from the empirical-Bayes side: we
select the tool that lets the work proceed on the resources available,
and do not judge whether a neural network could succeed.

\subsection{Limitations}
\label{sec:discussion:limitations}

\begin{table}[t]
\centering
\caption{Development choices made with data that included the test
patients, and where each is described (S: supplementary section).  The
bow-tie $B^*$ used no test-pool data.}
\label{tab:dev-provenance}
\small
\begin{tabular}{@{}>{\raggedright\arraybackslash}p{0.29\textwidth}>{\raggedright\arraybackslash}p{0.47\textwidth}l@{}}
\toprule
Choice & What informed it & Where \\
\midrule
V8c loss weights & PIS data & \ref{app:v8c:loss} \\
V8c per-bin loss weighting & judged partly on the test patients & \ref{app:v8c:loss} \\
V8c capacity variants & judged partly on the test patients & \ref{app:v8c:capacity} \\
Second training batch (seven high-$\WED$ patients) & added after an
earlier network's $X_2$ error reduction fell with WED across validation
and test patients, LIDC-0116 included; LIDC-0628 then moved from
training to validation & \ref{app:split} \\
T2bp iteration rule & written after test-cell errors had been inspected
& \ref{sec:t2bp:posterior} \\
TCM knot count & data that included test patients & \ref{app:n0fc:tcm} \\
Lower photon-count clip & data that included test patients & \ref{app:n0fc:clip} \\
T2bp prior form & data that included test patients & \ref{app:t2bp:r1r2} \\
T2bp grid & data that included test patients & \ref{app:t2bp:grid} \\
Acquisition, bias-structure and lesion tests & run on test patients;
each planned reading fixed in writing before the test ran, descriptive
readings marked & \ref{app:biasstruct}, \ref{app:t2bp:lesion} \\
Validation volumes (composition repeat) & outside the test pool, not
outside development: V8c's checkpoint was selected on them, LIDC-0766
is one of the four patients that set $K$, the bow-tie's reference slices
include LIDC-0628, and the finding behind the second batch was read on
three of them & \ref{app:valcomp}, \ref{app:split} \\
\bottomrule
\end{tabular}
\end{table}

The simulation is monoenergetic chest anatomy without scatter, spectral,
electronic-noise or detector-cross-talk modeling, and the training
corpus does not represent pediatric or obese patients.  Several sources
at once would add cross-scatter, as in dual-source CT
\citep{engel2008,petersilka2010}, which dilutes the dim rays'
information as a bright neighbor ray does.  Electronic noise would weigh
most at high attenuation.  The $5\times3$
bundle is a model problem (Section~\ref{sec:intro:background}),
and the flux is treated as known, without the air-scan noise of
Paper~1's bounds ($+5\%$ for the middle path at $(2,8,2)$;
\citealp[Section~3.5]{besson2026pmb}).  DIF reads as a dose factor only
where the estimator is efficient: for SNN2, in every $\NoFC$ bin through
$[5,6)$ of the patients checked.  Elsewhere SNN2's DIF overstates the
needed dose up to about $2.8$-fold (high-contrast baseline bins) and
understates it up to about $16$-fold (extreme bins, where the box acts
as a prior; \suppsec{app:dif:dose}).  For T2bp, whose shrinkage makes
the reading doubtful, it was not checked.

Where the likelihood is weak, T2bp pulls middle-path values that are
rare within the patient, a small lesion among them, toward the patient's
common anatomy.  A lesion-insertion test under $\NoFC$, on the four test
volumes and four non-test volumes, raised the true middle-path line
integral of a random $2\%$ of bundles by half its Cram\'er--Rao
standard deviation and measured each estimate's response (its mean
change per unit inserted) and detectability (response over the noise
standard deviation measured from a replicate scan, relative to the
box-constrained estimate's), with readings fixed before it ran
(\suppsec{app:t2bp:lesion}).  Below $X_2=6$, T2bp kept a small lesion's
contrast (response $0.99$--$1.00$), and its detectability changed by a
median $+3.8\%$, short of the $5\%$ fixed in advance for an information
gain.  At $X_2\ge6$, outside the readings, its response fell to $0.84$
and its detectability to $0.88$.  The baseline
acquisition was not tested.  The prior is built from clinical
reconstructions, noise and texture included
(Section~\ref{sec:library:pipeline}), so these results describe the
volumes used; a smoother volume would give a narrower prior that
plausibly shrinks harder, which was not tested.

Development used the test patients: Table~\ref{tab:dev-provenance} lists
every choice made with data that included them, the tests run on them,
and the roles of the validation volumes in development; the bow-tie
$B^*$ used no test-pool data.  Confirmation on volumes held out from
development is left to future work.

Several analyses were not run.  No estimator was evaluated under a
conventional bow-tie or under $B^*$ without the TCM:
Section~\ref{sec:results:floor} compares those designs on the floor
only.  T2bp ran on the exactly solved box-constrained estimate only on
the eight test cells (Table~\ref{tab:composition-10cell}); elsewhere its
base estimator is SNN2, and that estimate serves as a reference
(Section~\ref{sec:results:dif} and the lesion test above).  Under the
baseline acquisition, the bundles at
SNN2's iteration ceiling carry $3$--$39\%$ of its squared error
(Section~\ref{sec:snn2:gn}), so part of V8c's and T2bp's gains there may
be the repair of unconverged estimates; under $\NoFC$ that tail is
negligible.  V8c was trained once per acquisition, with one seed; its
run-to-run variation is unmeasured.  The rankings rest on four test
volumes, two of them (PIS and PIS\_NFLTD) one anatomy at two scales,
and their intervals resample slices within a volume, not patients: they
describe the cells evaluated.

\section{Conclusion}
\label{sec:conclusion}

In TCT, flux compensation is a prerequisite: on 23 simulated chest
scans, $\NoFC$ lowers the middle-path CRB floor $4.2$--$10.6$-fold, to a
per-volume median $\DIF_2$ of $3.3$--$6.3$, near the equal-attenuation
value $13/3$ and $1.4$--$1.7$-fold below GE-type bow-ties.  SNN2 attains
this floor for most bundles; T2bp$\,\circ\,$SNN2, without trained
weights, lowers the error further, through bias and at some cost in
small-lesion contrast where $X_2\ge6$, and beats V8c-alone on every test
cell; on the four $\NoFC$ validation volumes, T2bp$\,\circ\,$V8c leads or
ties it.  V8c's gains over SNN2 shrink with water-equivalent diameter;
it fails on a patient acquired outside its corpus's exposure rule.  The
remaining gap to dose parity and the high-attenuation tail point to
inter-bundle priors, the decisive test for TCT, which we will build on
T2bp because it needs no training.

\section*{Acknowledgments}
The author used Claude (Anthropic) for \LaTeX{} drafting and editing,
manuscript restructuring and copy-editing, and for engineering and
analysis support in the computational work: drafting the
data-generation, analysis and figure scripts, rendering figures from
the author's data files, and checking reported values against their
sources. All data were produced by code run by the author, who
verified the reported values, made all scientific decisions and takes
full responsibility for the content.
This research received no external funding.

The author acknowledges the National Cancer Institute and the
Foundation for the National Institutes of Health, and their critical
role in the creation of the free publicly available LIDC/IDRI Database
used in this study.

\section*{Ethical statement}
This study used 25 public LIDC-IDRI chest CT volumes and one
de-identified clinical CT volume as image sources. The volumes contain
no patient-identifying information and enter the analysis only through
attenuation line integrals computed from them. No data were collected
from human participants.

\section*{Data availability statement}
The 25 LIDC-IDRI chest CT volumes are publicly available from The Cancer
Imaging Archive under a Creative Commons Attribution 3.0 license
\citep{armato2011,armato2015data}. The PIS volume is a de-identified
clinical CT volume that cannot be made public because of the privacy and
institutional constraints governing patient imaging data; it enters the
analysis only as attenuation line integrals. The simulated bundle data,
the line integrals derived from PIS, and the analysis code (SNN2, the
$\NoFC$ design, V8c and T2bp) are available from the author upon
reasonable request.

\clearpage
\markboth{Supplementary material}{Supplementary material}
\setcounter{section}{0}
\setcounter{table}{0}
\setcounter{figure}{0}
\setcounter{equation}{0}
\renewcommand{\thesection}{S\arabic{section}}
\renewcommand{\thetable}{S\arabic{table}}
\renewcommand{\thefigure}{S\arabic{figure}}
\renewcommand{\theequation}{S\arabic{equation}}
\renewcommand{\thealgocf}{S\arabic{algocf}}
\setcounter{algocf}{0}
\renewcommand{\theHsection}{S\arabic{section}}
\renewcommand{\theHtable}{S\arabic{table}}
\renewcommand{\theHfigure}{S\arabic{figure}}
\renewcommand{\theHequation}{S\arabic{equation}}
\renewcommand{\theHalgocf}{S\arabic{algocf}}

\section*{Supplementary Material}
{\raggedright\noindent\textbf{Flux Compensation as a Prerequisite for
Temporal~CT: Per-Bundle Cram\'er--Rao Floors and Estimators That
Reach Them}\par}\vspace{0.4em}
\noindent G M Besson, ForeVision XRCT\\[0.4em]
\noindent\textbf{Description.} Implementation details of the SNN2
estimator, the $\NoFC$ ($\Nzero$ flux compensation) bow-tie and
tube-current design, the V8c network and the
T2bp post-processor; DIF derivations; per-patient evaluation tables;
and acquisition and bias-structure tests.\\[0.4em]
\noindent\textit{This document provides supplementary material for the
main paper. Sections, equations, tables, figures, and algorithms here
are prefixed `S'; equation numbers ``(n)'' and references of the form
``Table~n'', ``Figure~n'' or ``Section~n'' without the prefix refer to
the main paper.}

\bigskip
\suppressfloats[t]
\section{SNN2 implementation details}
\label{app:snn2}

This section recalls the SNN1 bottleneck that motivated SNN2. It
collects the SNN2 per-iteration formulas, the IRLS equivalence proof,
the bias-correction taper, and the 9-deviate Poisson sampling
convention. It closes with the structure of the residual that the V8c
corrector (Section~\ref{app:v8c}) is trained to reduce.  Notation
follows main Sections~\ref{sec:model} and~\ref{sec:snn2};
Algorithm~\ref{alg:snn2} summarizes the
iteration.

\begin{algorithm}[t]
\DontPrintSemicolon
\KwIn{$Y_1,\ldots,Y_5$; per-ray $\Nzeroray{k}$; $X_{\max}=9.5$,
      $\Delta_{\max}=1.5$, $\varepsilon=10^{-6}$, $N_{\mathrm{iter}}=20$.}
\KwOut{$\hat{X}_1,\hat{X}_2,\hat{X}_3$.}
\tcp{SVD warm start, $A^{+}$ the $3\times5$ pseudo-inverse of $A$}
$\hat t_k \gets \max\bigl((A^{+}Y)_k/\max(\Nzeroray{k},1),\,10^{-10}\bigr)$;\quad
$X_k \gets \min\bigl(\max(-\ln\hat t_k,\,0),\,9.2\bigr)$, $k=1,2,3$\;
\lIf{$Y_1>0.5$}{$X_1 \gets \min\bigl(\max(-\ln(Y_1/\Nzeroray{1}),0),X_{\max}\bigr)$}
\lIf{$Y_5>0.5$}{$X_3 \gets \min\bigl(\max(-\ln(Y_5/\Nzeroray{3}),0),X_{\max}\bigr)$}
\For{$i = 1,\ldots,N_{\mathrm{iter}}$}{
  $p_{jk} \gets \Nzeroray{k}A_{jk}e^{-\min(X_k,20)}$;\quad
  $y_j \gets \max(\sum_kp_{jk},\,1)$;\quad $\Delta Y_j \gets Y_j-y_j$\;
  Form $F$, $\svec{s}$ via~\eqref{eq:fim}, \eqref{eq:score}\;
  \eIf{$|\det F|<10^{-30}$}{$\svec{\delta} \gets 10^{-4}\,\svec{s}/\mathrm{tr}\,F$}{solve
    $F\svec{\delta}=\svec{s}$ (Cram\'er's rule)}
  Clamp: scale $\svec{\delta}$ if
    $\|\svec{\delta}\|_\infty>\Delta_{\max}$\;
  $\svec{X}^{\mathrm{prev}} \gets \svec{X}$;\quad
  $\svec{X} \gets \min\bigl(\max(\svec{X} + \svec{\delta},\,0),\,X_{\max}\bigr)$\;
  \lIf{$\|\svec{X}-\svec{X}^{\mathrm{prev}}\|_\infty<\varepsilon$}{\textbf{break}}
}
\tcp{one-shot bias correction (Section~\ref{app:snn2:bias})}
Recompute $p_{jk}$, $y_j$, $F$ at $\svec{X}$; $\sigma_k^2\gets[F^{-1}]_{kk}$;
$\svec{s}^{\mathrm{corr}}\gets$ score~\eqref{eq:score} with residuals
$\Delta Y_j+b_j$, $b_j$ from~\eqref{eq:bias_meas};
$\alpha\gets$ taper~\eqref{eq:taper}\;
\If{$|\det F|>10^{-30}$ and $\alpha>0.001$}{
  $\svec{X} \gets \min\bigl(\max(\svec{X} + \alpha F^{-1}\svec{s}^{\mathrm{corr}},\,0),\,X_{\max}\bigr)$\;
}
\textbf{return} $\svec{X}$\;
\caption{SNN2: Gauss--Newton bundle inversion, as implemented (double
precision; inputs and outputs stored in single precision).}
\label{alg:snn2}
\end{algorithm}

\subsection{The SNN1 bottleneck}
\label{sec:snn2:snn1}

SNN1, Paper~1's iteration, distributes each row residual across the
active rays in
proportion to the partial intensities $p_{jk} =
\Nzeroray{k}\,A_{jk}\,e^{-X_k}$.  In the high-middle-attenuation regime
($X_2{=}6$, $X_1{=}X_3{=}3$, uniform $\Nzero$) the fraction reaching the
middle path, $w_{j2}=p_{j2}/y_j$, is $0.047$ on rows~2 and~4 and $0.024$
on row~3, under $5\%$. Combined with the annealing schedule, the
effective per-iteration $X_2$ update falls below $1\%$ of the residual,
while the well-determined endpoints shift it again next iteration.
The iteration limit-cycles instead of converging, producing a
positive bias of up to $+0.31$ at bins~6--8 on Paper~1's patient
data \citep[Section~S2.3]{besson2026pmb}.

\subsection{Cram\'er's-rule solve for the $3{\times}3$ Newton step}
\label{app:snn2:cramer}

Let $F$ be the symmetric FIM~\eqref{eq:fim} and $\svec{s}$ the
score~\eqref{eq:score}.  The adjugate (cofactor matrix) elements are
\begin{align}
  \mathcal{A}_{11} &= F_{22}F_{33} - F_{23}^2,
  &\mathcal{A}_{22} &= F_{11}F_{33} - F_{13}^2,
  \notag\\
  \mathcal{A}_{33} &= F_{11}F_{22} - F_{12}^2,
  &\mathcal{A}_{12} &= F_{13}F_{23} - F_{12}F_{33},
  \notag\\
  \mathcal{A}_{13} &= F_{12}F_{23} - F_{13}F_{22},
  &\mathcal{A}_{23} &= F_{12}F_{13} - F_{11}F_{23}.
  \label{eq:adjugate}
\end{align}
The determinant is
$\det F = F_{11}\mathcal{A}_{11}+F_{12}\mathcal{A}_{12}+
F_{13}\mathcal{A}_{13}$, and
\begin{equation}
  \delta_k = \frac{1}{\det F}\,\sum_{m=1}^{3}\mathcal{A}_{km}\,s_m,
  \qquad k=1,2,3.
  \label{eq:cramer}
\end{equation}
Total cost is of order $10^2$ arithmetic operations and three
exponentials per bundle per iteration, spread over forming $F$ and
$\svec s$ and the Cram\'er solve.  No pivoting is required: $F$ is symmetric positive
definite by construction (a sum of rank-1 outer products with
positive weights).

\subsection{Trust-region clamp and convergence criterion}
\label{app:snn2:clamp}

If $\|\svec{\delta}\|_\infty > \Delta_{\max}$ at any iteration, the
entire step is rescaled,
$\svec{\delta} \gets \svec{\delta}\,\Delta_{\max}/\|\svec{\delta}\|_\infty$,
preserving the Newton direction.  We use $\Delta_{\max}=1.5$.  Each
updated iterate, including the bias-correction step of
Section~\ref{app:snn2:bias}, is then projected onto the box
$0\le X_k\le9.5$; the endpoint initializations are projected onto the
same box, and the SVD warm start, used for $X_2$ and for an endpoint
whose dedicated reading is zero, is clipped to $[0,9.2]$.
This
allows convergence from SVD warm-starts that may be several
attenuation units away from the truth without divergent excursions
in early iterations.  The box excludes some true values: in the $23$
non-test corpus volumes a path exceeds $9.5$ in $30{,}715$ of
$1.90\times10^9$ bundles ($1.6\times10^{-5}$; $X_2$ reaches $11.5$ in
LIDC-0766), and SNN2's estimate of that path is capped at $9.5$.
The iteration terminates when the change of
the projected iterate satisfies
$\|\svec{X}-\svec{X}^{\mathrm{prev}}\|_\infty<\varepsilon$
($\varepsilon=10^{-6}$) or after $N_{\mathrm{iter}}=20$ iterations.
On the corpus runs logged (eight patients, both acquisitions), the
mean iteration count per patient is $3.5$--$4.5$.  The share of
bundles reaching the iteration ceiling differs sharply between the two
acquisitions: under $\NoFC$ the per-volume median is
$1.3\times10^{-7}$ (at most $6.9\times10^{-4}$), whereas under the
flat-fluence baseline it is $5.3\times10^{-3}$ (at most
$1.8\times10^{-2}$), for a pooled validation share of $0.87\%$.  The
capped bundles lie mostly at $X_2\ge6$: at $X_2<6$ the baseline median
falls to $4\times10^{-4}$.  Under the baseline acquisition they also
carry a disproportionate share of SNN2's squared error, $2.8$--$39\%$
by validation volume and $2.9$--$26\%$ on the test cells, so they are
not a negligible tail in that regime; under $\NoFC$ they are.

\subsection{Equivalence to iteratively reweighted least squares (IRLS) in intensity coordinates}
\label{app:snn2:mle}

Define $\beta_k = \Nzeroray{k}\,e^{-X_k}$.  The forward model becomes
\begin{equation}
  y_j = \sum_{k=1}^{3} A_{jk}\,\beta_k,
  \label{eq:linfwd}
\end{equation}
linear in $\bm{\beta}$.  The Poisson log-likelihood, regarded as a
function of $\bm{\beta}$, has score and Fisher matrix
\begin{align}
  \tilde{s}_k(\bm{\beta}) &=
    \sum_j A_{jk}\,(Y_j - y_j)/y_j,
  \\
  \tilde{F}_{km}(\bm{\beta}) &=
    \sum_j A_{jk}\,A_{jm}/y_j
  \;=\; (A^{T}WA)_{km},
  \label{eq:Fbeta}
\end{align}
with $W = \mathrm{diag}(1/y_j)$.  The IRLS update in $\bm{\beta}$ is
$\Delta\bm{\beta} = (A^{T}WA)^{-1}A^{T}W(\bm{Y}-\bm{y})$.  By the
chain rule with $\partial\beta_k/\partial X_k = -\beta_k$,
\begin{equation}
  s_k = -\beta_k\,\tilde{s}_k,
  \qquad
  F_{km} = \beta_k\,\beta_m\,\tilde{F}_{km},
  \label{eq:Fxbeta}
\end{equation}
so the Newton step in $\svec{X}$-coordinates is
\[
  \Delta\svec{X} = F^{-1}\svec{s} = -\mathrm{diag}(\bm{\beta})^{-1}
  (A^{T}WA)^{-1}A^{T}W(\bm{Y}-\bm{y}),
\]
which is exactly $-\Delta\bm{\beta}/\bm{\beta}$, componentwise; to first
order, the $\svec{X}$-step therefore moves $\bm\beta$ by the IRLS update
$\Delta\bm\beta$.  Both
parameterizations have the same interior fixed point, the Poisson MLE,
where the score vanishes.

\emph{Active bounds.}  Algorithm~\ref{alg:snn2} projects each step onto
the box $0\le X_k\le X_{\max}$.  Where a bound is active, its fixed point
is not the box-constrained MLE:
\begin{enumerate}[nosep, leftmargin=*]
  \item \emph{Fixed point.}  Let $\svec{X}^*$ be a fixed point with
        ray~$m$ at a bound, say $X_m^*=0$, and let
        $\svec{\delta}=F^{-1}\svec{s}$ be the Newton step there (the
        clamp of Section~\ref{app:snn2:clamp} rescales it without
        changing its direction).  The projection keeps $X_m^*=0$ if
        $\delta_m\le0$, and leaves a free ray $k\neq m$ unchanged only if
        $\delta_k=0$.
  \item \emph{Score at the fixed point.}  Hence
        $\svec{s}=F\svec{\delta}=\delta_m\,F_{\cdot m}$, the $m$th column
        of $F$ times $\delta_m$: for each free ray, $s_k=F_{km}\,\delta_m$.
  \item \emph{Optimality conditions.}  The box-constrained MLE satisfies
        the Karush--Kuhn--Tucker conditions: $s_k=0$ for every free ray,
        and $s_m\le0$ for a ray at its lower bound.  Every pair of rays
        shares reading~$3$, so $F_{km}>0$ in~\eqref{eq:fim}, and
        $s_k\neq0$ unless $\delta_m=0$: the fixed point is not a
        Karush--Kuhn--Tucker point.  The condition at the bound itself
        holds, $s_m=F_{mm}\,\delta_m\le0$; the free rays violate theirs.
        With several rays at bounds, $s_k=\sum_mF_{km}\,\delta_m$, the sum
        over the rays at bounds.
  \item \emph{What it is instead.}  To first order the Newton step
        reaches the unconstrained stationary point,
        $\svec{X}^{\mathrm{u}}\approx\svec{X}^*+\svec{\delta}$, which
        agrees with $\svec{X}^*$ on the free rays.  To first order SNN2
        therefore returns the unconstrained MLE with its out-of-box
        component clipped; the box-constrained MLE would also move the
        free rays to account for the clipped one.
\end{enumerate}
The upper bound is the same with the inequalities reversed.  Bounds can
be active for endpoint rays through air ($X_k$ near $0$) and for paths
near or beyond the upper limit (Section~\ref{app:snn2:clamp}).  On $12$
slices of four corpus volumes under $\NoFC$
(Section~\ref{app:dif:snn2tail}), SNN2's estimate has a coordinate at a
bound in $14$--$59\%$ of the bundles with $X_2<6$, almost all at the
lower bound.  A free-variable (reduced) Newton step, which holds a ray at
a bound fixed when its score points out of the box, reaches the
box-constrained MLE.  The log-likelihood is concave in $\bm\beta$ and the
box in $\svec X$ is a box in $\bm\beta$, so that point is the global
box-constrained maximum (away from the count floor $y_j\ge1$).  This
reduced iteration starts from SNN2's initial estimate, takes SNN2's step on the free rays,
and stops when no coordinate moves by more than $10^{-6}$, within $200$
iterations; on every volume and acquisition evaluated, the median scaled
Karush--Kuhn--Tucker residual over bundles with an active bound was at
most $10^{-6}$, and at most $0.1\%$ of bundles reached the $200$-iteration
cap.  On the
same slices the reduced step, followed by the same bias correction,
lowered SNN2's middle-path mean squared error over $X_2<6$ by
$4$--$14\%$, to $0.87$ and $0.90$ times the mean CRB variance on the
two volumes near the floor, at the cost of a small bias (at most $0.011$
in magnitude in a unit bin, negative on those two volumes).  The gain
comes from endpoint rays through air, where the bound acts as prior
information that the projected step does not use; the bias correction
was derived for an interior estimate.  A later check on the eight test
cells (every 8th slice) measured this.  Of the endpoint coordinates that the
box-constrained MLE holds at zero, $98.1\%$ over the eight cells have a
true line integral below $0.01$ ($87.1\%$ exactly zero); the share is
$100\%$ on every baseline cell, and under $\NoFC$ it is $66\%$ for
LIDC-0116 and $86\%$ for PIS\_NFLTD, the rest being thin paths
($0.01$--$0.1$).
Where an endpoint through air is held at zero ($22\%$ of the
sampled bundles), SNN2's middle path at $X_2<6$ is too high by $0.003$--$0.031$,
and the box-constrained MLE's root-mean-square error is $6$--$41\%$
lower ($20$--$24\%$ under $\NoFC$).  Where an endpoint is in air but not
held at zero, both are slightly low and agree within $1\%$; where no
endpoint is in air or held, within $0.1\%$.  Where thin paths are held at
zero ($0.3\%$ of the bundles, under $\NoFC$), the box-constrained MLE's
root-mean-square error is $16$--$26\%$ lower.  SNN2's results in this
paper use the projected step; the box-constrained MLE is reported as a
separate estimator (main Section~\ref{sec:results:dif} and
Table~\ref{tab:composition-10cell}; Tables~\ref{tab:t2bp:chain}
and~\ref{tab:kktcomp}).

The unweighted SVD inverse (Paper~1's linear baseline
and SNN2's warm-start) is the same
formula with $W=I$ instead of $W=\mathrm{diag}(1/y_j)$, and is
therefore not Fisher-efficient at finite signal levels even when
the linear $\bm{\beta}$ system is well-conditioned.

\subsection{Bias correction with CRB-gated taper}
\label{app:snn2:bias}

At convergence, write $\bm{\eta}=\hat{\svec X}-\svec X$ for the
estimation error, with covariance close to $F^{-1}$.  In the expansion
of the fitted row mean $y_j(\hat{\svec X})=\sum_kp_{jk}e^{-\eta_k}$, with
$p_{jk}$ at the truth, to second order in $\bm\eta$, the curvature term
contributes in expectation the per-measurement bias
\begin{equation}
  b_j \;=\; \tfrac{1}{2}\sum_{k:\,A_{jk}\neq 0} p_{jk}\,\sigma_k^2,
  \label{eq:bias_meas}
\end{equation}
with $\sigma_k^2 = [F^{-1}]_{kk}$ evaluated at $\hat{\svec{X}}$.
A~one-shot correction adds $b_j$ to each residual and applies one
additional Newton step,
$\hat{\svec{X}} \gets \hat{\svec{X}} + \alpha\,F^{-1}\svec{s}^{\mathrm{corr}}$,
where $\svec{s}^{\mathrm{corr}}$ uses corrected residuals
$\Delta Y_j + b_j$ and $\alpha\in[0,1]$ is a CRB-gated taper.
Defining $\sigma_{\max} = \max_k \sigma_k$ and the normalized gate
coordinate $t = (\sigma_{\max}-g+w/2)/w$,
\begin{equation}
  \alpha(\sigma_{\max}) =
  \begin{cases}
    1, & t \le 0,\\[2pt]
    \tfrac{1}{2}\bigl(1+\cos \pi t\bigr), & 0 < t < 1,\\[2pt]
    0, & t \ge 1,
  \end{cases}
  \label{eq:taper}
\end{equation}
with gate center $g=0.10$ and transition width $w=0.10$, so that
$\alpha=1$ for $\sigma_{\max}\le0.05$ and $\alpha=0$ for
$\sigma_{\max}\ge0.15$ (the limit quoted in main
Section~\ref{sec:snn2:bias}).  The
raised-cosine form is $C^1$-continuous, avoiding discontinuities in
the estimator output across the gate.  On uniform-flux i.i.d.\
synthetic data the correction removes ${\sim}30$--$40\%$ of the small
residual positive bias at bins~4--5, tapers off through bin~6, and is
inactive from bin~7 on, where $\sigma_{\max}$ exceeds the gate.  There
the second-order model fails, and an ungated correction overshoots by
a factor of $5$--$10$ (the bin-8 $X_2$ bias goes from $+0.15$ to
$-1.18$).  Under $\NoFC$ the CRB is smaller, so the gate stays open to
higher bins.

\emph{Closed form.}  At an interior fixed point the score vanishes,
$s_k=0$, so the corrected score is
$s^{\mathrm{corr}}_k=-\sum_jp_{jk}b_j/y_j
=-\tfrac12\sum_m\bigl(\sum_jp_{jk}p_{jm}/y_j\bigr)\sigma_m^2
=-\tfrac12\sum_mF_{km}\sigma_m^2$.  The step is then
$F^{-1}\svec s^{\mathrm{corr}}=-\tfrac12(\sigma_1^2,\sigma_2^2,\sigma_3^2)$,
and the correction reads
\begin{equation}
  \hat X_k \;\gets\; \hat X_k-\tfrac{\alpha}{2}\,[F^{-1}]_{kk}.
  \label{eq:bias_closed}
\end{equation}
The amount subtracted, $\tfrac12[F^{-1}]_{kk}$, is the first-order bias
of the interior MLE, so with $\alpha=1$ the step removes it.  The row means are linear in the intensities $\beta_k$
(Section~\ref{app:snn2:mle}), so in the Cox--Snell bias
\citep{coxsnell1968S} of $\hat{\bm\beta}$, written as in
Section~\ref{app:biasstruct:shift} with $\bm\beta$ in place of $\svec X$,
the cumulants satisfy $\kappa_{km,n}=-\tfrac12\kappa_{kmn}$ and the bias
vanishes; the delta
method for $X_k=-\ln(\beta_k/\Nzeroray{k})$ then gives a bias
$+\tfrac12[F^{-1}]_{kk}$ for $\hat X_k$.  For a single ray with $n$ detected photons on
average, $[F^{-1}]_{kk}=1/n$ and the bias is $+1/(2n)$, that of the
single-ray estimate $-\ln(Y/\Nzero)$ (main Section~\ref{sec:snn2:mle}).

\subsection{9-deviate Poisson sampling convention}
\label{app:snn2:deviates}

The bundle file format stores nine independent Poisson deviates
$d[j,k]\sim\mathrm{Poisson}(\Nzeroray{k}\,A_{jk}\,e^{-X_k})$ per
bundle, one per nonzero $(j,k)$ pair of the system matrix.  The
five row sums consumed by SNN2 are reconstructed at runtime,
$Y_j = \sum_k A_{jk}\,d[j,k]$.  Storing the nine individual deviates
rather than the five sums makes the bundle file format physically
faithful (each deviate corresponds to a distinct source-firing
event in the real TCT geometry) and supports downstream analyses
that need the per-source counts.  By the additivity of independent
Poissons, the $Y_j$ obtained this way are statistically identical
to those obtained by sampling
$Y_j\sim\mathrm{Poisson}(\sum_k A_{jk}\Nzeroray{k}e^{-X_k})$ directly.

\subsection{SNN2 residual structure: training pool}
\label{app:snn2_residual:structure}

V8c is a residual corrector layered on top of SNN2's output. Its
training signal is the per-bundle residual
\begin{equation}
  \delta_k \;=\; X^{\mathrm{true}}_k - \hat X^{\mathrm{SNN2}}_k,
  \qquad k \in \{1, 2, 3\}.
  \label{eq:delta_def}
\end{equation}%
\footnote{\textit{Correction-target convention.} $\delta_k$ is the
additive value V8c is trained to produce. The conventional
estimator-bias sign is
$\mathrm{Bias} \equiv \E[\hat X^{\mathrm{SNN2}}_k] - X^{\mathrm{true}}_k$;
the two conventions differ by a sign: $\overline{\delta_k} = -\mathrm{Bias}$.}
Before evaluating V8c's per-cell performance, we characterize the
residual distribution itself at each acquisition regime. Direct
measurement settles two questions. (i)~How does $\delta_2$ partition
between systematic bias and random spread across the line-integral (LI)
bins of the training corpus? (ii)~How much of that residual is linearly
explainable from the 26 input features V8c consumes?

\textit{CRB context.} The residual analysis below operates inside an
already-compressed bound. Evaluating the per-ray CRB
(Section~\ref{app:dif:crb}) at the true line integrals of the $19$
training patients and pooling it per bin, the middle-path ratio is
\begin{equation}
  \frac{[\bar\sigma^{\mathrm{CRB}}_2(\NoFC)]^2}{[\bar\sigma^{\mathrm{CRB}}_2(\mathrm{baseline})]^2}
  \;=\; 0.057,\ 0.018,\ 0.0063,\ 0.0056
  \quad\text{at bins }7, 8, 9, 10,
  \label{eq:crb_ratio}
\end{equation}
an $18$--$180\times$ reduction in the $\hat X_2$ variance lower bound
($4$--$13\times$ in standard deviation) at the bins where V8c works
hardest, achieved by the $\NoFC$ co-design (Section~\ref{sec:n0fc})
before any estimator is applied.  The SNN2 residual variance contracts
by $22$, $20$ and $8.5\times$ at bins~7, 8 and~10
(Table~\ref{tab:snn2_residual_structure}): about as much as the bound at
bin~7, and much less at bins~8 and~10. There, the box
(Section~\ref{app:dif:dose}) already holds SNN2's baseline spread below
the bound.  Any V8c gain under
$\NoFC$ is therefore improvement on top of a tighter Fisher envelope,
not a separate effect.

Table~\ref{tab:snn2_residual_structure} reports three summary numbers
per LI bin and acquisition regime, computed on stratified 10\,000-bundle
subsamples drawn from the shuffled training cache ($1.58 \times 10^9$
bundles across 19 patients). They are the empirical residual mean
$\overline{\delta_2}$ and standard deviation $\sigma_{\delta_2}$, and
the coefficient of determination $R^2_{26}$ of the ordinary
least-squares (OLS) regression of $\delta_2$ onto all 26 input features
($n=10^4\gg p=26$). Less its null, $R^2_{26}$ measures the linear signal V8c could in
principle extract from per-bundle inputs alone; it neither lower- nor
upper-bounds V8c's nonlinear capacity.

Because the bins are of the true $X_2$, $R^2_{26}$ is positive even for a
residual of pure noise.  With $X_2$ spread uniformly over a unit-width bin
and an error $e$ independent of it, regressing $\delta_2=-e$ on the SNN2
estimate $X_2+e$ alone gives
$R^2=\sigma_{\delta_2}^2/(\sigma_{\delta_2}^2+1/12)$, the null listed
in Table~\ref{tab:snn2_residual_structure}.  It is a lower bound
\emph{(approximation)}: bin populations fall with $X_2$, which narrows
the within-bin spread below $1/12$, and the other inputs include
nonlinear functions of the same estimate. A tenfold population fall across the bin
raises the null at bins~7 and~8 from $0.86$ and $0.91$ to $0.89$ and
$0.93$ under the baseline, and from $0.22$ and $0.34$ to $0.26$ and
$0.40$ under $\NoFC$.  Only $R^2_{26}$ in excess of the true null measures
the structure that the features add; an empirical null, from the same
regression on residuals simulated at the true line integrals, would
settle the high bins.

\begin{table}[t]
\centering
\caption{SNN2 residual structure on the $X_2$ middle path, train
pool, at headline LI bins. Bins as in~\eqref{eq:x2bins} (main text):
bin~$b$ collects $X^{\mathrm{true}}_2\in[b-1,b)$, $b=1,\dots,9$, and
bin~10 collects $X^{\mathrm{true}}_2\ge9$ (true $X_2$ reaches $11.5$ in the corpus).
$R^2_{26}$: OLS regression of $\delta_2$ on the 26 input features.
Null: $\sigma_{\delta_2}^2/(\sigma_{\delta_2}^2+1/12)$, the $R^2$ that a
residual of pure noise gives within a unit-width bin of the true $X_2$
(approximation; in parentheses for the open bin~10).}
\label{tab:snn2_residual_structure}
\begin{tabular}{@{}clrrrr@{}}
\toprule
Regime & Bin & $\overline{\delta_2}$ & $\sigma_{\delta_2}$ & $R^2_{26}$ & Null \\
\midrule
\multirow{5}{*}{Baseline}
  & 2  & $+0.000$ & $0.008$ & $0.13$ & $0.00$ \\
  & 5  & $-0.005$ & $0.104$ & $0.35$ & $0.11$ \\
  & 7  & $-0.138$ & $0.714$ & $0.91$ & $0.86$ \\
  & 8  & $-0.195$ & $0.937$ & $0.95$ & $0.91$ \\
  & 10 & $+0.900$ & $1.330$ & $0.98$ & ($0.96$) \\
\midrule
\multirow{5}{*}{$\NoFC$}
  & 2  & $+0.000$ & $0.007$ & $0.06$ & $0.00$ \\
  & 5  & $-0.000$ & $0.039$ & $0.11$ & $0.02$ \\
  & 7  & $-0.009$ & $0.152$ & $0.42$ & $0.22$ \\
  & 8  & $-0.019$ & $0.208$ & $0.59$ & $0.34$ \\
  & 10 & $+0.159$ & $0.457$ & $0.83$ & ($0.71$) \\
\bottomrule
\end{tabular}
\end{table}

Three findings emerge.

\textit{Finding 1: learnable signal at high attenuation, mostly under
$\NoFC$.}  $R^2_{26}$ exceeds the lower-bound null in every bin, but at
the baseline's bins~7 and~8 only by $0.05$ and $0.04$. A falling
population within the bin alone reduces them to about $0.02$. There,
most of $R^2_{26}$ is the selection effect of binning on the true $X_2$.  Under
$\NoFC$ the excess is larger ($0.20$ and $0.25$, or $0.16$ and $0.19$
against the tenfold-fall null).  Where the signal is real, under
$\NoFC$, SNN2's spread is near the CRB in aggregate (main
Section~\ref{sec:results:snn2}), but its residual depends on features
the network sees, most where bias dominates (main
Section~\ref{sec:snn2:mle}), so V8c has structure to correct.

\textit{Finding 2: $\NoFC$ contracts the residual scale 3--5$\times$ at
high bins, bias faster than spread.} $\sigma_{\delta_2}$ drops from 0.71
to 0.15 at bin~7 (4.7$\times$), 4.5$\times$ at bin~8, 2.9$\times$ at
bin~10. The bias $|\overline{\delta_2}|$ contracts faster still
(15$\times$, 10$\times$, 5.7$\times$). The residual signature flips sign
between bins~8 and~10. SNN2 over-predicts $X_2$ where the Jensen-MLE
bias dominates (bins~6--8) and under-predicts at photon starvation
(bin~10).\footnote{\textit{Mechanism (hypothesis).} At bins~6--8 the
Newton iteration reaches the Poisson MLE with its finite-sample
positive bias on $\hat X_2$ (Jensen), the second-order correction being
gated off there. At bin~10 the box caps the estimates at $9.5$ for the
true values beyond it (Section~\ref{app:snn2:clamp}), the SVD warm-start
is clipped at $9.2$, and the middle path contributes only about
$20$--$50$ photons to the most-attenuated row under baseline
acquisition. $\NoFC$'s tube-current
schedule raises $N_{0,2}$ on these rays (the bow-tie only attenuates),
compressing the bin-10 residual bias $5.7\times$ without flipping its
sign (Table~\ref{tab:snn2_residual_structure}).} Equivalently, the
bias-to-noise ratio $|\overline{\delta_2}|/\sigma_{\delta_2}$ collapses
from 0.19 to 0.06 at bin~7 and 0.68 to 0.35 at bin~10. Before V8c is
invoked, $\NoFC$ has removed most of the residual bias and much of the
variance V8c was designed to correct.

\textit{Finding 3: at the high bins, $R^2_{26}$ falls under $\NoFC$
because its null falls.}  With the residual three to five times
narrower, the null falls from $0.86$--$0.91$ to $0.22$--$0.34$ at
bins~7--8, while the excess rises.  There the share of the residual the
features predict is larger under $\NoFC$, but the residual is about
twenty times smaller in variance. What the features can predict in
absolute terms (excess times $\sigma_{\delta_2}^2$) is two to five times
smaller. At bins~7 and~8 it is $0.010$--$0.035$ under the baseline
and $0.004$--$0.011$ under $\NoFC$, the ranges spanning the two nulls
above.
A corrector can thus remove a larger share of a smaller residual;
whether that helps end to end is settled by the cross-evaluation
(Section~\ref{app:v8c:cross}).

Four of the five features carrying this signal at bins 7--10 are the
same in both regimes---the SNN2 middle-path estimate, two middle-path
FIM entries, and the predicted CRB---and are listed and physically
identified in Section~\ref{app:snn2_residual:features}. On the full
pool of the four validation patients the bin-10 bias-to-noise ratio
falls from $0.62$ to $0.11$ under $\NoFC$
(Section~\ref{app:n0fc:emp:val}; training-pool subsample: $0.68\to0.35$);
$R^2_{26}$ was computed on the training pools only. The test pool is examined patient-wise in
Section~\ref{sec:results:v8c}.

\subsection{Most-correlated features per LI bin}
\label{app:snn2_residual:features}

We rank the 26 input features by $|\mathrm{Pearson}(f_j,\delta_2)|$ in
the stratified training subsample. Four features lead at bins~7--10 in
\emph{both} regimes, signs preserved: the SNN2 middle-path estimate
$\hat X^{\mathrm{SNN2}}_2$ ($f_{08}$), the FIM entries $F_{12}$ and
$F_{23}$ ($f_{11},f_{14}$), and the predicted CRB
$[F^{-1}]_{22}$ ($f_{16}$). $F_{22}$ ($f_{13}$) completes the
top five everywhere except baseline bin~7, where $F_{13}$ ($f_{12}$)
takes its place.  Within
bins of the true $X_2$ the SNN2 estimate correlates with $\delta_2$ by
construction, since $\delta_2$ is minus its error, so part of $f_{08}$'s
lead is the selection effect behind the null of
Section~\ref{app:snn2_residual:structure}.  The Pearson magnitudes
mostly contract under $\NoFC$ (consistent with the fall of $R^2_{26}$
and of its null in Table~\ref{tab:snn2_residual_structure}; $f_{13}$ at
bin~10 is the exception). The structure, however, is unchanged: $\NoFC$
scales the residual, it does not re-pattern it.  These single-feature
correlations suggest that, where the residual is predictable, the SNN2
operating point and its local Fisher information carry the signal
(\emph{hypothesis}); they do not show what V8c itself uses.

\section{Bow-tie and TCM parameterization}
\label{app:n0fc}

This section documents the parameterization of the bow-tie $B(\gamma)$
and of the tube-current modulation (TCM) $\tau(\theta)$, the objectives
of the bow-tie and TCM fits, the calibration of $N_{\mathrm{base}}$, the
clipping convention, the theory behind equalization, and a comparison
with conventional bow-ties.

\subsection{Bow-tie family}
\label{app:n0fc:bowtie}

\emph{Supports the bow-tie design of main Section~\ref{sec:n0fc:param}: equation~\eqref{eq:bowtie}, with the fitted parameters of Section~\ref{app:n0fc:objective}, reproduces it.}

$B(\gamma)$ is a 4-parameter family with a flat center, an
exponential ramp on the wings, and a transmission floor:
\begin{equation}
  B(\gamma) =
  \begin{cases}
    1, & |\gamma| \le \gamma_{\mathrm{flat}},\\[3pt]
    \max\!\Bigl(B_{\min},\,
      \exp\!\bigl(-\alpha\,\bigl(\tfrac{|\gamma|-\gamma_{\mathrm{flat}}}{\gamma_{\max}-\gamma_{\mathrm{flat}}}\bigr)^{n}\bigr)
    \Bigr), & |\gamma| > \gamma_{\mathrm{flat}}.
  \end{cases}
  \label{eq:bowtie}
\end{equation}
The four parameters are bounded as
$\gamma_{\mathrm{flat}}\in[0,\gamma_{\max}/2]$,
$\alpha\in[0.1,6.0]$ (natural-log attenuation at $\gamma_{\max}$),
$n\in[1.5,5.0]$ (ramp exponent), and
$B_{\min}\in[0.005,0.05]$ (peripheral floor).
$\gamma_{\max}=\arcsin(R_M/R_S)=\arcsin(250/500)\approx0.524$~rad
($30^\circ$), with $R_M=250$~mm the radius of the measured field of view
and $R_S=500$~mm that of the source trajectory
(Section~\ref{sec:library:pipeline}).  The family covers the design space from a nearly flat filter
(small $\alpha$) through Hsieh-style cylinder profiles
($n\approx 2$) to aggressive edge attenuation (large $\alpha$, small
$n$); the fitted design sits at $n=1.5$.

\subsection{TCM parameterization}
\label{app:n0fc:tcm}

\emph{Supports main Section~\ref{sec:n0fc:param}, and records a Section~\ref{sec:discussion:limitations} disclosure: the knot count was chosen on a test patient's slice.}

$\tau(\theta)$ is piecewise-linear on $N_\tau{=}72$ knots uniformly
spaced on $[0,2\pi)$ and periodic at $2\pi$, with knot values clipped
to $[-\ln 10,+\ln 10]$, a tube-current factor
$\exp(\tau)\in[0.1,10]$.  The knot count was chosen empirically:
$N_\tau{=}36$ produced interpolation errors up to $20\%$ of the
$\tau$-range on PIS~slice~100 (PIS is a test patient) with uniform
bow-tie, and $N_\tau{=}72$ reduces this to ${\sim}10\%$.  Each
(patient, slice) pair has its own independently optimized $\tau$
vector, with no $z$-smoothness regularization: the unconstrained
per-slice solution bounds the equalization achievable under the
parameterization (the objective $J_2$ of
Section~\ref{app:n0fc:objective}) without committing to a hardware
$z$-smoothness constraint, which is a deployment-side concern.  It
does not bound the CRB floor.  Over all $24$ volumes (PIS\_NFLTD
included), the per-volume median floor changes by $-4$ to $+3\%$ with
the TCM taken from a slice up to ten slices away or replaced by one
angular profile per patient, and with the bow-tie alone it is no higher
($-16$ to $-0.5\%$).  The TCM acts on the tail: the per-volume mean
floor rises by at most $4\%$ with a TCM from up to ten slices away, by
$1.4$--$14\%$ with a single profile, and by $1.5$--$33\%$ with no TCM
(main Section~\ref{sec:results:floor}).

Over the $6{,}413$ slices of the $27$ volumes with $\NoFC$ designs
(the $23$ non-test volumes, PIS\_NFLTD and the test slices of the three
test patients), the per-slice range $e^{\tau_{\max}-\tau_{\min}}$ has
median $3.7{\times}$, 95th percentile $26{\times}$ and maximum
$89{\times}$ (LIDC-0766, whose $\tau_{\max}$ reaches the clip);
$21\%$ of slices, in $23$ of the $27$ volumes, exceed $10{\times}$.
A bundle's three source rays lie in one slice, so their tube-current
ratio is at most that slice's range.  The schedules therefore show
what a modulation range of up to $100{\times}$ allows; a narrower
hardware range would constrain the slices that use more than it.

\emph{Angles.}  The arguments $\gamma_k$ and $\theta_k$ of main
equation~\eqref{eq:n0fc} follow from reading the simulated fan beam
reciprocally.  A simulated source at angle $\theta_d$ on the circle of
radius $R=500$~mm emits $999$ rays at fan angles
$\gamma_r=(r-500)\Delta\gamma$, $\Delta\gamma=60^\circ/998$ (main
Section~\ref{sec:library:pipeline}).  That source position is read as a
TCT detector cell and the other end of each ray, on the same circle, as
a TCT source position (\emph{assumption}: equal source and detector
radii).  The isocenter and the ray's two ends form an isosceles
triangle with base angles $|\gamma_r|$, so the ray meets the circle
again at $\theta_r=\theta_d+\pi+2\gamma_r$ --- which the bundle
generator uses --- and $\gamma_r$ is also the ray's angle at its TCT
source, where $B(\gamma)$ acts.  A bundle takes rays $r$, $r+333$ and
$r+666$ ($r=1,\dots,333$), whose fan angles differ by
$333\,\Delta\gamma=20.02^\circ$, so its three source positions sit
$40.04^\circ$ apart and $\tau$ is evaluated at three angles.

\subsection{Bow-tie and TCM fit objectives}
\label{app:n0fc:objective}

\emph{Supports main Section~\ref{sec:n0fc:param}, and records a Section~\ref{sec:discussion:limitations} disclosure: the fit population and that no test-pool patient contributes to $B^*$.}

Both fits act on the deterministic per-ray detected intensity
$I_k=\Phi_k=\Nzeroray{k}\,e^{-X_k}\propto B(\gamma_k)\,e^{\tau(\theta_k)}\,e^{-X_k}$
(the detected fluence of main Section~\ref{sec:n0fc:problem};
$N_{\mathrm{base}}$ cancels in every quantity below) and weight each
bundle $b$ by $w_b=(\Delta X_b)^2$, where
$\Delta X_b=\max_k X_k-\min_k X_k$ is its within-bundle line-integral
contrast, so that high-contrast bundles dominate.  Let
$S_2=\{1,2\}$, $S_3=\{1,2,3\}$, $S_4=\{2,3\}$ be the active rays of the
multi-ray rows and
$\mathrm{DR}_j=\max_{k\in S_j}I_k\,/\min_{k\in S_j}I_k$.  The two
objectives are
\begin{align}
  J_1(B) &= \frac{\sum_b w_b\,\tfrac13\sum_{j=2}^{4}
            \ln^2\mathrm{DR}_j^{(b)}}{\sum_b w_b},
  \label{eq:J1}\\
  J_2(\tau_{p,z}) &= \sum_{b\in(p,z)} w_b\sum_{j=2}^{4}
            \mathrm{Var}_{k\in S_j}\bigl[\ln I_k^{(b)}\bigr].
  \label{eq:J2}
\end{align}
Both vanish exactly at within-row equalization; the variance in $J_2$
is the smooth surrogate of $\ln^2\mathrm{DR}_j$ that makes the TCM fit
quadratic in the knot values.

The bow-tie fit minimizes $J_1$ over the four parameters, within the
bounds of Section~\ref{app:n0fc:bowtie}, by differential evolution.
Its reference population is $15$ slices (axial quantiles $0.2$, $0.35$,
$0.5$, $0.65$, $0.8$), $5.1\times10^6$ bundles, from three non-test
patients drawn from the low, mid and high water-equivalent diameter
(WED) strata of Table~\ref{tab:corpus-dif-stratum}: LIDC-0810
($225$~mm), LIDC-0039 ($258$~mm) and LIDC-0628 ($317$~mm).  The
extreme stratum's only in-distribution patient, LIDC-0127, joined the
corpus after this fit.  The fit alternates once with the TCM fit --- the
$B^*$ obtained at $\tau\equiv0$ is refitted given the resulting $\tau$
on the same slices --- and the refit moved no parameter by more than
$1\%$ of its bound range, so the alternation was stopped.  The TCM fit
minimizes $J_2$ independently for each (patient, slice) over the $72$
knot values with $B^*$ fixed, by an iterated per-knot weighted-mean
(Jacobi-type) update with the knots clipped as above.

The objectives were fixed in advance; no ranking over alternative
objectives or candidate designs was performed, and no test-pool patient
contributes to $B^*$.  Two of the four fitted parameters lie at their
bounds ($\alpha=6.0$ upper, $n=1.5$ lower) and a third nearly so
($\gamma_{\mathrm{flat}}=0.16^\circ$, lower bound $0$), so short of the
floor the ramp attenuates nearly as strongly as the family allows; the
floor itself is an interior optimum ($B_{\min}=0.0062$ and $0.0065$ in
the two alternation steps, against a bound of $0.005$).  $B^*$
attenuates $26\times$ at $|\gamma|=20^\circ$, $95\times$ at $25^\circ$
and $155\times$ beyond $26.7^\circ$.  For comparison, the one clinical
bow-tie whose peripheral thickness we found reported numerically, that
of a Siemens SOMATOM Definition~64, reaches $6.05$~cm (large field of
view) and $8.57$~cm (small) of Teflon at the fan
edge~\citep{hassan2018}; for a $120$~kVp beam with $12$~mm Al
filtration we estimate attenuations of about $12\times$ and $32\times$.
Whether wider bounds would equalize further (lower $J_1$) is not
tested here.

\subsection{$N_{\mathrm{base}}$ calibration}
\label{app:n0fc:nbase}

\emph{Supports the per-patient exposure calibration of main Section~\ref{sec:library:split} and the photon budgets quoted in Section~\ref{sec:n0fc:param}.}

$N_{\mathrm{base}}$ is calibrated per patient, like automatic exposure
control but for a fixed mean detected count.  Under the \emph{baseline}
acquisition (uniform $\Nzero$, $B\equiv1$, $\tau\equiv0$) it makes the
mean of $N_{\mathrm{base}}\sum_{k=1}^{3}e^{-X_k}$, the detected count
of one exposure of a bundle's three source rays, equal a target
$\bar{N}_{\mathrm{det}}=3\times10^{5}$~photons, i.e.\ $10^{5}$ per ray;
each ray is read in three of the five rows, so a bundle's five readings
record three times this count.  In closed form,
\begin{equation}
  N_{\mathrm{base}} \;=\;
  \frac{\bar{N}_{\mathrm{det}}}
       {\bigl\langle\, \sum_{k=1}^{3} e^{-X_k} \,\bigr\rangle},
  \label{eq:nbase}
\end{equation}
the average running over all bundles of every line-integral sinogram
generated for the patient.\footnote{LIDC-0628 was first a training
patient.  Its training-set bundle files, used by the corpus floors,
predate its validation-set files and carry the earlier calibration
($N_{\mathrm{base}}$ $0.17\%$ higher) and a separately optimized
$\tau$.}  This is a one-shot computation, and because the denominator
uses only the attenuation field the baseline and $\NoFC$ designs share
the same $N_{\mathrm{base}}$: the calibration is independent of
$(B,\tau)$, which reshape the per-ray fluence about this fixed scale
rather than re-normalizing it.  The single global
$\bar{N}_{\mathrm{det}}$ across patients is what makes the estimators'
DIF figures of Section~\ref{sec:results} comparable (the CRB floor does
not depend on the photon level).  Because larger patients
attenuate more, $N_{\mathrm{base}}$ rises with patient size: over the
$26$ corpus volumes it runs from $1.80$ to $5.25\times10^{5}$, with a
Spearman correlation of $0.97$ with WED.  PIS\_NFLTD is the exception,
its $N_{\mathrm{base}}$ copied from PIS (main
Section~\ref{sec:library:split}).

\paragraph{Realized photon budget.}  Because the baseline and $\NoFC$
share $N_{\mathrm{base}}$ and $B\le1$, they differ in total photon count.
Counted over $24$ evenly spaced slices per volume (about
$8\times10^{6}$ bundles each), the $\NoFC$/baseline ratio of incident
photons on rays crossing the patient ($X_k>0.1$) is $0.35$--$0.62$ over
the $23$ non-test volumes (median $0.43$), $0.36$--$0.46$ for the three
test patients and $0.30$ for PIS\_NFLTD.  Over all rays it is
$0.28$--$0.48$ in the corpus (median $0.30$; standard error at most
$0.011$), because $13$--$48\%$ of rays miss the patient and $\NoFC$
gives those $0.7$--$3.5\%$ of their baseline photons.  Before clipping
a slice's all-ray ratio is $\bar{B}\,\langle e^{\tau}\rangle_\theta$,
with $\bar{B}\approx0.278$ the bow-tie's mean transmission over the
TCT ray positions; since each slice's knots average close to zero,
$\langle e^{\tau}\rangle_\theta$ grows with the modulation range
(Jensen's inequality), reaching $1.01$--$1.72$ per volume after
clipping (median $1.09$).  Dividing it out, the bow-tie alone would
deliver about $0.32$--$0.51$ of the baseline's incident photons on rays
through the patient.  The mean detected count per ray is
$1.3$--$7.4\times10^{3}$ under $\NoFC$ (PIS\_NFLTD $970$) against
$10^{5}$ under the baseline ($99{,}950$--$100{,}163$ measured;
PIS\_NFLTD $38{,}341$): the bow-tie removes the thin and air rays that
dominate the baseline's detected count.  Photons that interact in the
patient, $\Nzeroray{k}(1-e^{-X_k})$ summed over rays, are a closer
proxy for absorbed energy in this monoenergetic model.  Their
$\NoFC$/baseline ratio is $0.38$--$0.69$ over the $23$ non-test volumes
and $0.33$--$0.52$ on the four test volumes (LIDC-0951 $0.519$, PIS
$0.464$, LIDC-0116 $0.405$, PIS\_NFLTD $0.333$).  It exceeds the
all-ray incident ratio in every volume: with $f$ the baseline's mean
of $1-e^{-X_k}$ over rays, the incident ratio is $f$ times the
interacting ratio plus $1-f$ times the detected ratio, and the
detected ratio, $0.013$--$0.074$, lies below the incident ratio, so the
interacting ratio lies above it.

\subsection{Clipping}
\label{app:n0fc:clip}

\emph{Supports main Section~\ref{sec:n0fc:param}, and records a Section~\ref{sec:discussion:limitations} disclosure: the lower bound was relaxed on observation over volumes that included test patients.}

After~\eqref{eq:n0fc} and~\eqref{eq:nbase} are evaluated, the
resulting $\Nzeroray{k}$ is clipped to a hardware-feasible range
$[10^3, 10^6]$ photons per ray per bundle.  The lower bound was
relaxed from an initial $7.5\!\times\!10^4$ on observation, over the
volumes then available (test patients included), that the original
value caused $60\text{--}73\%$ of rays to clip on the floor,
suppressing the design's intended peripheral attenuation.
At the relaxed bounds, the bundles of the $27$ volumes with $\NoFC$
designs clip $0.6$--$8.2\%$ of rays at the floor (median $2.0\%$; overall $2.7\%$; LIDC-0127 lowest, LIDC-0766 highest). At most $0.3\%$ clip at the ceiling (overall $0.27\%$), except LIDC-0127 ($0.8\%$), LIDC-0766 ($3.2\%$) and LIDC-0116 ($3.3\%$).

\subsection{Theoretical analysis of $\NoFC$ equalization}
\label{app:n0fc:theory}

\emph{Supports the design rationale of main Sections~\ref{sec:n0fc} and~\ref{sec:disc:design}: why equalization is the target, and what it leaves on the table.}

This subsection collects the theoretical claims summarized in
Section~\ref{sec:n0fc}: ideal equalization, the exponential barrier that
prevents direct attainment, and the dependence of the CRB on the
within-bundle fluence pattern.

\subsubsection{Ideal equalization.}

If per-ray fluence could be set freely, the prescription
$\Nzeroray{k}\propto e^{+X_k}$ yields $\Phi_k\equiv C$ for all~$k$.
Substituting into~\eqref{eq:fim} gives
\begin{equation}
  F_{km}^{\mathrm{ideal}} \;=\;
  \sum_j p_{jk}p_{jm}/y_j \;=\; C\,M_{km},
  \label{eq:fim_ideal}
\end{equation}
where $M_{km} = \sum_j A_{jk}A_{jm}/(\sum_{k'}A_{jk'})$ is the
equal-attenuation reference matrix from Paper~1's CRB analysis.  Every
bundle's contrast factor $\DIF_k^{\mathrm{contrast}}$ collapses to
unity and the per-bundle CRB attains its equal-attenuation value
regardless of anatomy.  Whether this equalized allocation is also the
variance-optimal per-ray design is examined in
Section~\ref{app:n0fc:opt}.

\subsubsection{Exponential barrier.}

The ideal prescription requires a per-bundle tube-current ratio
$\exp(\Delta X)$ between adjacent sources, where $\Delta X$ is the
within-bundle attenuation contrast.  At $\Delta X=6.5$~units (an air
endpoint adjacent to a dense middle path) it demands $665{\times}$
between the rays of one bundle, far beyond the range our unconstrained
schedules span even over a full rotation (median $3.7{\times}$, at most
$89{\times}$; Section~\ref{app:n0fc:tcm}).  Linear current modulation
alone gives only logarithmic relief against exponential contrast; the
bow-tie $B(\gamma)$ contributes a second multiplicative factor of up to
$155{\times}$ in passive attenuation ($B^*$, fitted $B_{\min}=0.0065$;
Section~\ref{app:n0fc:objective}).

\subsubsection{CRB and the within-bundle fluence pattern.}

Define the achieved detected-fluence ratio per row $j$ as
$q_j = \min_{k:A_{jk}\neq 0}\Phi_k / \max_{k:A_{jk}\neq 0}\Phi_k$,
with $q_j = 1$ at perfect equalization.  Each FIM element
of~\eqref{eq:fim} is
\begin{equation}
  F_{km} \;=\;
  \sum_j A_{jk}A_{jm}\,\frac{\Phi_k\Phi_m}{\sum_{k'}A_{jk'}\Phi_{k'}},
\end{equation}
which depends on the whole pattern $(\Phi_1,\Phi_2,\Phi_3)$ and not on
the $q_j$ alone; the CRB is computed from it exactly throughout this
paper.  Two one-parameter families show the range
(Table~\ref{tab:dif_pattern}).  With the middle ray dimmest,
$\Phi\propto(1,q,1)$, $\DIF_2=3\,\Phi_2\,[F^{-1}]_{22}$ rises from
$13/3$ at $q=1$ somewhat more slowly than $1/q$ (tending to $3/q$ as
$q\to0$), while $[F^{-1}]_{22}$ itself rises faster than $1/q$; with
the middle ray brightest, $\Phi\propto(q,1,q)$, $\DIF_2$ falls below
$13/3$.  No scaling in $\langle q_j\rangle$ describes both, so the
paper reports measured reductions instead
(Section~\ref{sec:n0fc:empirical}: $\sigCRB_2$ falls $12$--$14$-fold at
bins~9--10).

\begin{table}[ht]
\centering
\caption{Middle-path $\DIF_2=3\,\Phi_2\,[F^{-1}]_{22}$ for two
one-parameter fluence patterns ($\DIF_2$ is invariant to the overall
fluence scale).}
\label{tab:dif_pattern}
\small
\begin{tabular}{@{}lccccc@{}}
\toprule
$q$ & $1$ & $0.9$ & $0.7$ & $0.5$ & $0.4$ \\
\midrule
Middle ray dimmest, $\Phi\propto(1,q,1)$   & $4.33$ & $4.68$ & $5.66$ & $7.40$ & $8.92$ \\
Middle ray brightest, $\Phi\propto(q,1,q)$ & $4.33$ & $4.02$ & $3.39$ & $2.75$ & $2.42$ \\
\bottomrule
\end{tabular}
\end{table}

\subsubsection{Optimality of the equalized allocation.}
\label{app:n0fc:opt}

The equalization objective minimizes the within-row coefficient of
variation of $\Phi_k$, a surrogate for the per-bundle estimation
variance rather than a minimizer of it, and the two minimizers differ
even at equal attenuation: the reference matrix $M$
of~\eqref{eq:fim_ideal} is not path-uniform
($[M^{-1}]_{22}=13/9>[M^{-1}]_{11}=7/9$, the middle being the coupled
bottleneck), so the variance-optimal allocation over-serves the middle
($\Phi_2^\star>\Phi_1^\star=\Phi_3^\star$).  To quantify the cost of
this suboptimality across realistic anatomy rather than on a symmetric
slice, we solved the per-bundle A-optimal allocation for
every bundle of the $26$ corpus volumes and the PIS\_NFLTD
probe under the $\NoFC$ allocation, $2.19\times10^{9}$ bundles.
LIDC-0628 entered twice through a duplicated patient list, so the
statistics below are over $2.28\times10^{9}$ bundle instances;
counting it once moves the bundle-weighted mean reduction by $0.004$
percentage points and
no other reported figure at the precision quoted.

\paragraph{The allocation problem.}  For one bundle with line
integrals $\svec X$, take the detected fluences
$\Phi_k=\Nzeroray{k}e^{-X_k}$ as the design variables and solve
\begin{equation}
  \min_{\svec\Phi>0}\ \tr F^{-1}(\svec\Phi)
  \quad\text{subject to}\quad
  \svec c^{\!\top}\svec\Phi=D,
  \qquad c_k=e^{X_k},
  \label{eq:aopt}
\end{equation}
with $F(\svec\Phi)=\sum_{j}\svec g_j\svec g_j^{\!\top}/y_j$,
$g_{jk}=A_{jk}\Phi_k$ and $y_j=\sum_k g_{jk}$ as
in~\eqref{eq:fim}.  The budget is the bundle's incident photon total
$D=\sum_k\Nzeroray{k}$, held at its realized value, so the comparison
is at fixed bundle dose; the attenuations enter the objective not at
all and the constraint only through the weights $c_k$, and the
back-map is $\Nzeroray{k}^\star=\Phi_k^\star e^{X_k}$.  The
only other constraint is positivity: the hardware range of
Section~\ref{app:n0fc:tcm} and the clip of
Section~\ref{app:n0fc:clip} are \emph{not} imposed, which is why the
result below is an oracle upper bound rather than a deliverable
design.  Imposing them would only shrink the feasible set, so the
oracle reduction bounds the constrained one from above,
bundle by bundle and hence in every quantile; at extreme contrast the
constrained set can be empty, since equalization itself demands up to
$665{\times}$ (Section~\ref{app:n0fc:theory}), and there the binding
quantity is the shortfall \emph{from} equalization rather than any
gain beyond it.  The objective is homogeneous of degree $-1$ in $\svec\Phi$,
so only the allocation direction is at issue.  The problem is
non-convex --- the Hessian of $\tr F^{-1}$ is positive definite at
equalization but indefinite elsewhere --- so it is solved by multistart
interior point in three variables under one constraint, without
assuming unimodality; multistart does not guarantee the global
optimum, so the reductions below are lower estimates of the oracle's
headroom.  At equal attenuation it has a closed-form
anchor used as the solver's self-test: equalization gives
$\tr F^{-1}=7/3+13/3+7/3=9$ exactly, while the optimum is the root
$s^\star\approx0.28398$ of $8s^3-11s^2+6s-1=0$, a middle-to-endpoint
ratio $\Phi^\star_{\mathrm{mid}}/\Phi^\star_{\mathrm{end}}\approx1.521$
and $\tr F^{-1}\approx8.2214$, a gap of $9.47\%$.

The headroom is real but modest, and---being the gap to a per-bundle
minimum---nonnegative by construction.  Relative to equalization the
achievable $\tr F^{-1}$ reduction has a corpus median of $2.2\%$ and
mean of $7.0\%$ ($90$th and $99$th percentiles $20\%$ and $43\%$;
per-patient means $4.8$--$10.1\%$).  The gap between median and mean
reflects a heavy upper tail: the decile of bundles with the largest
achievable reduction carries $44\%$ of the total, and that tail is not
confined to high contrast, bundles with $\Delta X\ge5$ holding $34\%$
of the bundles and $31\%$ of the total reduction.  The typical
reduction is largest near equal attenuation (median $7.9\%$ for
$\Delta X<0.5$, against $2$--$3\%$ for $\Delta X\ge3$); at exactly equal
attenuation it is $8.7\%$, the optimal reallocation over-serving the
middle by ${\sim}1.5{\times}$ and lowering $\DIF_2$ from $13/3$ to
${\approx}3.3$.  Over the corpus as a whole, though, the A-optimal
allocation leaves the bottleneck essentially untouched: its corpus-mean
$\DIF_2=4.44$ is marginally \emph{above} equalization's
$13/3\approx4.33$---trace optimization buys total variance by trading
the worst path, not by improving it.  A worst-ray (minimax-$\DIF$)
criterion would instead cut the bottleneck to ${\approx}2.54$
(${\sim}41\%$), but that optimum is geometry-only and demands per-ray
ratios more extreme than equalization itself, beyond the hardware range
of the exponential barrier above.  These are idealized stakes, not
deliverable gains.

Two findings indicate that re-optimizing the within-bundle flux
objective is not a productive lever.  First, the headroom is an
\emph{oracle}: the A-optimal allocation depends on the per-bundle line
integrals the scan is acquiring, so it cannot be prescribed in advance.
A deployable over-modulation of the equalization target,
$\Phi_k\propto e^{\kappa X_k}$ with a single scout-estimable exponent,
recovers only ${\sim}22\%$ of the oracle headroom at \emph{every}
granularity tested (global, per-patient, per-slice, and
contrast-scheduled $\kappa(\Delta X)$), each patient's optimal exponent
being $\kappa\approx-0.05$, i.e.\ effectively equalization; even an
oracle-informed per-bundle exponent reaches only $45\%$, because the
variance-optimal allocation is a two-degree-of-freedom direction, not
an exponential function of ray attenuation.  Second, under the
\emph{realized} $\NoFC$ allocation the median bundle's $\tr F^{-1}$ is
$1.31{\times}$ its value at exact equalization ($1.16$--$1.66{\times}$
per patient), with a heavy upper tail (corpus mean $2.7{\times}$).
This realization gap---a ratio of CRBs, set by the achieved
allocation's shortfall from equalization---dwarfs the ${\sim}2\%$
median beyond-equalization headroom, and changing the allocation
objective does not address it.  The remaining gains therefore lie with
constrained acquisition co-design, with the prior-level corrector (V8c,
T2bp), and with realizing equalization more faithfully---not with
surpassing it; equalization is retained as the $\NoFC$ design target on
this basis.

\subsection{Single-source and TCT fluence profiles}
\label{app:n0fc:profiles}

\emph{Supports main Sections~\ref{sec:n0fc:problem}, \ref{sec:results:floor} and~\ref{sec:disc:design}: equalization's doubled exponent, and that a conventional bow-tie captures most, but not all, of $\NoFC$'s floor reduction.  Table~\ref{tab:bowtie_compare} is the control behind that claim.}

Single-source CT and TCT call for different fluence profiles, which is
why a conventional bow-tie captures most, but not all, of $\NoFC$'s floor
reduction (main Section~\ref{sec:disc:design}).

\emph{Single-source optimum.}  A ray $i$ with $N_i$ incident photons
and line integral $X_i$ has detected mean $N_ie^{-X_i}$, so
$-\ln(\text{counts}/N_i)$ has variance $e^{X_i}/N_i$ at large counts.
Minimizing $V=\sum_iw_ie^{X_i}/N_i$ over $N_i>0$ subject to
$\sum_iN_i=D$, with reconstruction weights $w_i>0$ and $D$ the photon
total as dose proxy, has stationarity
$-w_ie^{X_i}/N_i^2+\lambda=0$, hence
$N_i\propto\sqrt{w_i}\,e^{X_i/2}$ with
$\sqrt{\lambda}=\sum_i\sqrt{w_ie^{X_i}}/D$; the Lagrangian Hessian is
diagonal with entries $2w_ie^{X_i}/N_i^3>0$, so $V$ is convex and this
is the minimum, $V_{\min}=(\sum_i\sqrt{w_ie^{X_i}})^2/D$.  The detected
counts at the optimum, $N_ie^{-X_i}\propto\sqrt{w_i}\,e^{-X_i/2}$, are
\emph{not} equalized: this is the square-root law that tube-current
modulation follows \citep{gies1999}.  For a static bow-tie giving every
ray at fan angle $\gamma$ the same $N\,B(\gamma)$ photons, the same
steps with one unknown per angle give
$B(\gamma)\propto\langle w\,e^{X}\rangle_\gamma^{1/2}$, and with equal
weights $B(\gamma)\propto\langle e^{X}\rangle_\gamma^{1/2}$ --- the
per-patient design of Table~\ref{tab:bowtie_compare}.

\emph{TCT target.}  A TCT reading sums the detected fluences of up to
three rays, and the information it carries about a dim ray scales as
$\Phi_{\mathrm{dim}}^2/y_j$, which collapses when
$\Phi_{\mathrm{dim}}\ll\Phi_{\mathrm{bright}}$ (main
Section~\ref{sec:n0fc:problem}).  Equalizing within the bundle needs
$\Nzeroray{k}\propto e^{X_k}$ (equation~\eqref{eq:fim_ideal}):
exponent~$1$ instead of~$1/2$.  From center to edge the attenuation is
therefore $e^{\Delta X/2}$ under the square-root law against
$e^{\Delta X}$ under equalization --- $12$ against $148$-fold for
$\Delta X=5$, $33$ against $1{,}097$-fold for $\Delta X=7$
(\emph{assumption}: a center-to-edge $\Delta X$ of $5$--$7$ for adult
chests) --- where the one clinical bow-tie quantified in
Section~\ref{app:n0fc:objective} attenuates about $12$--$32$-fold and
$\NoFC$'s fitted $B^*$ $155$-fold.

\emph{Comparison on the corpus.}  Table~\ref{tab:bowtie_compare} sets
$\NoFC$ against conventional designs at the same reference tube
current.  The GE-type profiles capture most of $\NoFC$'s log reduction
of the median floor, $0.67$--$0.77$, against $0.21$--$0.60$ for the
S2.1-family ramps and $0.59$ for the square-root law, because they ramp
early and starve the outer rays at $10$--$25^\circ$ where a bundle's
outer rays sit; the remaining gain is the depth of $B^*$, which alone
matches $\NoFC$ at the median ($4.8$ against $5.1$), while $\NoFC$'s
mean floor is about three times below the GE-type profiles'.  The
square-root law is not optimal for TCT even on the quantity it
minimizes for independent rays: at equal mean CRB variance over all
rays through the patient (a criterion not tabulated; the last column of
Table~\ref{tab:bowtie_compare} equates the median middle-path variance
instead), the per-patient square-root-law bow-tie needs
$0.10$ of the baseline's photons, $B^*$ alone $0.03$ and $\NoFC$
$0.02$.  The ratio of each conventional design's median floor to
$\NoFC$'s falls with WED (Spearman $-0.81$ to $-0.89$ over the $23$
volumes): in large patients a conventional bow-tie comes closer to
$\NoFC$.

\begin{table}[ht]
\centering
\caption{Conventional bow-ties against $\NoFC$ on the $23$ non-test
volumes: medians over volumes of each volume's value, from $24$
sampled slices per volume (so values differ slightly from the
all-slice values of main Section~\ref{sec:results:floor}; for example,
$\NoFC$'s median floor is $5.06$ here against $5.0$ there).  Every
design uses the volume's own $N_{\mathrm{base}}$, the same reference
tube current, and no tube-current modulation except $\NoFC$.  S2.1-family
ramps: the family of~\eqref{eq:bowtie} with a $3^\circ$ flat center,
no floor, edge attenuation $12\times$ or $32\times$ at the $30^\circ$ fan
edge and ramp exponent $n$.  Square-root law: per patient,
$B(\gamma)\propto\langle e^{X}\rangle_\gamma^{1/2}$ (derivation above).
GE-type: the three bow-tie files of the XCIST simulator
\citep{wu2022}, two of which XCIST documents as approximations of GE
LightSpeed VCT bow-ties, converted to transmission at $120$~kVp
(photon-count weighted) and mapped onto this fan; edge attenuation
$11.5$--$11.8\times$.  Share: $\ln(\text{baseline}/\text{design})$ over
$\ln(\text{baseline}/\NoFC)$ for each volume's median floor $\DIF_2$.
Ratio: the design's median floor over $\NoFC$'s, per volume.  Mean
$\DIF_2$: each volume's mean floor over its bundles.  Dose: incident
photons on rays through the
patient, relative to the baseline.  Equal precision: the dose at which
the median middle-path CRB variance equals the baseline's.}
\label{tab:bowtie_compare}
\footnotesize
\setlength{\tabcolsep}{4pt}
\begin{tabular}{@{}lrrrrrr@{}}
\toprule
 & \multicolumn{2}{c}{Median $\DIF_2$} & & & \multicolumn{2}{c}{Dose} \\
\cmidrule(lr){2-3}\cmidrule(lr){6-7}
Design & Value & Share & Ratio & Mean $\DIF_2$ & Incident & Equal precision \\
\midrule
Baseline ($B\equiv1$)            & $27.9$ & $0$    & ---    & $145.1$ & $1.00$ & $1.00$ \\
Ramp, $12\times$, $n=3$          & $19.8$ & $0.21$ & $3.69$ & $79.6$  & $0.86$ & $0.58$ \\
Ramp, $12\times$, $n=2$          & $16.3$ & $0.34$ & $3.01$ & $59.2$  & $0.76$ & $0.43$ \\
Ramp, $12\times$, $n=1.5$        & $14.1$ & $0.42$ & $2.62$ & $48.5$  & $0.67$ & $0.34$ \\
Ramp, $32\times$, $n=3$          & $16.7$ & $0.32$ & $3.16$ & $61.8$  & $0.82$ & $0.44$ \\
Ramp, $32\times$, $n=2$          & $12.5$ & $0.49$ & $2.41$ & $40.9$  & $0.69$ & $0.30$ \\
Ramp, $32\times$, $n=1.5$        & $10.5$ & $0.60$ & $2.00$ & $31.3$  & $0.60$ & $0.23$ \\
Square-root law                  & $10.2$ & $0.59$ & $2.01$ & $26.3$  & $0.58$ & $0.23$ \\
GE-type, large                   & $9.0$  & $0.67$ & $1.72$ & $25.9$  & $0.57$ & $0.19$ \\
GE-type, medium                  & $8.0$  & $0.74$ & $1.52$ & $22.7$  & $0.50$ & $0.17$ \\
GE-type, small                   & $7.7$  & $0.77$ & $1.44$ & $22.5$  & $0.42$ & $0.16$ \\
$B^*$ alone                      & $4.8$  & $1.03$ & $0.96$ & $9.3$   & $0.39$ & $0.09$ \\
$\NoFC$                          & $5.1$  & $1$    & $1$    & $8.0$   & $0.43$ & $0.11$ \\
\bottomrule
\end{tabular}
\end{table}

\section{V8c implementation details}
\label{app:v8c}

V8c descends from V4, a single-patient middle-path corrector written
during the Paper~1 work and not used in that paper, through V7, a
corrector of the SNN2 residual, and V8, which added the three-path
head.  V4 is documented nowhere else, so it is described here from its
code, as far as V8c changes it.

\subsection{Architecture deltas from V4}
\label{app:v8c:arch}

\emph{Supports main Section~\ref{sec:library:v8c}; this is what lets the network be rebuilt.}

Table~\ref{tab:v8c_arch} gives the network as coded.  The four changes
from V4 are: a three-path output head, zero-initialized so an untrained
network reproduces SNN2 exactly, in place of V4's single $\hat\delta_2$
with single-row endpoints; the strict per-ray FIM~\eqref{eq:fim} in
closed form as six features, which is what informs V8c of the per-ray
counts when the $\Nzeroray{k}$ differ (V4 had none, Paper~1
having used uniform $\Nzero$); the 15-column bundle format, so
the normalized row count divides by $\sum_kA_{jk}\Nzeroray{k}$ and not
by a scalar $\Nzero$; and a trunk widened from $(256,256,128,64)$,
about $0.81$M parameters.  The analytic per-ray FIM is unit-tested
against closed forms and against the uniform-$\Nzero$ formula, and the
evaluation loader re-verifies the stored FIM on the first $1000$
bundles of every file.

\begin{table}[ht]
\centering
\caption{V8c as coded (PyTorch, $3\,008\,323$ parameters); feature
indices from~1.  LN: LayerNorm \citep{ba2016S}.  Half precision
resolves $2^{-8}\approx0.004$ on $[4,8)$, against a smallest SNN2
residual spread of $0.039$ over the $X_2$ bins in that range.}
\label{tab:v8c_arch}
\small
\begin{tabular}{@{}ll@{}}
\toprule
\multicolumn{2}{@{}l}{\emph{Input features} (26)} \\
\midrule
1--5   & normalized row counts $I_{\mathrm{norm},j}=Y_j/\sum_kA_{jk}\Nzeroray{k}$ \\
6      & count level, $\log_{10}$ of the bundle-mean $\Nzeroray{k}$ \\
7--9   & warm start $\hat{\svec{X}}^{\mathrm{SNN2}}$ \\
10--15 & per-ray FIM~\eqref{eq:fim} at $\hat{\svec{X}}^{\mathrm{SNN2}}$: $F_{11},F_{12},F_{13},F_{22},F_{23},F_{33}$ \\
16     & middle-path CRB variance $[F^{-1}]_{22}$ \\
17--21 & log-domain measurements $-\ln I_{\mathrm{norm},j}$ \\
22--26 & Fisher shares $\phi_j=F^{\mathrm{tot}}_j/\sum_{j'}F^{\mathrm{tot}}_{j'}$, $F^{\mathrm{tot}}_j=\sum_kp_{jk}^2/y_j$ \\
\addlinespace
\multicolumn{2}{@{}l}{\emph{Layers and conventions}} \\
\midrule
Scaling & standardized per feature; moments from $10^6$ random training bundles \\
Entry   & linear $26\to512$, LN, ELU \\
Trunk   & gated residual blocks, three at $512$, two at $256$, one at $128$ \\
        & transitions (LN, linear, ELU) $512\to256\to128\to64$ \\
Block   & $\mathbf x\mapsto\mathbf x+\mathbf g\odot T(\mathbf x)$,
          $T(\mathbf x)=W_2\,\mathrm{ELU}(W_1\mathrm{LN}_1\mathbf x+\mathbf b_1)+\mathbf b_2$, \\
        & $\mathbf g=\sigma(W_g\mathrm{LN}_2\mathbf x+\mathbf b_g)$;
          $W_1,W_2,W_g$ square; $\mathbf b_g$ init $-2$ \\
        & no inner expansion, no dropout (both in V4); gate input normalized \\
Head    & LN, linear $64\to32$, ELU, linear $32\to3$, zero-initialized \\
Output  & $\hat{\svec{X}}^{\mathrm{SNN2}}+(\hat\delta_1,\hat\delta_2,\hat\delta_3)$, no clipping \\
Cache   & standardized features, SNN2 estimates, true $X_k$ in half precision; \\
        & network computes in single precision \\
\bottomrule
\end{tabular}
\end{table}

\subsection{Loss function}
\label{app:v8c:loss}

\emph{Supports main Section~\ref{sec:library:v8c}, and records two Section~\ref{sec:discussion:limitations} disclosures: the loss setting was informed by a test patient's data, though not by its test evaluation, and a per-bin weighting was dropped partly on the test patients' errors.}

V8c is trained on the residuals
$\delta_k=X_k^{\mathrm{true}}-\hat X_k^{\mathrm{SNN2}}$ of all three
paths.  For a minibatch of $B$ bundles with errors
$e_{ik}=\hat\delta_{ik}-\delta_{ik}$, the loss is
\begin{equation}
  \mathcal{L}=\frac{1}{3B}\sum_{i=1}^{B}\sum_{k=1}^{3}
  w_k\big[h_{\Delta}(e_{ik})+\lambda\ln(1+e_{ik}^{2})\big],
  \label{eq:v8c_loss}
\end{equation}
with the Huber penalty $h_\Delta(e)=e^2/2$ for $|e|\le\Delta$ and
$\Delta(|e|-\Delta/2)$ beyond, $\Delta=0.5$, $\lambda=0.3$, and path
weights $(w_1,w_2,w_3)=(0.5,2,0.5)$.  Both terms grow more slowly than
a squared error, which limits the pull of the high-attenuation tail,
and the $X_2$ weight concentrates the network on the middle path, where
SNN2 leaves the largest residual.  No per-bundle weight is applied.

An earlier setting weighted each bundle by its CRB, which with the
exact per-ray FIM suppressed the gradient of the high-attenuation
bundles the corrector exists to correct: the validation high-bin score
stalled at $0.023$ after $60$ epochs, against $0.010$ by epoch~19
without it.  Both runs used single-patient PIS data, and PIS is one of
the three test patients (Section~\ref{sec:library:split}), so this
choice was informed by a test patient's data, though not by its test
evaluation.  A per-bin weighting of the $X_2$ term, tried on the
earlier $13$-patient corpus ($13$ training patients, before the second
batch of Section~\ref{app:split}), scored worse on validation ($0.399$
against $0.334$) and in the $X_2\in[6,7)$ error on each test patient,
and was dropped.

\subsection{Capacity sensitivity finding}
\label{app:v8c:capacity}

\emph{Supports main Section~\ref{sec:bridge}'s statement that model size was not the limit, and records a Section~\ref{sec:discussion:limitations} disclosure: the decision rule involved test data.}

A capacity study on the earlier $13$-patient corpus compared the
$3.0$M-parameter trunk against $6.2$M and $12.0$M variants, all else
fixed.  The $3.0$M trunk scored best on validation ($0.338$ against
$0.343$ for both variants); the wide variant matched it within $1.7\%$
on the test patients and the balanced variant was worse, so the $3.0$M
trunk was kept (full table: arXiv:2610.19229v2, Table~S4).  The decision
rule, fixed before the runs, kept the $3.0$M trunk unless a larger
variant cut
the high-bin $X_2$ error on LIDC-0116 by more than $2\%$, and
LIDC-0116 is one of the three test patients, so the rule involved test
data.  Model size is not the bottleneck in this family on that corpus;
further gains should need information from between bundles (Paper~3).

\subsection{Training}
\label{app:v8c:train}

\emph{Supports main Section~\ref{sec:library:v8c}'s training description, and the epoch selection and the $2{\times}2$ cross-evaluation used in main Section~\ref{sec:results:v8c} (Table~\ref{tab:2x2_pool}).}

The 26 features are computed once per bundle and cached in globally
shuffled order, one cache per pool and acquisition regime: training,
$19$ patients and $1\,583\,225\,856$ bundles; validation, $4$ patients
and $319\,509\,504$.  At $340\,992$ bundles per slice a patient's
share of the pool is its share of slices, $2.3\%$ (LIDC-0707) to
$11.6\%$ (LIDC-0684), and no per-patient reweighting is applied.  Each
epoch draws $154\,611$ start positions uniformly at random and reads
$2048$ consecutive bundles from each as one minibatch; the cache order
being random, every minibatch is a uniform sample of the pool and an
epoch covers $316\,643\,328$ bundles, $20\%$ of it.

The optimizer is AdamW \citep{loshchilov2019S} with learning rate
$3\times10^{-4}$, weight decay $10^{-5}$ and gradient-norm clipping at
$1$, on cosine annealing with warm restarts \citep{loshchilov2017S},
$T_0=20$ epochs, $T_{\mathrm{mult}}=2$ and minimum $3\times10^{-6}$;
the rate restarts at epoch~21, the first of a $40$-epoch cycle that
the $30$-epoch budget truncates.  Validation uses the full validation
pool, every epoch under $\NoFC$ and at epoch~1, every third epoch and
the last epoch under the baseline.  The checkpoint score is the
weighted mean, with weights $1$, $2$ and $1.5$, of the validation
$X_2$ standard deviation in bins~6, 7 and~8; bias is not scored.

Table~\ref{tab:v8c_train} gives both trajectories.  Under the baseline
the score varies by $1.3\%$ over all validated epochs and epoch~3 is
selected; under $\NoFC$ epoch~1 is selected, the score rising to
$0.245$ by epoch~20 while the training loss falls by $37\%$, and the
warm restart recovering only $0.226$.  Validation loss selects the
same epoch in both regimes.  The $\NoFC$ run stopped after epoch~26 of
the planned $30$ without an error message; every later epoch scores
worse than epoch~1, so the stop is unlikely to have changed the
selected checkpoint.  Training time on the GTX~1080, validation
excluded, is $52$~min per epoch with the cache on an NVMe
solid-state drive ($\NoFC$; $22.6$~h for $26$ epochs) and $107$~min on
a hard disk (baseline; $53.5$~h for $30$).

\emph{Cross-regime transfer.}\label{app:v8c:cross}  In the
$2\times2$ cross-evaluation a
baseline-trained V8c degrades sharply on $\NoFC$ data, while an
$\NoFC$-trained V8c transfers to baseline data (Table~\ref{tab:2x2_pool}):
cell~2's pool RMSE is $5.2\times$ (LIDC-0951), $5.1\times$ (PIS) and
$2.1\times$ (LIDC-0116) that of SNN2-$\NoFC$, and cell~3 improves on
SNN2-baseline by $30\%$, $22\%$ and $4\%$, the first two within $0.4\%$
and $7\%$ of the matched diagonal.  Two explanations are
plausible and neither is tested: the baseline-trained corrector may
apply corrections sized for a residual scale that $\NoFC$ removes; or
\emph{(hypothesis)} the per-ray Fisher features make baseline
acquisition (uniform $\Nzeroray{\cdot}$) a special case of what an
$\NoFC$-trained network learns, while a baseline-trained network never
sees the $\Nzeroray{\cdot}$ variation that $\NoFC$ inference requires.

\begin{table}[ht]
\centering
\small
\caption{$X_2$ pool $\mathrm{RMSE}$ for the full $2{\times}2$ cross-evaluation,
val-best checkpoint.
\texttt{diag} cells (1, 4) are matched train/eval regime; \texttt{off}
cells (2, 3) are mismatched. SNN2 columns are the prior-free
regime-specific references.  The baseline acquisition uses about twice
$\NoFC$'s incident photons (main Section~\ref{sec:n0fc:param}).}
\label{tab:2x2_pool}
\begin{tabular}{lcccccc}
\toprule
Patient & C1 & C1 & C4 & C4 & SNN2 & SNN2 \\
        & baseline & $\to\NoFC$ & $\to$baseline & $\NoFC$ & baseline & $\NoFC$ \\
        & (diag.) & (off.) & (off.) & (diag.) & & \\
\midrule
LIDC-0951 & 0.118 & 0.140 & 0.118 & 0.0244 & 0.168 & 0.0268 \\
PIS       & 0.115 & 0.160 & 0.123 & 0.0296 & 0.158 & 0.0312 \\
LIDC-0116 & 0.325 & 0.339 & 0.398 & 0.1865 & 0.415 & 0.1637 \\
\bottomrule
\end{tabular}
\end{table}

\begin{table}[ht]
\centering
\caption{V8c validation trajectories on the full validation pool.
Losses are $\times10^{3}$; the score is the weighted mean of the
validation $X_2$ standard deviation in bins~6--8 (lower is better).
The baseline run is validated at epoch~1, every third epoch and the
last epoch; ``last'' is epoch~30 (baseline) and epoch~26 ($\NoFC$).
The learning rate restarts after epoch~20.  Bold: selected
checkpoint.}
\label{tab:v8c_train}
\small
\begin{tabular}{@{}rrrrrrr@{}}
\toprule
& \multicolumn{3}{c}{Baseline} & \multicolumn{3}{c}{$\NoFC$} \\
\cmidrule(lr){2-4}\cmidrule(lr){5-7}
Epoch & train & val & score & train & val & score \\
\midrule
1    & 13.51 & 25.70 & 0.294 & 1.679 & \textbf{6.29} & \textbf{0.192} \\
3    & 12.70 & \textbf{25.67} & \textbf{0.292} & 1.379 & 7.66 & 0.210 \\
6    & 12.49 & 26.87 & 0.295 & 1.259 & 8.69 & 0.223 \\
9    & 12.38 & 25.91 & 0.293 & 1.189 & 8.87 & 0.225 \\
12   & 12.30 & 26.29 & 0.294 & 1.138 & 9.63 & 0.234 \\
15   & 12.22 & 26.59 & 0.296 & 1.096 & 10.07 & 0.240 \\
18   & 12.16 & 26.59 & 0.296 & 1.070 & 10.46 & 0.245 \\
21   & 12.43 & 26.00 & 0.293 & 1.188 & 8.97 & 0.226 \\
24   & 12.40 & 26.47 & 0.293 & 1.166 & 9.68 & 0.233 \\
last & 12.33 & 26.24 & 0.294 & 1.150 & 9.64 & 0.234 \\
\bottomrule
\end{tabular}
\end{table}

\subsection{Corpus split and its history}
\label{app:split}

\emph{Records the corpus signature and the Section~\ref{sec:discussion:limitations} disclosures on how the corpus split was formed.  It supports no result; it is here so the corpus and its split can be audited.}

The corpus spans the chest-anatomy variation that drives TCT bundle
statistics, measured by a six-component signature: $\WED$, ellipticity,
line-integral bin entropy and 95th percentile, and mean and
95th-percentile intra-bundle variance, which capture body habitus and
the attenuation and contrast distributions that drive the per-bundle
Fisher penalty.  It spans $\WED\in[206.9, 381.2]$~mm, ellipticity
$[1.18, 1.81]$ and bin entropy $[2.18, 2.92]$; PIS sits near the low end
at $244.4$~mm.

The placement of the validation and test patients (main
Section~\ref{sec:library:split}) is deliberate.  High and low
ellipticity sit in validation (LIDC-0766, $1.81$; LIDC-0709, $1.23$) so
angular overfitting surfaces in the validation-loss diagnostic;
LIDC-0052 is the entropy maximum for $240$--$285$~mm; and LIDC-0628
($316.8$~mm) brackets the validation/test high-$\WED$ gap so
high-$\WED$ overfit stays detectable.  LIDC-0303, -0579 and -0127, the patients above $320$~mm
other than the test patient LIDC-0116 (up to $381.2$~mm), are kept in
training because the paper does not test extrapolation in $\WED$.
Seven training patients (LIDC-0004, -0086, -0684, -1008, -0303, -0579
and -0127; $296.5$--$381.2$~mm) were added in a second batch,
motivated by the finding that an earlier $13$-patient V8c's $X_2$
error reduction fell with WED across its validation and test patients
(to $1.3\times$ in $X_2\in[6,7)$ on LIDC-0116); LIDC-0628 then moved
from training to validation.

Main Table~\ref{tab:dev-provenance} gathers every development choice that
used data from the test patients, with where each is described.

\section{DIF derivation and closed-form expressions}
\label{app:dif}

The closed-form Cram\'er--Rao DIF under uniform $\Nzero$ was
derived in Paper~1 \citep{besson2026pmb}: the per-ray FIM factorizes as
$F = \Nzero\,G(\alpha_1, \alpha_2, \alpha_3)$ with
$\alpha_k = e^{-X_k}$, and the resulting DIF
\begin{equation}
  \DIF_k^{\mathrm{CRB}}(X_1, X_2, X_3) \;=\;
  3\,\alpha_k\,[G^{-1}]_{kk}
  \label{eq:dif_crb}
\end{equation}
is $\Nzero$-independent.  Paper~1 also gives the explicit
$3 \times 3$ entries of $G$ and verifies the equal-attenuation
reference values $\DIF_1^{\mathrm{eq}} = \DIF_3^{\mathrm{eq}} = 7/3$,
$\DIF_2^{\mathrm{eq}} = 13/3$.  This section records the two extensions
new in Paper~2 --- the per-ray $\Nzeroray{k}$ generalization required by
$\NoFC$, and the empirical $\DIF$ statistics --- then the dose reading of
a DIF and the two volumes on which SNN2 exceeds the floor.

\subsection{Per-ray $\Nzeroray{k}$ generalization}
\label{app:dif:crb}

\emph{Supports the per-ray $\DIF$ definition of main
Section~\ref{sec:model:dif}, used throughout Section~\ref{sec:results}.}

$\NoFC$ (Section~\ref{sec:n0fc}) breaks the uniform-$\Nzero$ assumption
of Paper~1.  Generalizing the per-ray FIM~\eqref{eq:fim} to
distinct $\Nzeroray{k}$ leaves the Poisson score form intact:
with per-ray mean detected counts
$p_{jk} = \Nzeroray{k}\,A_{jk}\,\alpha_k$ ($\alpha_k = e^{-X_k}$)
and predicted row counts $y_j = \sum_k p_{jk}$,
\begin{equation}
  F_{km} \;=\; \sum_{j=1}^{5}\frac{p_{jk}\,p_{jm}}{y_j}
            \;=\; \Nzeroray{k}\,\Nzeroray{m}\,\alpha_k\,\alpha_m
            \sum_{j=1}^{5}\frac{A_{jk}\,A_{jm}}{y_j}.
  \label{eq:fim_perray_app}
\end{equation}
Two changes follow.  First, $F$ no longer factors as a scalar
$\Nzero$ times the dimensionless $G(\alpha_1,\alpha_2,\alpha_3)$
of the uniform case; the per-ray counts enter the rows of $F$
non-uniformly, so the $\Nzeroray{k}$ do not cancel out of
$[F^{-1}]_{kk}$.  Second, the equal-dose single-source floor in the
denominator of $\DIF$ is the conventional-CT CRB at the same
incident photon budget along ray~$k$.  Ray~$k$ is illuminated once by
each of the three sources (appearing in three of the five rows, the
column sum of $A$), each at the same nominal count $\Nzeroray{k}$, so
that the incident photon budget along ray~$k$ is $3\,\Nzeroray{k}$
and
\begin{equation}
  \sigma^2_{\mathrm{fair}}(X_k) \;=\; \frac{e^{X_k}}{3\,\Nzeroray{k}},
  \label{eq:sigfair_perray}
\end{equation}
reducing to $e^{X_k}/(3\Nzero)$ under uniform $\Nzero$.  The
structural form of~\eqref{eq:dif_crb} then carries over with
$G\to F$ and an explicit per-ray factor,
\begin{equation}
  \DIF_k \;=\; 3\,\Nzeroray{k}\,\alpha_k\,[F^{-1}]_{kk},
  \label{eq:dif_crb_perray}
\end{equation}
which loses the $\Nzeroray{k}$-independence of the uniform case:
the $\NoFC$ flux profile (Section~\ref{sec:n0fc:param}) is the entry
point through which dose-design choices flow into per-bundle DIF.

\emph{Single noise realizations.}\label{app:dif:single}  The estimator
DIF~\eqref{eq:dif_mse} and the floor~\eqref{eq:dif_def} (main text) are
expectations over the measurement noise.  A simulated bundle carries a
single realization, so its squared error
$(\hat X_k-X_k)^2/\sigma^2_{\mathrm{fair}}$ is one draw around
$\DIF_k^{\mathrm{MSE}}$; for an efficient estimator with Gaussian
errors it is distributed as
$\DIF_k\,\chi^2_1$, whose median is $0.455\,\DIF_k$.  The paper
therefore estimates~\eqref{eq:dif_mse} from squared errors averaged over
many bundles, pooled per bin (main
Section~\ref{sec:model:dif}) or, for per-volume summaries, within cells
of near-constant CRB floor (the unit $X_2$ bins crossed with bins $0.05$
decade wide in the floor $\DIF_2$; each bundle takes its cell's mean),
and does not summarize single-draw squared errors by their median.

\subsection{Empirical statistics on the corpus}
\label{app:dif:empirical}

\emph{Supports the contrast penalty motivating $\NoFC$ in main
Section~\ref{sec:n0fc} and the floor gains reported in
Section~\ref{sec:results:floor}.}

Two statistics, evaluated with~\eqref{eq:dif_crb} under uniform
$\Nzero$, illustrate the contrast penalty that the $\NoFC$ design
objective of Section~\ref{sec:n0fc} addresses.  First, over the bin-6
bundles ($X_2\in[5,6)$) of PIS\_NFLTD, Paper~1's PIS volume rescaled
\citep{besson2026pmb}, the pooled known-flux floor is
$5.69^2\approx32.4$ (Table~\ref{tab:paper1-floor}). It decomposes
multiplicatively as $(13/3)\,\cdot\,7.5$: a $7.5{\times}$ contrast
factor on top of the $4.33{\times}$ equal-attenuation term.  The mean of
the per-bundle $\DIF_2$ over the same bin, a different statistic, is
$40.6$ on $30$ slices.  Paper~1's pooled value on the PIS slices it held
out for testing,
$6.59^2\approx43.4$ (its Table~3), includes air-scan noise and Paper~1's
own slice selection, so it is not a like-for-like counterpart.
Second, on the highest-WED test patient (LIDC-0116) the per-bundle ratio
$\sigCRB_2/\sigfair$ at bin~9 has a median of about $36$ and exceeds
$10$ for $95\%$ of bundles (three slices, uniform $\Nzero$). Nearly all
of them have a within-bundle contrast $\Delta X\ge6$ (a weakly
attenuated endpoint next to a dense middle path).  Both show the penalty
that the equalization objective of Section~\ref{app:n0fc:objective}
removes; the bow-tie and the TCM approximate that objective within
their hardware range, and the extreme tail lies beyond it
(Section~\ref{app:n0fc:theory}).

\begin{figure}[!t]
\centering
\includegraphics[width=0.75\linewidth]{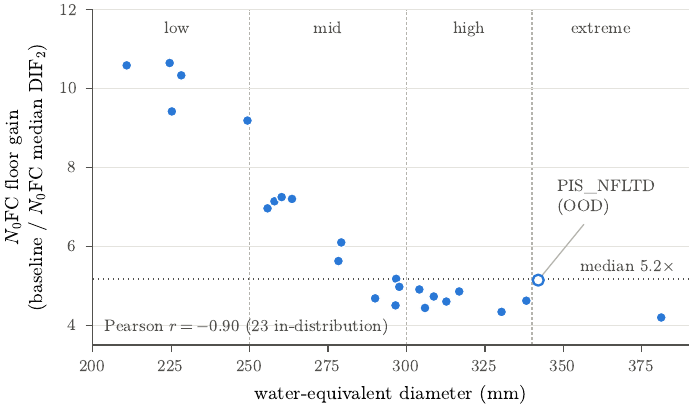}
\caption{Per-patient $\NoFC$ gain on the middle-path CRB floor,
$\DIF_2^{\mathrm{base}}/\DIF_2^{\NoFC}$ (exact median over each
patient's bundles), versus water-equivalent diameter, for the $23$
non-test volumes (filled) plus the out-of-distribution (OOD) probe PIS\_NFLTD (open);
dashed lines mark the WED strata of Table~\ref{tab:corpus-dif-stratum}.
The gain is largest for the smallest patients (Pearson $r=-0.90$):
their bright endpoints inflate the baseline floor worst, and $\NoFC$'s
intra-bundle equalization removes the most there.}
\label{fig:n0fc_floor_gain}
\end{figure}

Beyond these two statistics, the realized per-bundle CRB floor
quantifies what $\NoFC$ buys.  Per non-test volume, $\NoFC$
lowers the median middle-path floor $\DIF_2$ by $4.2$--$10.6\times$
(median $5.2\times$). The gain is largest at small WED, where bright
endpoints inflate the baseline floor most and $\NoFC$'s equalization has
the most to remove (Fig.~\ref{fig:n0fc_floor_gain}). The
out-of-distribution (OOD) probe PIS\_NFLTD ($5.1\times$ at $342$\,mm)
lies above the non-test volumes above $300$\,mm
($4.2$--$4.9\times$).  Under $\NoFC$ the floor lands at a per-patient
median $\DIF_2$ of $3.3$--$6.3$ (per-patient mean $5.2$--$16.0$, the
spread set by the high-attenuation tail of
Section~\ref{sec:results:floor}); the baseline absolutes follow as this
level times the gain.

\subsection{Dose reading of DIF}
\label{app:dif:dose}

\emph{Supports the reading of a DIF as a dose factor, used in main
Sections~\ref{sec:results} and~\ref{sec:discussion}, and disclosed in
Section~\ref{sec:discussion:limitations}.}

At the CRB, a DIF of $d$ means exactly that $d$ times the per-ray photon
budget restores $\sigfair$: under Poisson counting noise the
FIM~\eqref{eq:fim_perray_app} is linear in the per-ray counts, so
$(\sigCRB)^2$ and $\sigfair^2$~\eqref{eq:sigfair_perray}
both scale as $1/\Nzero$.  For an estimator the same reading also
requires its mean squared error to scale as $1/\Nzero$.  We checked this
for SNN2 on three corpus patients spanning the WED range (LIDC-0709,
$225$~mm; LIDC-0052, $264$~mm; LIDC-0127, $381$~mm) under both
acquisitions.  Per unit $X_2$ bin, $500$ bundles were drawn from each
patient's bundle file; for each, $400$ fresh Poisson realizations of the
five row counts were generated at budgets $s\,\Nzeroray{k}$,
$s\in\{1,2,4,8\}$, and inverted with the production SNN2 settings
(Section~\ref{app:snn2}).  Over the $44$ (patient, acquisition, bin)
cells:
\begin{itemize}
\item Raising each bundle's budget by its CRB DIF brings SNN2's
  middle-path mean squared error to $\sigfair^2$ at the original budget
  in every cell (median ratio $0.99$--$1.01$). At the higher count SNN2
  attains the bound, so the floor reads as a dose factor for an actual
  estimator.
\item SNN2's own DIF reads the same way where SNN2 is efficient (DIF
  within $5\%$ of the floor: all $\NoFC$ bins through $[5,6)$, and
  baseline bins through $[3,4)$, or $[5,6)$ for LIDC-0127). Rerunning at
  that DIF gives $0.97$--$1.01$, and the error scales as $1/\Nzero$ to
  within $8\%$ at $s=8$.  On LIDC-0766, not among these three, rare
  bundles keep the $\NoFC$ bins $[4,5)$ and $[5,6)$ above the floor
  (Section~\ref{app:dif:snn2tail}).
\item Under the baseline acquisition it fails in two opposite
  directions.  In the high-contrast bins of the two smaller patients
  ($X_2\in[4,7)$) SNN2's error exceeds the floor $1.1$--$2.8$-fold and
  the excess vanishes faster than $1/\Nzero$, so SNN2's DIF overstates
  the needed dose by the same factor (the floor DIF suffices).  In the
  extreme bins, where estimates are confined to $[0,9.5]$, the box acts
  as a prior that fades as the counts rise: SNN2's DIF is $0.06$--$0.64$
  times the floor and understates the needed dose up to about $16$-fold.
\item Under $\NoFC$ both effects are small: reruns at SNN2's DIF give
  $0.79$--$1.01$.
\end{itemize}
T2bp, whose prior is built from the patient's own bundles at the same
dose, and electronic noise were not tested.

\subsection{Where SNN2 exceeds the floor}
\label{app:dif:snn2tail}

\emph{Supports main Section~\ref{sec:results:dif}: SNN2 sits at the
$\NoFC$ floor on $21$ of the $23$ non-test volumes, and this is why the
other
two differ.}

Under $\NoFC$, SNN2's middle-path mean squared error over $X_2<6$ is
$1.01$--$1.07$ times the mean CRB variance on $21$ of the $23$ non-test
volumes, but $1.14$ on LIDC-0510 and $1.69$ on LIDC-0766 (main
Section~\ref{sec:results:dif}).  To find out why, SNN2 was recomputed on
$12$ evenly spaced slices of these two volumes and of two validation
patients near the floor (LIDC-0709 and LIDC-0052): $3.1$--$4.1$
million bundles with $X_2<6$ per volume, from the $\NoFC$ files.  The
possible readings were written down before the run.

\emph{Checks.}  The recomputed estimates agree with the stored ones to
within $5\times10^{-7}$ in every bundle.  The sampled ratios reproduce
the corpus values ($1.689$ and $1.158$ against $1.689$ and $1.140$).
The stored Poisson deviates are consistent with their stored
$\Nzeroray{k}$, including in bundles with a ray at the $10^3$ or $10^6$
clip (rays with expected count above $20$: mean standardized residual at
most $0.004$ in magnitude, mean square within $0.004$ of $1$).

\emph{Outcome.}  On LIDC-0766, neither more iterations ($200$ instead of
$20$: $1.689$) nor dropping the bias-correction step ($1.687$) changes
the ratio, and the box-constrained MLE of Section~\ref{app:snn2:mle},
followed by the same bias correction, lowers it only to $1.584$.
Bundles whose middle-path error exceeds five CRB standard deviations,
$0.03\%$ of the total, carry $76\%$ of the excess.  The $57$ such
bundles in the $200\,000$-bundle sample of Table~\ref{tab:snn2tail} have
a true $X_2$ of $4.5$--$6$ and short endpoint paths (both below $1.5$ in
$45$ of them, both in air in none), and $\NoFC$ leaves their middle ray
a median $44$ and $56$ times dimmer than its two neighbors (median
detected fluences about $10^4$ against $200$); $50$ of them have
$c\ge20$.  A reading then carries little information on $X_2$ (main
Section~\ref{sec:n0fc:problem}): the median CRB standard deviation of
$X_2$ in these bundles is $0.49$, so the middle ray's intensity
$\beta_2$ is known to about half its value, and $X_2=-\ln(\beta_2/\Nzeroray{2})$
turns the low side of that noise into a long upper tail.  All $57$
errors are positive, and in $32$ of them ($56\%$) the estimate runs to
the box limit $9.5$.  The CRB is a local, large-sample bound, and the
maximum-likelihood estimate does not attain it there.
Table~\ref{tab:snn2tail} sorts the bundles by that contrast: below
$c=20$ the ratio is $1.00$--$1.05$, as on the other volumes, and the
$3.8\%$ of bundles with $c\ge20$ carry $96\%$ of the excess.  LIDC-0510
shows the same pattern, weaker: its estimates at the upper limit carry
$30\%$ of its excess.  Why LIDC-0766 holds more such bundles was not
tested.  In its tail bundles $\NoFC$ gives the endpoint rays a median
$\Nzeroray{k}$ of $1.3\times10^4$, only about three times less than the
middle ray's $4.2\times10^4$.  LIDC-0766 also has the largest
ellipticity in the corpus ($1.81$), and in its thin direction endpoint
rays may cross little tissue while the middle ray crosses much
(\emph{hypothesis}).

\begin{table}[ht]
\centering
\caption{SNN2 on LIDC-0766 ($\NoFC$, bundles with true $X_2\in[3,6)$; a
random $200\,000$ of the $2.76$ million in $12$ sampled slices), by the
within-bundle contrast $c=\max(\Phi_1,\Phi_3)/\Phi_2$ of detected
fluences.  MSE/CRB: middle-path mean squared error over the mean CRB
variance at the truth, for SNN2 as stored and for the box-constrained
MLE ($200$ reduced steps of Section~\ref{app:snn2:mle}, then the same
bias correction).  Excess: share of
$\sum(e_2^2-\sigma^2_{\mathrm{CRB},2})$; tail: share of bundles with
$|e_2|>5\,\sigma_{\mathrm{CRB},2}$; both for SNN2 as stored.  The
sample's MSE/CRB is $1.56$, against $1.69$ for all $2.76$ million;
$50$ of its $57$ tail bundles fall in the last two rows.}
\label{tab:snn2tail}
\small
\begin{tabular}{@{}lrrrrr@{}}
\toprule
 & & \multicolumn{2}{c}{MSE/CRB} & & \\
\cmidrule(lr){3-4}
$c$ & Share & SNN2 & Box MLE & Excess & Tail \\
\midrule
$<2$     & $0.504$ & $1.001$ & $0.997$ & $0.000$ & $0$ \\
$2$--$5$   & $0.238$ & $1.027$ & $0.979$ & $0.006$ & $4\times10^{-5}$ \\
$5$--$10$  & $0.140$ & $1.043$ & $0.981$ & $0.016$ & $7\times10^{-5}$ \\
$10$--$20$ & $0.080$ & $1.051$ & $1.013$ & $0.018$ & $2\times10^{-4}$ \\
$20$--$40$ & $0.030$ & $1.419$ & $1.399$ & $0.126$ & $2\times10^{-3}$ \\
$\ge40$  & $0.008$ & $2.814$ & $2.517$ & $0.833$ & $2.3\times10^{-2}$ \\
\bottomrule
\end{tabular}
\end{table}

\section{Per-bin, per-patient evaluation tables}
\label{app:tables}

\subsection{$\NoFC$ empirical detail: validation pool}
\label{app:n0fc:emp:val}

The headline numbers in Section~\ref{sec:n0fc:empirical} are extracted
from the per-bin tables of Table~\ref{tab:n0fc_val_x2_appendix}.
The pattern is consistent across bins~6--10: $\NoFC$ delivers an
$\mathrm{SD}_2$ reduction of $3$--$3.7\times$ and a
$|\mathrm{Bias}_2|$ collapse of $5$--$19\times$, largest at bin~10
($19\times$; Table~\ref{tab:n0fc_val_x2_appendix}).  Bin~10 is sparse
and differs between pools: under $\NoFC$ its bias is $0.048$ in
magnitude here, on all bundles of the four validation patients, against
$0.159$ in Table~\ref{tab:snn2_residual_structure}, on a
$10\,000$-bundle subsample of the training pool.

\begin{table}[t]
\centering
\caption{SNN2 per-bin $X_2$ statistics on the validation pool used with
the $19$-patient training corpus (LIDC-0052, -0628, -0709, -0766; $\sim 320$\,M
bundles per regime).}
\label{tab:n0fc_val_x2_appendix}
\smallskip
\small
\begin{tabular}{@{}rcccc@{}}
\toprule
& \multicolumn{2}{c}{$\mathrm{SD}_2$} &
\multicolumn{2}{c}{$|\mathrm{Bias}_2|$} \\
\cmidrule(lr){2-3} \cmidrule(lr){4-5}
Bin & Baseline & $\NoFC$ & Baseline & $\NoFC$ \\
\midrule
4   & 0.042 & 0.025 & 0.0002 & 0.00004 \\
5   & 0.102 & 0.059 & 0.0038 & 0.0007 \\
6   & 0.352 & 0.116 & 0.038  & 0.0040 \\
7   & 0.731 & 0.197 & 0.142  & 0.014  \\
8   & 1.007 & 0.298 & 0.220  & 0.032  \\
9   & 1.473 & 0.469 & 0.343  & 0.065  \\
10  & 1.452 & 0.440 & 0.907  & 0.048  \\
\bottomrule
\end{tabular}
\end{table}

\subsection{$\NoFC$ empirical detail: test pool per-patient context}
\label{app:n0fc:emp:test}

Four points expand on the SNN2 test-patient results of main
Section~\ref{sec:results:snn2} (the SNN2 columns of
Table~\ref{tab:2x2_pool}).
\emph{Largest relative reduction is on the lowest-WED patient}
(LIDC-0951, $0.16\times$): its baseline acquisition is far from optimal
for its anatomy, so the redistribution removes the most inefficiency.
\emph{Largest absolute reduction is on the highest-WED patient}
(LIDC-0116, $0.415\!\to\!0.164$, a $0.25$-unit drop): the difficult
regime---high attenuation, sparse central-ray photons, populated
bins~9--10---is where $\NoFC$ delivers the largest absolute reduction.
There $\NoFC$ gives the middle ray $1.70$ and $2.16$ times the baseline's
incident photons at bins~9 and~10 (counted as in Section~\ref{app:n0fc:nbase}).
\emph{Endpoint paths:} peripheral flux is reduced
under $\NoFC$, so endpoint SDs grow at every populated bin (by
$1.3$--$16.5\times$), up to $0.103$ (LIDC-0116, $X_1\in[7,8)$), while the
endpoint floors rise only modestly (main
Section~\ref{sec:n0fc:fisher}).
\emph{Ratio of standard deviation to the CRB at the estimate} (main
Section~\ref{sec:results:snn2}; bins with at least $10^4$ bundles):
$0.90$--$1.03$ on the endpoint paths under baseline acquisition and
$0.78$--$1.19$ under $\NoFC$.  On the middle path it is $1.01$--$1.15$
through bin~5 under baseline acquisition and $1.02$--$1.40$ through
bin~8 under $\NoFC$, the largest values on LIDC-0116.  At baseline
bins~6--8 it ranges from $0.21$ to $1.36$, falling below $1$ where the
bias grows.

\subsection{V8c and SNN2 per bin on PIS\_NFLTD and PIS}
\label{app:tables:ood}

Table~\ref{tab:pis-nfltd-v8c-vs-snn2-base} lists the values that main
Figure~\ref{fig:v8c-ood} plots: per unit-width $X_2$ bin, the bias and
RMSE of V8c-alone and SNN2-alone on PIS\_NFLTD and PIS under both
acquisitions.

\begin{table}[t]
\centering
\caption{V8c-alone and SNN2-alone per unit-width $X_2$ bin (main
equation~\eqref{eq:x2bins}) on PIS\_NFLTD and PIS, both acquisitions:
bundle population $N$, middle-path bias and RMSE of each estimator, and
the ratio of their RMSEs (V8c over SNN2).  PIS has no bundle at
$X_2\ge7$.  On PIS\_NFLTD, V8c underestimates $X_2$ from bin~$7$
(baseline) or bin~$8$ ($\NoFC$) on, with biases reaching $-2.75$ and
$-1.20$; there SNN2's bias stays within $\pm0.17$, six to twenty-two
times smaller.}
\label{tab:pis-nfltd-v8c-vs-snn2-base}
\footnotesize
\setlength{\tabcolsep}{4pt}
\begin{tabular}{@{}rlrrrrrr@{}}
\toprule
 & & & \multicolumn{2}{c}{SNN2-alone} & \multicolumn{2}{c}{V8c-alone} & \\
\cmidrule(lr){4-5}\cmidrule(lr){6-7}
Bin & $X_2$ & $N$ & bias & RMSE & bias & RMSE & Ratio \\
\midrule
\multicolumn{8}{@{}l}{\textit{PIS\_NFLTD, baseline}} \\
3  & $[2,3)$          & $2.07\times10^{6}$   & $0.000$ & $0.0114$ & $+0.001$ & $0.0115$ & $1.00$ \\
4  & $[3,4)$          & $7.02\times10^{6}$   & $0.000$ & $0.0193$ & $+0.002$ & $0.0195$ & $1.01$ \\
5  & $[4,5)$          & $1.56\times10^{7}$   & $0.000$ & $0.0450$ & $+0.046$ & $0.119$ & $2.64$ \\
6  & $[5,6)$          & $2.39\times10^{7}$   & $+0.005$ & $0.117$ & $+0.031$ & $0.214$ & $1.84$ \\
7  & $[6,7)$          & $1.36\times10^{7}$   & $+0.037$ & $0.318$ & $-0.305$ & $0.515$ & $1.62$ \\
8  & $[7,8)$          & $5.50\times10^{6}$   & $+0.166$ & $0.757$ & $-1.002$ & $1.23$ & $1.63$ \\
9  & $[8,9)$          & $5.29\times10^{5}$   & $+0.131$ & $0.903$ & $-1.709$ & $1.87$ & $2.07$ \\
10 & $[9,\infty)$     & $248$                & $-0.164$ & $0.617$ & $-2.752$ & $2.87$ & $4.65$ \\
\midrule
\multicolumn{8}{@{}l}{\textit{PIS\_NFLTD, $\NoFC$}} \\
3  & $[2,3)$          & $2.07\times10^{6}$   & $0.000$ & $0.0118$ & $+0.006$ & $0.0142$ & $1.20$ \\
4  & $[3,4)$          & $7.02\times10^{6}$   & $0.000$ & $0.0177$ & $+0.009$ & $0.0236$ & $1.33$ \\
5  & $[4,5)$          & $1.56\times10^{7}$   & $0.000$ & $0.0309$ & $+0.015$ & $0.0413$ & $1.33$ \\
6  & $[5,6)$          & $2.39\times10^{7}$   & $+0.001$ & $0.0594$ & $+0.031$ & $0.0884$ & $1.49$ \\
7  & $[6,7)$          & $1.36\times10^{7}$   & $+0.007$ & $0.135$ & $+0.029$ & $0.171$ & $1.27$ \\
8  & $[7,8)$          & $5.50\times10^{6}$   & $+0.038$ & $0.316$ & $-0.270$ & $0.390$ & $1.23$ \\
9  & $[8,9)$          & $5.29\times10^{5}$   & $+0.081$ & $0.493$ & $-0.670$ & $0.728$ & $1.48$ \\
10 & $[9,\infty)$     & $248$                & $-0.054$ & $0.483$ & $-1.198$ & $1.23$ & $2.55$ \\
\midrule
\multicolumn{8}{@{}l}{\textit{PIS, baseline}} \\
2  & $[1,2)$          & $1.10\times10^{6}$   & $0.000$ & $0.00700$ & $0.000$ & $0.00697$ & $1.00$ \\
3  & $[2,3)$          & $1.61\times10^{7}$   & $0.000$ & $0.0141$ & $+0.001$ & $0.0140$ & $0.99$ \\
4  & $[3,4)$          & $2.71\times10^{7}$   & $0.000$ & $0.0514$ & $+0.010$ & $0.0526$ & $1.02$ \\
5  & $[4,5)$          & $1.92\times10^{7}$   & $+0.007$ & $0.135$ & $+0.027$ & $0.144$ & $1.07$ \\
6  & $[5,6)$          & $4.63\times10^{6}$   & $+0.081$ & $0.515$ & $+0.061$ & $0.302$ & $0.59$ \\
7  & $[6,7)$          & $3.99\times10^{4}$   & $+0.333$ & $1.16$ & $-0.094$ & $0.325$ & $0.28$ \\
\midrule
\multicolumn{8}{@{}l}{\textit{PIS, $\NoFC$}} \\
2  & $[1,2)$          & $1.10\times10^{6}$   & $0.000$ & $0.00605$ & $-0.001$ & $0.00584$ & $0.97$ \\
3  & $[2,3)$          & $1.61\times10^{7}$   & $0.000$ & $0.00959$ & $-0.001$ & $0.00912$ & $0.95$ \\
4  & $[3,4)$          & $2.71\times10^{7}$   & $0.000$ & $0.0179$ & $+0.001$ & $0.0171$ & $0.95$ \\
5  & $[4,5)$          & $1.92\times10^{7}$   & $0.000$ & $0.0369$ & $+0.004$ & $0.0353$ & $0.95$ \\
6  & $[5,6)$          & $4.63\times10^{6}$   & $+0.001$ & $0.0795$ & $+0.008$ & $0.0755$ & $0.95$ \\
7  & $[6,7)$          & $3.99\times10^{4}$   & $+0.006$ & $0.124$ & $-0.003$ & $0.0903$ & $0.73$ \\
\bottomrule
\end{tabular}
\end{table}

\section{T2bp implementation details}
\label{app:t2bp}

\subsection{R1 algorithm}
\label{app:t2bp:alg}

T2bp R1, the version used throughout, conditions its prior on $\hat X_3$
alone (the endpoint-conditioned prior~\eqref{eq:t2bp-prior} of main
Section~\ref{sec:t2bp:prior}; the variant R2 is in
Section~\ref{app:t2bp:r1r2}), and is applied per (patient, regime) cell.  Each iteration is
two-pass. Pass~1 accumulates the joint $(\hat X_2,\hat X_3)$ histogram
and, at the first iteration, the per-bin likelihood width $\sigma_L$ and
the grid. $\sigma_L$ (the median Cram\'er--Rao width at the SNN2
estimate, Section~\ref{app:t2bp:deploy}) sets the adaptive bandwidth
$h_2=\max(h_2^{\min},\,c_2\,\sigma_L)$, with $c_2=0.5$ and
$h_2^{\min}=0.02$. Pass~2 streams the bundles again and computes each
per-bundle posterior mean~\eqref{eq:t2bp-posterior}.  The conditioning variable
$\hat X_3$ is discretized on $40$ knots uniformly spaced on $[0,8]$.

\emph{Details}, from the analysis code:
\begin{enumerate}[nosep, leftmargin=*]
  \item \emph{Histogram.}  $H$ counts bundles in cells $0.05$ wide in
        $\hat X_2$ on $[0,10)$ (estimates outside go to the end cells)
        and, along $\hat X_3$, at the nearest knot
        (spacing $8/39\approx0.205$); estimates beyond $8$ go to the last
        knot.  $\hat X_3$ is always the base estimate; only $\hat X_2$
        is updated between iterations.
  \item \emph{Prior.}  The counts are smoothed across knots with
        Gaussian weights of width $h_3$, normalized over the knots.  Each
        knot's column then becomes a density on the integration grid,
        through a Gaussian kernel of width $h_2(x)$ taken at the
        evaluation point $x$ and centered on the cell centers, normalized
        with the trapezoidal weights.
  \item \emph{Likelihood.}  Rows~$1$ and~$5$ do not involve $X_2$, so,
        up to a constant,
        $\ln L(Y\mid x)=\sum_{j=2}^{4}\bigl(Y_j\ln y_j(x)-y_j(x)\bigr)$,
        with $y_j(x)=a_j+\Nzeroray{2}e^{-x}$.  Here $a_j$ collects the
        endpoint contributions $\Nzeroray{1}e^{-\hat X_1}$ (rows~$2$
        and~$3$) and $\Nzeroray{3}e^{-\hat X_3}$ (rows~$3$ and~$4$) at the
        base estimates.
  \item \emph{Posterior.}  The log prior at a bundle's $\hat X_3$
        (clipped to $[0,8)$) is interpolated linearly between the two
        bracketing knots.  The log posterior is shifted by its maximum
        over the grid before exponentiation, and the posterior mean is
        computed with the trapezoidal weights.
  \item \emph{Base estimator.}  For T2bp$\,\circ\,$V8c the V8c estimates
        replace SNN2's everywhere: they supply $\hat X_1$ and $\hat X_3$
        in the likelihood, the conditioning $\hat X_3$, the histogram, and
        $\sigma_L$, the Cram\'er--Rao width at the V8c estimate.
  \item \emph{Why $\hat X_3$.}  Over the bundles of a view the two
        endpoint rays play mirror roles: the first ray of bundle~$b$ and
        the third of bundle~$334-b$ sit at opposite fan angles.  $\hat X_1$
        would therefore serve equally, and conditioning on both (R2,
        Section~\ref{app:t2bp:r1r2}) gave no consistent gain.
\end{enumerate}

\begin{algorithm}[t]
\DontPrintSemicolon
\KwIn{bundle file (per-ray $\Nzeroray{k}$, counts $Y_j$); SNN2
      predictions $\bm{\mu}^{\mathrm{base}}$; iteration count $K=6$.}
\KwOut{refined per-bundle $\hat X_2^{\mathrm{T2bp}}$.}
$\bm{\mu}^{\mathrm{cur}} \gets \bm{\mu}^{\mathrm{base}}$\;
Set $\hat X_3$ knots: $40$ uniform on $[0,8]$\;
\For{$k = 1,\ldots,K$}{
  \textit{Pass 1:} stream bundles; accumulate $H[\hat X_2,\hat X_3]$
    from $\bm{\mu}^{\mathrm{cur}}$\;
  \lIf{$k=1$}{set $\sigma_L$ to the per-bin median of
    $([F^{-1}(\bm{\mu}^{\mathrm{base}})]_{22})^{1/2}$ in $0.5$-wide bins
    of $\hat X_2^{\mathrm{base}}$; compute $h_2(X_2)$; build the
    non-uniform grid (Section~\ref{app:t2bp:grid}) and trapezoidal weights}
  Build conditional priors: smooth $H$ along $\hat X_3$ ($h_3{=}0.15$),
    KDE along $X_2$ (adaptive $h_2$), normalize each $\hat X_3$-column\;
  \textit{Pass 2:} \For{each bundle}{
    evaluate $\log L(Y\mid x_i)$ on the grid\;
    interpolate $\log p(x_i\mid\hat X_3)$ between nearest knots\;
    $\hat X_2^{\mathrm{T2bp}} \gets
       \sum_i x_i\,p_i\,w_i \big/ \sum_i p_i\,w_i$\;
  }
  $\bm{\mu}^{\mathrm{cur}} \gets \hat X_2^{\mathrm{T2bp}}$\;
}
\textbf{return} $\hat X_2^{\mathrm{T2bp}}$\;
\caption{T2bp R1 (one cell).}
\label{alg:t2bp}
\end{algorithm}

\subsection{Truth-free likelihood width and iteration count}
\label{app:t2bp:deploy}

Algorithm~\ref{alg:t2bp} reads the bundle records (per-ray
$\Nzeroray{k}$ and counts $Y_j$) and the SNN2 estimates; the simulated
$X_2$ enters only the reported per-bin and pool errors.  Two settings
make this so: the likelihood width $\sigma_L$ and a fixed iteration
count $K=6$.  An earlier implementation used the ground truth at both
places.  Its $\sigma_L$ was the standard deviation of
$\hat X_2^{\mathrm{SNN2}}-X_2$ in $0.5$-wide bins of the true $X_2$. Its
loop stopped when the pool RMSE against the ground truth changed by less
than $0.5\%$ (at most $5$ iterations for the three test patients and
$10$ for the corpus and PIS\_NFLTD; the corpus sweep ran $2$--$7$).
This section defines the replacements, shows why the iteration must be
truncated, and compares the two settings.  Its error statistics use
$0.5$-wide bins of the true $X_2$ with an open top bin $[6,\infty)$,
finer than the unit bins of~\eqref{eq:x2bins} (main text).

\paragraph{Likelihood width.}  For each bundle, the Fisher matrix $F$
of~\eqref{eq:fim} (main text) is evaluated at the SNN2 estimate
$\hat{\svec X}^{\mathrm{SNN2}}$ with the bundle's known $\Nzeroray{k}$.
The expected counts are floored at one photon and the line integrals
capped at $20$. Bundles whose $F$ is numerically singular
($|\det F|\le10^{-30}$) are skipped.  In each $0.5$-wide bin of
$\hat X_2^{\mathrm{SNN2}}$ (bins $b=1,2,\dots$; the last is
$[6,\infty)$), $\sigma_L$ is the median of $([F^{-1}]_{22})^{1/2}$ over
the bin's bundles. It is read from a histogram of $\log_{10}$ widths
with $0.001$-decade cells and computed once, at the first iteration.
Bins holding at most $100$ bundles use the bandwidth floor
$h_2=h_2^{\min}$.  The median was chosen over the root mean square
before any comparison,
because near-singular $F$ dominate the mean in the photon-starved top
bin.  Table~\ref{tab:t2bp:sigmaL} compares the two statistics with the
earlier, truth-binned standard deviation on LIDC-0116 under the
baseline acquisition.  For $X_2\in[2,5)$ the root-mean-square width
equals the SNN2 error standard deviation to $0.1\%$, consistent with
SNN2 being efficient there, and the median is $5$--$18\%$ smaller, as
expected for a right-skewed distribution of widths.  In the three
highest bins the median is $26$--$55\%$ smaller, so $h_2$ and the grid
spacing are narrower there.

\begin{table}[!ht]
\centering
\caption{Likelihood width $\sigma_L$ per $X_2$ bin on LIDC-0116, baseline
acquisition (full data): the earlier truth-binned SNN2 error
standard deviation, and the median and root mean square of the
Cram\'er--Rao width at the SNN2 estimate in bins of
$\hat X_2^{\mathrm{SNN2}}$ (relative difference to the first in
parentheses).  The median is used.}
\label{tab:t2bp:sigmaL}
\small
\begin{tabular}{lrrr}
\toprule
$X_2$ bin & error SD (true bins) & CRB median & CRB r.m.s. \\
\midrule
$[2.0,2.5)$ & $0.00544$ & $0.00518$ ($-5\%$)  & $0.00543$ ($-0.1\%$) \\
$[3.0,3.5)$ & $0.0102$  & $0.0095$ ($-7\%$)   & $0.0102$ ($+0.1\%$) \\
$[4.0,4.5)$ & $0.0239$  & $0.0205$ ($-14\%$)  & $0.0239$ ($+0.1\%$) \\
$[4.5,5.0)$ & $0.0368$  & $0.0301$ ($-18\%$)  & $0.0368$ ($+0.1\%$) \\
$[5.0,5.5)$ & $0.0608$  & $0.0447$ ($-26\%$)  & $0.0615$ ($+1.1\%$) \\
$[5.5,6.0)$ & $0.126$   & $0.0707$ ($-44\%$)  & $0.116$ ($-8\%$) \\
$[6,\infty)$ & $0.683$  & $0.308$ ($-55\%$)   & $3.37$ ($+393\%$) \\
\bottomrule
\end{tabular}
\end{table}

\paragraph{Why the iteration must be truncated.}  R1 rebuilds its
prior from the histogram of the previous posterior means.  A scalar
Gaussian model shows what this does.  Within one $\hat X_3$ knot, let
the middle-path value $x$ vary over the patient's bundles with mean
$\mu$ and variance $\tau^2$ (the anatomical spread). Let each bundle
supply a Gaussian likelihood centered on its SNN2 estimate $y=x+e$,
where $e$ is independent of $x$, with mean zero and variance $\sigma^2$
(the likelihood width).  Iteration $k$ ($k=1,2,\dots$) uses the prior
$\mathcal N(\mu,s_k^2)$.  The posterior mean is
\begin{equation}
  m_k = \mu + B_k\,(y-\mu), \qquad B_k = \frac{s_k^2}{s_k^2+\sigma^2},
  \label{eq:t2bp:gauss-mean}
\end{equation}
and, because every iteration uses the same data $y$, its spread over
the bundles is $\Var(m_k)=B_k^2\,\Var(y)=B_k^2(\tau^2+\sigma^2)$.  R1
builds the next prior from the histogram of the $m_k$, smoothed by a
Gaussian kernel of width $h$; the first prior is the smoothed histogram
of the SNN2 estimates themselves.  Hence
\begin{equation}
  s_1^2 = \tau^2+\sigma^2+h^2, \qquad
  s_{k+1}^2 = B_k^2\,(\tau^2+\sigma^2) + h^2 .
  \label{eq:t2bp:gauss-rec}
\end{equation}
The mean squared error of $m_k$ follows from $m_k-x=(B_k-1)(x-\mu)+B_k e$:
\begin{align}
  \E\bigl[(m_k-x)^2\bigr]
  &= (1-B_k)^2\tau^2 + B_k^2\sigma^2 \nonumber\\
  &= (\tau^2+\sigma^2)\,(B_k-B^\circ)^2
    + \frac{\tau^2\sigma^2}{\tau^2+\sigma^2},
  \qquad B^\circ = \frac{\tau^2}{\tau^2+\sigma^2}.
  \label{eq:t2bp:gauss-mse}
\end{align}
The second form follows by expanding the first,
$\tau^2-2B_k\tau^2+B_k^2(\tau^2+\sigma^2)$, and completing the square in
$B_k$.  The error is smallest at $B_k=B^\circ$, that is, when the prior
variance equals the true spread, $s_k^2=\tau^2$.

The map $g(s^2)=\bigl(s^2/(s^2+\sigma^2)\bigr)^2(\tau^2+\sigma^2)+h^2$
of~\eqref{eq:t2bp:gauss-rec} increases with $s^2$, and
$g(s_1^2)<s_1^2$ because $B_1<1$.  By induction
($s_{k+1}^2<s_k^2$ implies $g(s_{k+1}^2)<g(s_k^2)$), $s_k^2$ decreases
at every iteration and converges to the largest fixed point of $g$
below $s_1^2$.  For $h=0$ the fixed points satisfy
$s^2\bigl[(s^2+\sigma^2)^2-s^2(\tau^2+\sigma^2)\bigr]=0$, that is,
$s^2=0$ or
\begin{equation}
  s^4-(\tau^2-\sigma^2)\,s^2+\sigma^4=0 ,
  \label{eq:t2bp:gauss-fp}
\end{equation}
whose discriminant is
$(\tau^2-\sigma^2)^2-4\sigma^4=(\tau^2-3\sigma^2)(\tau^2+\sigma^2)$.
Two regimes follow.
\begin{itemize}[nosep, leftmargin=*]
  \item $\tau^2\ge3\sigma^2$ (likelihood narrow against the spread).
    $s_k^2$ converges to the larger root
    $s_+^2=\tfrac12\bigl[\tau^2-\sigma^2
      +\sqrt{(\tau^2-3\sigma^2)(\tau^2+\sigma^2)}\,\bigr]$, which is
    below $\tau^2$ because the square root is below $\tau^2+\sigma^2$.
    The error passes its minimum as $s_k^2$ crosses $\tau^2$ and then
    settles slightly above it.  For $\tau^2\gg\sigma^2$,
    $s_+^2\approx\tau^2-\sigma^2-\sigma^4/\tau^2$ and the excess at the
    limit is about $(\sigma^2/\tau^2)^3$ of the minimum ($0.1\%$ at
    $\tau^2=10\sigma^2$); at $\tau^2=3\sigma^2$, $s_+^2=\sigma^2$ and the
    excess is $1/3$.
  \item $\tau^2<3\sigma^2$ (likelihood broad against the spread).  For
    $h=0$ the only fixed point is $s^2=0$; for $h>0$ the prior contracts
    to $s^2\approx h^2$, and every posterior mean is pulled toward
    $\mu$.  The error falls while $s_k^2>\tau^2$, is smallest near
    $s_k^2=\tau^2$, and then rises toward $\tau^2$, which exceeds the
    SNN2 error $\sigma^2$ itself when $\sigma^2<\tau^2<3\sigma^2$.  A
    smaller $h$ lowers the floor $h^2$ and deepens the rise.
\end{itemize}
In this model the best iteration count depends on $\tau^2/\sigma^2$,
which varies across bins and patients and is not known in advance
(R1 does not estimate it).  The real cells behave as the model predicts
\emph{(the transfer from this Gaussian model to R1's Poisson likelihood
and kernel prior is a hypothesis)}.  On LIDC-0116 under the baseline
acquisition, every 16th slice, the rise comes entirely from the top bin
$[6,\infty)$, which holds $37\%$ of the bundles.  Its RMSE is lowest at
$k=3$--$4$ ($0.50$) and reaches $0.514$ at $k=10$ with the earlier
truth-binned $\sigma_L$ and $0.553$ with the narrower Cram\'er--Rao
median (smaller $h_2$, as the model predicts).  The bins below $5.5$ change
by at most $0.1\%$ across iterations, and $[5.5,6)$ falls by $4$--$6\%$
(after a $0.2\%$ rise at $k=2$ with the earlier $\sigma_L$).  The corpus patient LIDC-0766 behaves the
same way (best $k=4$, $2.2\%$ above it at $k=10$;
Table~\ref{tab:t2bp:selk}).  Rebuilding the prior from the average
posterior \emph{density} instead of the posterior means (an EM update)
has the true spread as its fixed point in this model:
$\Var(m)+\E[\Var(x\mid y)]=B^2(\tau^2+\sigma^2)+B\sigma^2$, which equals
$\tau^2$ at $B=B^\circ$.  On two synthetic cells it ended $25$--$35\%$
above R1 and was not adopted.  A plausible cause \emph{(hypothesis)} is
R1's plug-in likelihood, which holds $X_1$ and $X_3$ at their SNN2
values. If so, where the middle path carries little of the signal, endpoint
errors dominate that likelihood, and maximizing the marginal likelihood
fits the prior to them.  The EM update is the classical
iteration for a nonparametric maximum-likelihood prior, which can also
be computed directly by convex optimization \citep{koenker2014S}.  Such a
prior avoids the narrowing of a prior rebuilt from posterior means, but
with R1's plug-in likelihood it would presumably fit the same endpoint
errors; it is left to future work.

\paragraph{Choice of $K$.}  The rule was written down before the
selection runs.  The truth-free $\sigma_L$ was run for $k=1,\dots,10$
iterations, on every 16th slice, on four corpus patients under the
baseline acquisition, chosen as the hardest non-test cases.
They are LIDC-0766 (validation) and LIDC-0004 (training), the
two highest corpus pool RMSEs, and LIDC-0303 and LIDC-0507 (training),
the slowest to converge in the earlier corpus runs.  With $R_c(k)$ the pool
RMSE of patient $c$ after $k$ iterations, its regret is
$L_c(k)=100\,[R_c(k)-\min_j R_c(j)]/\min_j R_c(j)$, and
$K=\arg\min_k\max_c L_c(k)$, ties going to the smaller $k$.
Table~\ref{tab:t2bp:selk} gives $K=6$, with a worst-case regret of
$0.83\%$.  The $\sigma_L$ statistic was fixed before any test run.
The rule for $K$ uses only corpus cells, but it was written after the
per-iteration test-cell errors had been inspected (the need for a
stopping rule was found on one of them, LIDC-0116 under the baseline
acquisition).  It did not select the test-favored count: at every 16th
slice, $k=4$ gives a worst-case change over the eight test cells of
$+1.4\%$, against $+3.6\%$ at $K=6$.

\begin{table}[!ht]
\centering
\caption{Choice of $K$: regret $L_c(k)$ (\%) of the truth-free T2bp
after $k$ iterations, relative to each corpus patient's own best $k$
($1\le k\le10$; baseline acquisition, every 16th slice).  Columns
$k=1$, $2$, $8$ and $9$ are omitted; their worst cases exceed $0.83\%$
(Figure~\ref{fig:t2bp-iter}(b) of the main text).}
\label{tab:t2bp:selk}
\small
\begin{tabular}{llrrrrrrr}
\toprule
Patient & Split & $k=3$ & $4$ & $5$ & $6$ & $7$ & $10$ & best $k$ \\
\midrule
LIDC-0004 & training   & $4.58$ & $1.47$ & $0.41$ & $0.08$ & $0.01$ & $0.02$ & $8$ \\
LIDC-0303 & training   & $7.96$ & $2.90$ & $0.71$ & $0.00$ & $0.05$ & $1.60$ & $6$ \\
LIDC-0507 & training   & $9.03$ & $3.84$ & $1.62$ & $0.67$ & $0.27$ & $0.00$ & $10$ \\
LIDC-0766 & validation & $1.10$ & $0.00$ & $0.26$ & $0.83$ & $1.36$ & $2.22$ & $4$ \\
\midrule
Worst case &          & $9.03$ & $3.84$ & $1.62$ & $\mathbf{0.83}$ & $1.36$ & $2.22$ & \\
\bottomrule
\end{tabular}
\end{table}

\paragraph{Test cells.}  Table~\ref{tab:t2bp:deploytest} compares the
two settings on the eight test cells.  On every 16th slice, the
truth-free settings change pool RMSE by $-1.70$ to $+3.64\%$.  All four
$\NoFC$ cells are equal or better ($-1.70$ to $-0.01\%$); the one
increase is the semi-convergent LIDC-0116 baseline cell, where the
truth-based stop had halted near the minimum.  Changing $\sigma_L$ alone
(truth-based stop kept) gives $-1.70$ to $+0.02\%$; changing the stop
alone ($K=6$ with the truth-binned $\sigma_L$) gives $-0.47$ to
$+1.16\%$.

At full data (last two columns of Table~\ref{tab:t2bp:deploytest}, and
Table~\ref{tab:composition-10cell} of the main text) the change is
$-1.69$ to $+2.53\%$, and the $\NoFC$ cells are again equal or better
($-1.69$ to $-0.02\%$).  LIDC-0116 baseline rises by $2.53\%$: its error
is lowest at $k=4$ ($0.2935$, below the earlier $0.2950$) and rises to
$0.3024$ at $K=6$, the semi-convergence described above.  LIDC-0951
baseline rises by $1.37\%$, which the every-16th-slice comparison did
not predict ($-0.31\%$).  Its error falls at every iteration, and at
equal $k=4$ the difference is the same ($+1.39\%$), so the cause is the
likelihood width, not the iteration count.  The change is confined to
$X_2\ge4.5$: the narrower Cram\'er--Rao median, and with it $h_2$, lowers
the error in $[4.5,5)$ by $2\%$ and raises it by $2$--$9\%$ above
$X_2=5$.  On every 16th slice the same bins move the same way, but
$[5,5.5)$ rises by $0.4\%$ instead of $3.3\%$, which reverses the sign
of the pool change.

\begin{table}[!ht]
\centering
\caption{Pool $X_2$ RMSE of T2bp$\,\circ\,$SNN2 on the eight test cells
with the earlier truth-based settings ($\sigma_L$ binned on the true
$X_2$; stop at $<0.5\%$ change of the RMSE against the ground truth,
after $k$ iterations) and with the truth-free settings (Cram\'er--Rao
$\sigma_L$, $K=6$).  Columns 3--7: every 16th slice; ``$\sigma_L$
only'' and ``$K$ only'' change one setting.  Columns 8--9: full data.}
\label{tab:t2bp:deploytest}
\footnotesize
\setlength{\tabcolsep}{4pt}
\begin{tabular}{llrrrrrrr}
\toprule
 & & \multicolumn{5}{c}{every 16th slice} & \multicolumn{2}{c}{full data} \\
\cmidrule(lr){3-7}\cmidrule(lr){8-9}
Patient & Regime & earlier ($k$) & truth-free & change & $\sigma_L$ only & $K$ only & earlier & truth-free \\
\midrule
PIS        & baseline & $0.1106$ ($4$) & $0.1088$ & $-1.67\%$ & $-1.56\%$ & $-0.04\%$ & $0.1074$ & $0.1065$ \\
LIDC-0951  & baseline & $0.1051$ ($3$) & $0.1048$ & $-0.31\%$ & $-0.29\%$ & $-0.07\%$ & $0.1050$ & $0.1064$ \\
LIDC-0116  & baseline & $0.3105$ ($4$) & $0.3218$ & $+3.64\%$ & $+0.02\%$ & $+1.16\%$ & $0.2950$ & $0.3024$ \\
PIS\_NFLTD & baseline & $0.1808$ ($7$) & $0.1790$ & $-1.01\%$ & $-1.25\%$ & $+0.42\%$ & $0.1767$ & $0.1752$ \\
PIS        & $\NoFC$  & $0.02929$ ($2$) & $0.02928$ & $-0.03\%$ & $-0.00\%$ & $-0.02\%$ & $0.02896$ & $0.02895$ \\
LIDC-0951  & $\NoFC$  & $0.02442$ ($2$) & $0.02441$ & $-0.01\%$ & $-0.00\%$ & $-0.01\%$ & $0.02409$ & $0.02409$ \\
LIDC-0116  & $\NoFC$  & $0.1345$ ($5$) & $0.1323$ & $-1.70\%$ & $-1.70\%$ & $-0.47\%$ & $0.1296$ & $0.1274$ \\
PIS\_NFLTD & $\NoFC$  & $0.09741$ ($6$) & $0.09699$ & $-0.43\%$ & $-0.43\%$ & $+0.00\%$ & $0.09492$ & $0.09453$ \\
\bottomrule
\end{tabular}
\end{table}

\paragraph{Truth-blind check.}  Replacing the stored ground truth by a
constant leaves the $\sigma_L$ table and the first-iteration estimates
bit-identical, on synthetic cells and on the first two retained slices
($681{,}984$ bundles each) of LIDC-0116, PIS, LIDC-0507 and LIDC-0766; on
LIDC-0766 one earlier run differed in $1$ of the $681{,}984$ bundles (by
$5\times10^{-3}$), and five repeats were bit-identical.

\paragraph{Tuning disclosure.}  For $K$, the four selection patients
are tuning patients; they are among the $23$ non-test volumes of main
Section~\ref{sec:results:dif}.  $K$ was selected with SNN2 as the base
estimator and is used unchanged for V8c.
Without these four patients, the corpus summaries of T2bp$\,\circ\,$SNN2 in
Section~\ref{sec:results:dif} (main text) under $\NoFC$ become as
follows (values for all $23$ patients in parentheses). The per-patient mean squared error over $X_2<6$ is $0.74$--$1.04$ times the floor, median $0.88$ ($0.74$--$1.26$, $0.89$). Across the unit bins through $[6,7)$ (one value per bin over the patients) it is $0.84$--$0.97$ ($0.86$--$0.97$). The per-volume median MSE-DIF is
$2.9$--$5.9$, median $4.5$ ($2.9$--$5.9$, $4.6$). The RMSE reduction
baseline$\to\NoFC$ is $45$--$78\%$, median $67\%$ ($45$--$78\%$,
$64\%$), with Spearman $\rho=-0.92$ against WED ($-0.90$).  The four
were chosen as the hardest cases; their ratio to the floor is
$0.90$--$1.26$.
T2bp's gain over SNN2 is not systematically larger for them: its
median is $46\%$ for the four against $49\%$ for the $19$ never tuned
under the baseline acquisition, and $16\%$ against $15\%$ under $\NoFC$
(Mann--Whitney $p=0.41$ and $0.37$).  At $K=6$ the pool RMSE of no
never-tuned cell is more than $0.12\%$ above its minimum over $k\le6$.
For LIDC-0766, a tuning patient whose minimum is at $k=4$, the excess is
$0.77\%$ under the baseline (full data; $0.83\%$ in
Table~\ref{tab:t2bp:selk}, every 16th slice and $k\le10$).

\subsection{Non-uniform integration grid}
\label{app:t2bp:grid}

The posterior mean~\eqref{eq:t2bp-posterior} is a numerical integral,
and the integration grid is the one nontrivial numerical choice.  A uniform grid at the
inherited spacing $dx=0.05$ \emph{aliases} at low attenuation. Where the
likelihood is sharper than the spacing ($\sigma_L \ll dx$), the
posterior collapses onto a single cell and its empirical standard
deviation floors at $dx/\sqrt{12}\approx0.014$, and no quadrature on that
grid resolves a narrower likelihood.  Data that included test patients
informed the grid (main Table~\ref{tab:dev-provenance}).  The symptom
is direct --- with the uniform grid, the posterior mean's RMSE on PIS
$\NoFC$ bundles with $X_2\in[1.0,1.5)$ is about $3.5$ times that of
SNN2, the base estimator it is meant to refine, purely as a
discretization artifact.

T2bp R1 instead uses a non-uniform grid whose local spacing tracks the
narrower of the likelihood width and the prior bandwidth,
\begin{equation}
  dx(X_2) = \min\!\bigl(\sigma_L(X_2),\, h_2(X_2)\bigr),
  \quad \text{clipped to } [0.002, 0.05],
  \label{eq:t2bp_grid}
\end{equation}
walking from $X_2=0$ to $X_2=10$ and placing a point at each local
midpoint.  The clip bounds are wide numerical guardrails, not modeling
constants: the rule adds no tunable parameter beyond those of $h_2$
($c_2$, $h_2^{\mathrm{min}}$), the spacing being fixed by $\sigma_L$
and $h_2$.
Representative grids hold $574$ points for PIS $\NoFC$ ($dx$
ranging $0.004$--$0.05$, densest near $X_2{=}1$) and $536$ for the
broader LIDC-0116 $\NoFC$ distributions.
  Integration uses trapezoidal
weights $w_i=(x_{i+1}-x_{i-1})/2$ for both prior normalization and the
per-bundle posterior.  On a synthetic
Gaussian-prior\,$\times$\,Gaussian-likelihood test the non-uniform grid
recovers the analytic posterior mean to $10^{-6}$ absolute, versus
$10^{-3}$ for the uniform grid at $dx=0.05$.

\subsection{Bias-to-noise regimes}
\label{app:bsregimes}

T2bp's per-bin effect within a (patient, regime) cell can be indexed by
the base estimator's bias-to-noise ratio
\begin{equation}
r_b \;\coloneqq\; \frac{|b_{\text{base}}|}{\sigma_{\text{base}}},
\label{eq:rho-b-def}
\end{equation}
a diagnostic computed against the ground truth and not used by T2bp.
Over the eight test cells with SNN2 as the base estimator (truth-free
settings, $K=6$,
$0.5$-wide bins of the true $X_2$ holding at least $10^4$ bundles), $74$
bins have $r_b<0.1$. They change by a median of $-2.8\%$ (from $-27.8\%$
to $+4.1\%$; $11$ bins worse). All $8$ bins with $r_b\ge0.1$ improve, by
$16$--$65\%$. Where the base estimator is biased relative to its noise, T2bp
corrects it.  The split at $r_b=0.1$ is descriptive: it was not fixed
in advance, and only $8$ bins lie at or above it.  With V8c as the base estimator the per-bin
response is mixed, with gains and losses at large $r_b$; on
LIDC-0116 baseline T2bp raises the pool RMSE of V8c by
$11.8\%$, a residual high-bin bias deferred to Paper~3.

\paragraph{Composition under the baseline acquisition.}
Under the baseline acquisition V8c shows its trained in-distribution
advantage: T2bp$\,\circ\,$V8c wins PIS by $7\%$ and leads LIDC-0951 by
$5\%$, a tie (main Table~\ref{tab:composition-10cell}).  It loses on
LIDC-0116, where T2bp raises V8c's RMSE at $X_2\geq6$, and ties
on PIS\_NFLTD, where V8c starts far behind ($0.471$ against
SNN2's $0.279$).  T2bp improves V8c's pool RMSE on seven of
the eight cells, from a few percent on the two lowest-WED patients'
$\NoFC$ cells, already near the SNN2 floor, to $60\%$ on PIS\_NFLTD
baseline.  That gain reflects V8c's large pre-composition OOD error,
not a stronger signal: the composed RMSE is still $0.013$ above
T2bp$\,\circ\,$SNN2's.

\subsection{R1 versus R2}
\label{app:t2bp:r1r2}

All T2bp results in the main paper (Sections~\ref{sec:t2bp:posterior}, \ref{sec:results:floor}
and~\ref{sec:results:composition}--\ref{subsec:paper1-floor}) use R1, the
prior~\eqref{eq:t2bp-prior} conditioned on $\Xhat_3$.  A variant
conditioned on both endpoint estimates (R2), with prior $p(X_2\mid\Xhat_1,\Xhat_3)$, was evaluated and
deferred on data that included test patients (main
Table~\ref{tab:dev-provenance}).  On a $13$-cell $\NoFC$ check ($100\,000$-bundle subsamples,
run with the earlier settings of Section~\ref{app:t2bp:deploy}) its
first sweep matches the pool $X_2$ RMSE of the $\Xhat_3$-only prior to
within $2\%$ on $11$ cells. LIDC-0116 favors the $\Xhat_3$-only prior by
$8\%$ and PIS\_NFLTD favors R2 by $2\%$.  Its iteration is unstable past the
first sweep.  R1 is the faster engine and is used throughout.

\subsection{Lesion-contrast retention: prediction and measurement}
\label{app:t2bp:lesion}

Main Section~\ref{sec:discussion:limitations} reports how much of a
small lesion's line-integral contrast T2bp keeps.  A linear-Gaussian
model gave a prediction, kept here because the measurement that
followed contradicts it below $X_2=6$.

\paragraph{The prediction.}
\begin{enumerate}[nosep, leftmargin=*]
  \item \emph{Model.}  Locally the T2bp$\,\circ\,$SNN2 estimate is
        $m+s\,(\Xhat_2-m)$ (Section~\ref{app:biasstruct:method},
        equation~\eqref{eq:t2bp-split}).  A small
        change $\Delta$ of the true middle-path line integral moves the
        SNN2 estimate by $\Delta$ and T2bp's by $s\Delta$, and T2bp's
        noise is $s$ times SNN2's, so the contrast-to-noise ratio is
        unchanged and $s$ is the retained fraction of contrast.
  \item \emph{Estimate of $s$.}  Under this linear-Gaussian
        approximation, $s$ is the ratio of the T2bp to the SNN2 residual
        standard deviation.  It is computed from the corpus run (truth-free
        settings, $K=6$, both acquisitions), per patient and per
        $0.5$-wide bin of the true $X_2$ with an open top bin
        $[6,\infty)$:
        $s_b=\mathrm{SD}(\Xhat_2^{\mathrm{T2bp}}-X_2)/\mathrm{SD}(\Xhat_2^{\mathrm{SNN2}}-X_2)$.
        The main text quotes medians over the $23$ non-test patients ($22$
        below $X_2=3$, which LIDC-0127 does not reach).  The ratio equals
        the local slope only where T2bp's bias is nearly constant within a
        bin.  A bias that varies across the bin, such as $(1-s)(m-X_2)$,
        raises $s_b$, so $s_b$ was taken as an upper estimate of the
        retained contrast; Section~\ref{app:biasstruct:t2bp} fits the
        slope directly.  Across the $23$ non-test patients the median of
        $s_b$ is $0.92$--$0.95$ for $X_2\in[2,6)$ and $0.80$ (range
        $0.57$--$0.86$) for $X_2\ge6$ under $\NoFC$; under the baseline
        acquisition it is $0.93$--$0.98$ for $X_2\in[2,5.5)$, $0.63$ for
        $[5.5,6)$ and $0.43$ ($0.26$--$0.67$) for $X_2\ge6$.
  \item \emph{Rarity.}  The correlation with rarity is Spearman's $\rho$
        between $s_b$ and the bin's share of the patient's bundles, over
        the $23$ patients, in the bins $[5.5,6)$ and $[6,\infty)$:
        $0.73$ and $0.70$ under $\NoFC$, $0.84$ and $0.77$ under the
        baseline.  In the same bins the share and SNN2's error standard
        deviation are correlated at $-0.47$ and $-0.25$ ($\NoFC$) and
        $-0.85$ and $-0.97$ (baseline), so rarity and the photon level
        are confounded.  In summary, $s_b$ is lowest where the bin is
        rarest (Spearman $\rho=0.70$--$0.84$), as the empirical-Bayes
        mechanism predicts; for the slope of T2bp's mapping the relation
        varies by bin ($\rho=0.32$--$0.78$) and separates from the photon
        level only in part (Section~\ref{app:biasstruct:t2bp}).
\end{enumerate}

\paragraph{The measurement.}
A lesion-insertion test measured the retained contrast directly, with
readings fixed in writing before it ran, on eight $\NoFC$ cells at every
16th slice ($96$ slices): the four test volumes, LIDC-0001, LIDC-0127
and LIDC-0310 from training, and LIDC-0766 from validation.  On a random
$2\%$ of the bundles the true middle-path line integral was raised by
half its Cram\'er--Rao standard deviation, and in a second arm by two,
by thinning the counts of the readings that ray enters, which keeps them
exactly Poisson; a replicate scan redrew every count; and every
estimator, T2bp rebuilding its prior, was rerun on each arm
($653\,056$ lesioned bundles).  The response $\hat s$ is the mean change
of an estimate per unit of inserted change; the noise $\hat V$ is half
the mean squared difference between the scan and its replicate, in units
of the CRB variance, so that bias cancels bundle by bundle;
$\hat\eta=\hat s^2/\hat V$; and $r=(\hat\eta/\hat\eta_{\mathrm{KKT}})^{1/2}$
compares an estimator's detectability with that of the exactly solved
box-constrained maximum-likelihood estimate (KKT), $r^2$ being a
dose-equivalent factor.

The readings, over $X_2<6$ and as medians over cells, were these.
\emph{V1}: the box-constrained estimate must respond fully and be
efficient on interior bundles, $\hat s\in[0.98,1.02]$ and
$\hat\eta\in[0.95,1.05]$; it gave $1.004$ and $1.005$, a pass.
\emph{V2}: no interior $\hat\eta$ above $1.05$ for SNN2 or KKT; both gave
$1.005$.  \emph{D1}: $r\ge1.05$ for T2bp$\,\circ\,$SNN2 would mean an
information gain and $r\le0.95$ lower detectability; it gave $1.038$
($0.969$--$1.085$ across cells, a change of $-3.1$ to $+8.5\%$; $7$ of $8$ above $1$), neither.
\emph{D2}, descriptive: split by whether a path lies within $3\sigma$ of
a box bound, $r$ is $0.998$ on interior bundles and $1.094$ near a bound
(Table~\ref{tab:t2bp:lesion}).  \emph{D3}: a ratio
$\hat s(2\sigma)/\hat s(0.5\sigma)$ below $0.95$ would mean extra
shrinkage of larger lesions; it gave $0.9995$.  \emph{D4}: $\hat s$
within $\pm0.03$ of $s_b$.  It fails: $\hat s$ is $0.990$--$1.002$ in
every half-unit bin below $X_2=6$ (medians over the cells with at least
$200$ lesioned bundles in the bin), against $s_b$ of $0.92$--$0.95$ over
$X_2\in[2,6)$.  Below $X_2=6$ the ratio of residual standard deviations
therefore falls for a reason other than a lower slope.

\begin{table}[ht]
\caption{Lesion-insertion test: T2bp$\,\circ\,$SNN2 against the exactly
solved box-constrained estimate, eight $\NoFC$ cells at every 16th
slice.  $\hat s$: response to an inserted change of the middle path;
$\hat V$: noise in units of the CRB variance; $r$: detectability ratio.
Medians over cells with ranges in brackets.  Bundles are split by
whether a path lies within $3\sigma$ of a box bound; LIDC-0310 has no
interior bundles.  The last row, outside the readings fixed in advance,
is \emph{post hoc}.}
\label{tab:t2bp:lesion}
\footnotesize
\setlength{\tabcolsep}{3pt}
\begin{tabular}{@{}llcccr@{}}
\toprule
\multicolumn{2}{@{}l}{Bundles} & $\hat s$ & $\hat V$ & $r$ & cells \\
\midrule
$X_2<6$ & all & $0.997$ ($0.966$--$1.034$) & $0.920$ ($0.799$--$1.044$) & $1.038$ ($0.969$--$1.085$) & $8$ \\
        & interior & $0.998$ ($0.990$--$1.053$) & $1.005$ ($0.994$--$1.100$) & $0.998$ ($0.935$--$1.000$) & $7$ \\
        & near a bound & $1.005$ ($0.970$--$1.064$) & $0.857$ ($0.763$--$0.996$) & $1.094$ ($1.051$--$1.104$) & $8$ \\
\addlinespace
$X_2\ge6$ & all & $0.836$ ($0.783$--$0.922$) & $0.947$ ($0.933$--$0.963$) & $0.879$ ($0.854$--$0.941$) & $4$ \\
\bottomrule
\end{tabular}
\end{table}

\emph{Post hoc.}  The readings stopped at $X_2=6$, but the lesioned
bundles extend beyond it.  On the four cells with at least $1000$ of
them there (LIDC-0116, LIDC-0127, LIDC-0766 and PIS\_NFLTD), the
response of T2bp$\,\circ\,$SNN2 falls to $0.836$ while its noise falls
only to $0.947$, so its detectability is $0.879$ of the box-constrained
estimate's: in the deep bins the prediction, $s_b\approx0.80$, is
roughly right, and T2bp's variance reduction does cost contrast.
\emph{Hypothesis:} near a box bound at $X_2<6$, T2bp's noise falls at
full response because the box and its endpoint-conditioned prior supply
information about the endpoints; nothing here tests it.  The baseline
acquisition was not tested.

\subsection{Where the gain comes from, and how little it depends on the settings}
\label{app:t2bp:sensitivity}

Main Sections~\ref{sec:snn2:mle}, \ref{sec:t2bp:prior}
and~\ref{sec:results:dif} point here for three things: the decomposition
of T2bp's advantage at
$X_2<6$, the readings that were fixed before each run and what they
returned, and the sensitivity of the result to the three constants of
Section~\ref{app:t2bp:deploy}.

\paragraph{The chain, step by step.}
The mechanism corpus, on which readings (2) and (3) below were made, is
fourteen cells: the eight test cells and, under $\NoFC$, LIDC-0001,
LIDC-0086, LIDC-0127, LIDC-0310, LIDC-0766 and LIDC-1008.  Eleven pass
every check fixed in advance; PIS and LIDC-0951 under the baseline and
PIS\_NFLTD under $\NoFC$ each failed a reproduction check, in one
bundle.  Table~\ref{tab:t2bp:chain} takes the median over the corpus's
ten $\NoFC$ cells (not over the $23$ non-test volumes of main
Section~\ref{sec:results:dif}, whose medians differ) and attributes each
step.  Two of the five arms use
the true endpoints and are diagnostics, not deployable estimators.

\begin{table}[ht]
\caption{Median ratio of mean squared error to the unknown-endpoint
Cram\'er--Rao floor over $X_2<6$, ten $\NoFC$ cells, and what each step
buys.  Arms marked $\dagger$ condition on the true $X_1$ and $X_3$.}
\label{tab:t2bp:chain}
\small
\begin{tabular}{@{}lrl@{}}
\toprule
Estimator & MSE/floor & What the step buys \\
\midrule
SNN2                                   & $1.023$ & --- (sits at the unknown-endpoint bound) \\
Box-constrained MLE, no correction     & $0.901$ & box information \\
T2bp$\,\circ\,$SNN2                    & $0.906$ & nothing beyond the box MLE here \\
Oracle-endpoint MLE$^\dagger$          & $0.514$ & the endpoint nuisance, removed \\
T2bp with true endpoints$^\dagger$     & $0.532$ & nothing beyond the oracle MLE \\
\bottomrule
\end{tabular}
\end{table}

\paragraph{What was fixed in advance, and what it returned.}
Three readings bear on the attribution, each written down with its
thresholds before its run.
\begin{enumerate}[nosep, leftmargin=*]
  \item \emph{The share of T2bp's gain that the corrected box estimate
        reproduces}, at $X_2<6$, over $22$ of the $23$ non-test volumes
        under $\NoFC$ (LIDC-0507 failed a pool-sum check fixed in
        advance), with pass thresholds $0.70$ and $0.30$:
        $g=0.0912/0.1439=0.634$.  By its own rule the answer was
        \textbf{mixed}, and it is reported as such in main
        Section~\ref{sec:results:dif}.
  \item \emph{The gain over the uncorrected box MLE},
        $G=1-\mathrm{MSE}(\mathrm{T2bp}\circ\mathrm{Box})/\mathrm{MSE}(\mathrm{Box})$,
        with Box the uncorrected box-constrained MLE, as in
        Table~\ref{tab:kktcomp}: at $X_2<6$, $+0.019$ over the eleven
        cells passing every check ($4$ up, $4$ down, counting cells whose
        bootstrap interval excludes zero) and $+0.020$ over all fourteen ($6$ up,
        $5$ down) --- \textbf{largely reproduced by the box MLE}.  Over
        all bins, $+0.123$ and $+0.290$, up on $10$ of $11$ and $13$ of
        $14$ with none down: \textbf{T2bp keeps an incremental gain}
        there.
  \item \emph{The gain once the endpoints are known},
        $G_h=1-\mathrm{MSE}(\mathrm{T2bp}^\dagger)/\mathrm{MSE}(\text{oracle}^\dagger)$,
        with $\ge 0.05$ meaning the prior still adds and $<0.02$ meaning
        the classical estimator reproduces it: $-0.023$ at $X_2<6$ over
        all fourteen cells and over the ten $\NoFC$ cells alike, so
        \textbf{the endpoint effect is entirely classical}.  At
        $X_2\ge6$ the same reading gives $+0.152$ over the eleven cells
        and $+0.208$ over all fourteen, positive on every cell
        ($+0.051$ to $+0.872$); split by acquisition after the run, it
        is $+0.145$ under $\NoFC$ and $+0.690$ under the baseline.
\end{enumerate}
So T2bp's prior contributes nothing measurable at $X_2<6$, at either
step, and earns its keep in the deep bins and under weak data.  The
flat-prior arm is reported in Section~\ref{app:bsregimes} as a
descriptive contrast only: it is worse than SNN2 at $X_2<6$ in seven of
the fourteen cells, so the ratio of differences that would express it as
a share changes sign across the corpus and is not quoted as one.

\paragraph{The endpoint cost is the nuisance penalty, measured.}
Replacing the estimated endpoints with the truth lowers the MSE by a
factor $1.73$ ($\NoFC$) to $2.21$ (baseline) at $X_2<6$.  That factor
reflects the nuisance-parameter penalty $\kappa_N=[F^{-1}]_{22}F_{22}$,
the ratio of the unknown- to the known-endpoint Cram\'er--Rao variance
($F$ of~\eqref{eq:fim}, main text): measured at the truth,
$\kappa_N$ has median $2.080$ over the corpus ($1.738$--$2.476$) and
$2.00$ under $\NoFC$ ($1.74$--$2.30$), against the equal-attenuation
value $52/27=1.926$.  Against the \emph{known}-endpoint bound
$1/F_{22}$, both arms are efficient on truth-interior bundles: the
oracle MLE gives $1.008$ and T2bp with true endpoints $1.017$, both
inside $[0.97,1.05]$.  The classical estimator captures essentially all
of that gain: $[\mathrm{MSE}(\mathrm{Box})-\mathrm{MSE}(\text{oracle})]/[\mathrm{MSE}(\mathrm{Box})-\mathrm{MSE}(\mathrm{T2bp}^\dagger)]=1.027$
($0.833$--$1.215$).  The exceptions are where the middle ray is swamped
--- LIDC-0766 and the baseline cells, at $1.05$--$1.48$ --- where no
estimator attains its bound.

\paragraph{Two biased estimators, not one unbiased comparison.}
\emph{Post hoc; this decomposition was chosen after the readings above
had been run.}  Splitting $X_2<6$ by whether a path lies within $3\sigma$
of a box bound separates the two sub-floor estimates cleanly
(Tables~\ref{tab:t2bp:bias} and~\ref{tab:t2bp:biasbin}).  The box MLE's advantage is entirely
near-boundary, and it is bought with bias: its pooled $X_2$ bias runs
from $-0.2$ to $-6.6$ across bins~1 to~6, in units of $10^{-3}$, and
T2bp's from $-0.2$ to $-2.4$ --- the same sign and a similar size
through bin~4.
SNN2's bias correction, applied to the box MLE, returns about $40\%$ of
the near-boundary gain ($0.845\to0.920$) and with it the bias
($-6.6\to-1.9$ in bin~6), leaving an unbiased estimate that sits
at the floor.  The mechanism is that the correction is one joint
Gauss--Newton step on the bias-shifted residuals: at an interior maximum
the score vanishes and the step reduces to the first-order bias of the
interior MLE, but \emph{at an active bound the score does not vanish},
so the step also carries the Newton step the box had blocked and undoes
most of the box's effect.  On SNN2's own projected iterate the same
correction is neutral at $X_2<6$ ($1.024$ either way).

\begin{table}[ht]
\caption{Median ratio of mean squared error to floor over $X_2<6$ under
$\NoFC$, the $22$ non-test volumes of reading~(1), split by whether a
path lies within $3\sigma$ of a box
bound.  \emph{Post hoc.}}
\label{tab:t2bp:bias}
\small
\begin{tabular}{@{}lrrr@{}}
\toprule
Estimator & all & near & interior \\
\midrule
SNN2                   & $1.024$ & $1.024$ & $1.019$ \\
Box MLE $+$ correction & $0.939$ & $0.920$ & $1.019$ \\
Box MLE, no correction & $0.875$ & $0.845$ & $1.021$ \\
T2bp$\,\circ\,$SNN2    & $0.885$ & $0.858$ & $1.041$ \\
\bottomrule
\end{tabular}
\end{table}

\begin{table}[ht]
\caption{Pooled $X_2$ bias by half-unit bin, in units of $10^{-3}$, same
$22$ patients.  \emph{Post hoc.}}
\label{tab:t2bp:biasbin}
\small
\begin{tabular}{@{}lrrrrrr@{}}
\toprule
Estimator & bin $1$ & $2$ & $3$ & $4$ & $5$ & $6$ \\
\midrule
Box MLE $+$ correction & $-0.00$ & $-0.00$ & $-0.01$ & $-0.04$ & $-0.25$ & $-1.86$ \\
Box MLE, no correction & $-0.24$ & $-0.74$ & $-1.68$ & $-3.01$ & $-4.65$ & $-6.62$ \\
T2bp$\,\circ\,$SNN2    & $-0.21$ & $-0.57$ & $-1.38$ & $-2.22$ & $-2.97$ & $-2.35$ \\
\bottomrule
\end{tabular}
\end{table}

\paragraph{Sensitivity to the three constants.}
The kernel bandwidth $h_3$, the adaptive-width coefficient $c_2$ and the
width floor $h_2^{\min}$ were set when T2bp was designed and were not
changed afterwards; $c_2$ and $h_2^{\min}$ are declared numerical floors
rather than fitted values, and the whole production corpus ran with all
three unchanged.  Halving and
doubling each, one at a time, on ten cells (the eight test cells, and
LIDC-0310 and LIDC-0766 under $\NoFC$): $h_3$ and $h_2^{\min}$ move
the $X_2<6$ error by under $1\%$ everywhere, in every bin range.  So
does $c_2$ on eight of the ten.  On the two cells whose base estimator
is furthest from the floor --- the two baseline cells with
MSE/floor $1.82$ and $1.92$ --- halving $c_2$ \emph{improves} the
$X_2<6$ error by $13\%$ and $19\%$ and doubling it costs $28\%$ and
$38\%$.  So $c_2=0.5$ is untuned but not optimal: a smaller value is
better where the prior does the most work, and worse in the deep bins.
Iteration count is not a sensitive choice either: $K=6$ matches $K=8$ to
four decimals at $X_2<6$, and one pass already reaches $99.5\%$ of the
gain there, though it costs $11\%$ over all bins.

\paragraph{Self-influence of the per-patient prior.}
The prior is built from the same bundles it scores.  Scoring every kept
slice with a prior built only from the complementary slices --- so that
each slice is scored exactly once, by a prior it did not contribute to
--- changes the $X_2<6$ error by $+0.6\%$ on LIDC-0116 under $\NoFC$ and
by $-1.2\%$ on PIS\_NFLTD under the baseline acquisition; over all bins,
by $+1.3\%$ and $+4.2\%$.  The out-of-sample prior is built from about
half as many slices as the comparison prior, so these are upper bounds
on self-influence rather than estimates of it.

\subsection{Endpoint and three-path errors}
\label{app:t2bp:threepath}

\emph{Supports main Section~\ref{sec:results:composition}.}

T2bp replaces only the middle-path estimate, so each
T2bp$\,\circ\,$base estimator keeps its base estimator's endpoint errors.
Table~\ref{tab:t2bp:threepath} gives, on the eight test cells, the
endpoint error $[(\mathrm{MSE}_1+\mathrm{MSE}_3)/2]^{1/2}$ and the
three-path error $(\mathrm{MSE}_1+\mathrm{MSE}_2+\mathrm{MSE}_3)^{1/2}$,
each $\mathrm{MSE}_k$ pooled over the populated bins of path~$k$ as in
the main text.  Under the baseline acquisition the endpoints carry
$0.1$--$0.8\%$ of the base estimators' summed squared error, so T2bp's
three-path and middle-path gains agree to within $0.4$ percentage
points.  $\NoFC$ takes photons from the endpoint rays, and the base
estimators' endpoint errors are $2.7$--$4.8$ times the baseline's.  The
endpoints then carry $9$--$38\%$ of the summed squared error, and T2bp's
three-path gain is $58$--$89\%$ of its middle-path gain, the upper end
on LIDC-0116, whose middle path dominates.  For LIDC-0116 the $X_2$
errors include the $24{,}719$ bundles with $X_2\ge9.2$ that the per-bin
evaluation leaves out, as main Table~\ref{tab:composition-10cell}
does.

\begin{table}[ht]
\caption{Endpoint and three-path pool RMSE of the four estimators on
the eight test cells.  T2bp leaves the endpoints unchanged, so each
T2bp$\,\circ\,$base estimator has its base estimator's endpoint error.  The last
two columns give the middle path's share of each base estimator's
summed squared error.}
\label{tab:t2bp:threepath}
\small
\setlength{\tabcolsep}{4pt}
\begin{tabular}{@{}lrrrrrrrr@{}}
\toprule
 & \multicolumn{2}{c}{Endpoint} & \multicolumn{4}{c}{Three-path} & \multicolumn{2}{c}{$X_2$ share (\%)} \\
\cmidrule(lr){2-3}\cmidrule(lr){4-7}\cmidrule(l){8-9}
Patient & SNN2 & V8c & SNN2 & T2bp$\circ$SNN2 & V8c & T2bp$\circ$V8c & SNN2 & V8c \\
\midrule
\multicolumn{9}{l}{\textit{Flat-fluence baseline}} \\
PIS        & $0.00547$ & $0.00545$ & $0.15834$ & $0.10674$ & $0.11540$ & $0.09947$ & $99.8$ & $99.6$ \\
LIDC-0951  & $0.00417$ & $0.00418$ & $0.16845$ & $0.10658$ & $0.11808$ & $0.10175$ & $99.9$ & $99.7$ \\
LIDC-0116  & $0.01095$ & $0.01118$ & $0.41551$ & $0.30283$ & $0.32585$ & $0.36421$ & $99.9$ & $99.8$ \\
PIS\_NFLTD & $0.01775$ & $0.01836$ & $0.28050$ & $0.17699$ & $0.47221$ & $0.18976$ & $99.2$ & $99.7$ \\
\midrule
\multicolumn{9}{l}{\textit{$\NoFC$}} \\
PIS        & $0.01634$ & $0.01464$ & $0.03882$ & $0.03704$ & $0.03615$ & $0.03578$ & $64.6$ & $67.2$ \\
LIDC-0951  & $0.01470$ & $0.01141$ & $0.03388$ & $0.03182$ & $0.02925$ & $0.02839$ & $62.4$ & $69.6$ \\
LIDC-0116  & $0.03740$ & $0.04394$ & $0.17199$ & $0.13795$ & $0.19653$ & $0.15346$ & $90.5$ & $90.0$ \\
PIS\_NFLTD & $0.06149$ & $0.08833$ & $0.15045$ & $0.12844$ & $0.20252$ & $0.16848$ & $66.6$ & $62.0$ \\
\bottomrule
\end{tabular}
\end{table}

\section{Acquisition and bias-structure tests}
\label{app:biasstruct}

This section supports main Sections~\ref{sec:snn2:mle},
\ref{sec:results:v8c-ood} and~\ref{sec:bridge}.  The tests reuse test patients; each test's
expected outcomes, or its descriptive status, were fixed in writing
before it ran (post hoc readings are marked as such), and all data
checks passed.  $X_2$ bins are those of main
equation~\eqref{eq:x2bins}.  Four tests are reported here --- E1 (the
shift law at matched cells), E2, E3b with the override test, and E5 ---
and their readings and outcomes are collected in
Table~\ref{tab:tests}.  E4, the corpus-wide version of E1, is left to
Paper~3, where SNN2's per-bundle bias is the reference for estimators
that pool neighboring bundles.

\subsection{The shift law for per-bundle estimators}
\label{app:biasstruct:shift}

\emph{Supports main Sections~\ref{sec:snn2:mle} and~\ref{sec:bridge}:
SNN2's bias as a function of detected-fluence depth.}

Write $Z_k=X_k-\ln\Nzeroray{k}=-\ln\Phi_k$ for the detected-fluence
depth of ray~$k$.
\begin{enumerate}[nosep, leftmargin=*]
  \item \emph{The data depend on $(X,\Nzero)$ only through $Z$.}  The
        mean of row~$j$ is $y_j=\sum_kA_{jk}\Phi_k=\sum_kA_{jk}e^{-Z_k}$,
        and the counts of each reading are Poisson with a mean set by
        the $\Phi_k$ (Section~\ref{app:snn2:deviates}).  The
        distribution of the five row counts $Y$ is therefore a function
        of $Z$ alone.
  \item \emph{SNN2 is equivariant.}  Every step of SNN2 depends only on
        $Y$ and on the current fluence estimates
        $\hat\Phi_k=\Nzeroray{k}e^{-\Xhat_k}=e^{-\hat Z_k}$: the count
        floor $y_j\ge1$, the endpoint start, the SVD warm start
        $-\ln\bigl((A^{+}Y)_k/\Nzeroray{k}\bigr)$, the Gauss--Newton step (the Fisher
        information and the score are functions of $\hat\Phi$), the
        trust region (on the step), the stopping rule (on the change of
        the estimate), the CRB gate (on $\sigma_k$) and the second-order
        correction (on $\sigma_k^2$).  Only the box $0\le\Xhat_k\le9.5$
        and the warm-start clip $[0,9.2]$ are fixed in $X$.  Where
        neither binds, $\hat Z_k=\Xhat_k-\ln\Nzeroray{k}$ is a function
        of $Y$ alone.
  \item \emph{One bias function.}  Hence
        $\E[\Xhat_k-X_k\mid X,\Nzero]=\E[\hat Z_k-Z_k\mid Z]
        =G_k(Z_1,Z_2,Z_3)$, the same function for every patient and for
        both acquisitions.
  \item \emph{The shift law.}  Under the baseline acquisition
        $Z=X-u_p(1,1,1)$ for patient~$p$, with $u_p=\ln N_{\mathrm{base},p}$,
        so two
        patients compared at the same $X$ differ by a uniform shift of
        all three depths: SNN2's bias at $X$ under $N_0'$ equals its
        bias at $X-\ln(N_0'/N_0)\,(1,1,1)$ under $N_0$, wherever the box
        does not bind.
  \item \emph{First-order scaling.}  The log-likelihood is
        $\ell=\sum_j\bigl(Y_j\ln y_j-y_j\bigr)$.  With
        $\Nzeroray{k}=c\,r_k$, every $y_j$ scales as $c$ at fixed $X$,
        and every derivative of $\ell$ with respect to the $X_k$ is a sum
        of terms, each a $c$-free coefficient times $Y_j$ or $y_j$.
        Since $\E[Y_j]=\Var(Y_j)=y_j$, the Fisher information $F$ and
        the cumulants $\kappa_{kmn}=\E[\partial^3\ell/\partial X_k\partial
        X_m\partial X_n]$ and $\kappa_{km,n}=\E[(\partial^2\ell/\partial
        X_k\partial X_m)(\partial\ell/\partial X_n)]$ all scale as $c$.
        The first-order bias of the maximum-likelihood estimate
        \citep{coxsnell1968S},
        $b_s=\sum_{k,m,n}[F^{-1}]_{sk}[F^{-1}]_{mn}
        (\kappa_{km,n}+\tfrac12\kappa_{kmn})$, therefore scales as $1/c$:
        doubling the photons halves the bias, to first order.  Where
        SNN2's gated second-order correction acts, the measured bias
        falls somewhat faster, presumably because the correction opens
        more at higher counts (\emph{hypothesis}).
\end{enumerate}
The stored outputs obey the construction: recomputing the SVD warm
start and SNN2 from the stored counts and $\Nzero$ on $199{,}980$
sampled bundles ($19{,}998$ per file, ten files) reproduced every stored
estimate exactly in single precision.  V8c also works per bundle, but it
is not equivariant: it takes the estimates and $\log_{10}$ of the
bundle's mean $\Nzero$ as separate inputs (main
Section~\ref{sec:library:v8c}), so the shift law does not constrain it.

\subsection{Volumes, cells and statistics}
\label{app:biasstruct:method}

\emph{The machinery shared by the four tests, stated once.}

\emph{Volumes.}  E1 uses SNN2 estimates and bundle records of five
volumes in both regimes, every second slice: PIS, PIS\_NFLTD,
LIDC-0116 and LIDC-0951 (test files) and LIDC-0709 (validation file),
$934$ slices of $340{,}992$ bundles each.  E2 adds \emph{PIS\_NFLTDC},
PIS\_NFLTD acquired as every corpus volume is: it shares PIS\_NFLTD's
sinograms, so every bundle carries the same true $X$, and its
$N_{\mathrm{base}}=602{,}461$ comes from the calibration rule of
Section~\ref{app:n0fc:nbase} applied to PIS\_NFLTD's sinograms,
$2.63$ times the $229{,}266$ copied from PIS.  Its tube-current knots
come from its own TCM fit with the global $B^*$, and its noise is
independent (seeds derived from the patient identifier), so under the
baseline the two volumes differ only in $N_{\mathrm{base}}$ and under
$\NoFC$ in $N_{\mathrm{base}}$ and $\tau$.  E3b adds \emph{PIS\_N602},
PIS acquired with PIS\_NFLTDC's $N_{\mathrm{base}}$: PIS's sinograms,
PIS's bow-tie and TCM with $\Nzero$ scaled by $2.63$ on every ray (none
at the $10^3$ or $10^6$ clip), and independent noise.  For E2 and E3b
the bundle files cover all $200$ slices at $1024$ views ($68{,}198{,}400$
bundles per file); SNN2-alone was rerun, and V8c-alone with the paper's
regime-matched checkpoints.

\emph{Cells.}  Bundles are binned in three dimensions by $X$ (cells
$0.25$ wide on $[0,9.5)$ for each path, with a separate bin for an
exact air path) or by $Z'_k=X_k-\ln(\Nzeroray{k}/N_{\mathrm{ref}})$,
with $N_{\mathrm{ref}}=229{,}266$ (PIS's $N_{\mathrm{base}}$, so that
$Z'=X$ at that count).  Removal fractions, and the comparisons between
the two regimes of one patient, use reduced cells. These keep $X_2$ (or
$Z'_2$) at $0.25$ and bin the outer paths only through $X_1-X_2$ and
$X_3-X_2$ (or the $Z'$ differences) at $0.5$: $\NoFC$ shifts all three
depths, so the two regimes of one patient share almost no full
three-dimensional $Z'$ cells.  A cell enters a comparison if it holds
at least $200$ bundles in both members and fewer than $1\%$ of their
estimates sit at the upper box limit.

\emph{Statistics.}  Per cell, $b$ is the mean of $\Xhat_2-X_2$, with
standard error $\mathrm{SE}$, over its $n$ bundles.  For two members,
$q$ the first of a pair and $p$ the second, $d=b_p-b_q$,
$v=\mathrm{SE}_p^2+\mathrm{SE}_q^2$ and $w=2n_pn_q/(n_p+n_q)$; sums run
over the common cells.
\begin{enumerate}[nosep, leftmargin=*]
  \item \emph{Difference.}  The noise-corrected RMS difference is
        \[\Delta=\Bigl[\max\Bigl(\textstyle\sum w\,(d^2-v)\big/\sum w,\;0\Bigr)\Bigr]^{1/2}.\]
  \item \emph{Relative difference.}  $\Delta/b_{\mathrm{rms}}$, with
        \[b_{\mathrm{rms}}=\Bigl[\max\Bigl(\textstyle\sum w\,(\bar b^2-v/4)\big/\sum w,\;0\Bigr)\Bigr]^{1/2}\]
        the noise-corrected RMS of the pair's mean bias
        $\bar b=(b_p+b_q)/2$.
  \item \emph{Bias ratio.}  The median of $b_p/b_q$, weighted by $w$,
        over cells where $|b|\ge5\,\mathrm{SE}$ in both members.
  \item \emph{Removal.}  Binning by $Z'$ removes the fraction
        $1-\Delta_{Z'}/\Delta_X$ of a difference (reduced cells,
        $X_2<7$), after a first-order adjustment for the different
        distributions of $X$ inside the cells of the two members: $d$ is
        replaced by $d-\sum_kg_k\,(m_{p,k}-m_{q,k})$, where $m_{\cdot,k}$
        is a member's mean binning variable along axis~$k$ of the cell
        and $g_k$ the slope of the pooled bias along that axis, from the
        neighboring cells (a central difference where both neighbors
        hold at least $200$ bundles, otherwise one-sided, otherwise
        zero).  The rest is within noise when the adjusted $\Delta^2$ is
        at most three times its null standard deviation.
\end{enumerate}

\emph{E5's decomposition.}  T2bp's estimate (main
equation~\eqref{eq:t2bp-posterior}) combines the bundle's own likelihood
with the patient's prior.  Locally
$\Xhat_2^{\mathrm{T2bp}}\approx m+s\,(\Xhat_2-m)$, with $\Xhat_2$ the
SNN2 estimate, $s$ the weight on it and $m$ the local prior mean, so at
a fixed true $X_2$
\begin{equation}
  \E\bigl[\Xhat_2^{\mathrm{T2bp}}-X_2\bigr]\approx
  s\,\E\bigl[\Xhat_2-X_2\bigr]+(1-s)\,(m-X_2),
\label{eq:t2bp-split}
\end{equation}
the first term SNN2's bias carried through and the second shrinkage
toward the prior.  Each patient's T2bp output is fitted without the
ground truth in cells $0.25$ wide in $\Xhat_2$ and $\Xhat_3$: a line
$\alpha+s\,\Xhat_2$ for cells with at least $200$ bundles, the cell mean
where $\Xhat_2$ has no spread (SNN2's box at $9.5$), otherwise a plane
in $\Xhat_2$ and $\Xhat_3$ over the cell's $0.25$-wide, then unit-wide,
$\Xhat_2$ bin.  With $M$ that mapping and bars denoting means over the
bundles whose true $X_2$ lies in bin~$B$ (main
equation~\eqref{eq:x2bins}),
\begin{equation}
  b_2=\overline{M(\Xhat_2,\Xhat_3)}-\overline{X_2}
     =\underbrace{\overline{M(\Xhat_2,\Xhat_3)-M(X_2,\Xhat_3)}}_{\text{SNN2's error carried through}}
     +\underbrace{\overline{M(X_2,\Xhat_3)-X_2}}_{\text{shrinkage}}.
\label{eq:t2bp-terms}
\end{equation}
The shrinkage term evaluates the mapping where the truth lies;
evaluating each cell's line at the true $X_2$, outside its cell, moves
curvature into the first term instead (on synthetic data a shrinkage
share of $0.81$ against an exact $0.86$, and here $0.55$ against
$0.84$ under the baseline).  That choice was made in the design, after
the synthetic test and before the run.  The data are the corpus run's
per-bundle T2bp$\,\circ\,$SNN2 estimates (truth-free settings, $K=6$;
main Section~\ref{sec:results:dif}) for the $23$ non-test patients,
PIS\_NFLTD and PIS\_NFLTDC in both regimes --- $50$ volume--regime
combinations, $4.08\times10^{9}$ bundles --- with the SNN2 estimates
they started from; the ground truth enters only the evaluation.  Each
patient enters a bin with at least $3$ cells, and with each bin's mean
over patients removed the $R^2$ of the differences compares the
patients' deviations, while the $R^2$ with the level counts the whole
error.

\emph{Checks.}  All passed.  E1: the stored outputs reproduce exactly
(Section~\ref{app:biasstruct:shift}).  E2 and E3b: the same true $X$
bundle for bundle on all $200$ slices of both regimes,
$\Nzero=N_{\mathrm{base}}$ on every baseline ray, SNN2 recomputed
exactly on $19{,}998$ bundles per file, and the source volumes' printed
pool values reproduced; E2's $\NoFC$ incident $\Nzero$ stayed within
$[10^3,10^6]$, with $1.9\%$ of rays at the ceiling and $0.45\%$ at the
floor against $0\%$ and $0.85\%$ for PIS\_NFLTD.  The override test's
control, with the count input unchanged, reproduced V8c's stored
predictions exactly.  E5: T2bp's stored endpoint estimates equal SNN2's
in every bundle, and the pool RMSE of both estimators and T2bp's error
sums per bin reproduce the corpus records (to relative
$4\times10^{-6}$ and $5\times10^{-10}$).

\subsection{Readings fixed in advance, and outcomes}
\label{app:biasstruct:tests}

Table~\ref{tab:tests} gives, for each test, the reading fixed in
writing before it ran and the outcome.  The per-test tables of
version~2 of the preprint (arXiv:2610.19229v2, Tables~S12--S15) hold
the full per-cell and per-run results.

\begingroup
\setlength{\LTcapwidth}{\textwidth}
\setlength{\tabcolsep}{4pt}
\begin{longtable}{@{}>{\footnotesize\raggedright\arraybackslash}p{0.16\linewidth}>{\footnotesize}p{0.38\linewidth}>{\footnotesize}p{0.38\linewidth}@{}}
\caption{The four acquisition and bias-structure tests: the reading
fixed in writing before each ran, and its outcome.  $R$ is the ratio of
V8c-alone to SNN2-alone pool $X_2$ RMSE; $t$ the share of E2's change
in V8c's deep bias ($X_2\ge7$) that the count input alone moves;
$\xi_Q$ the ratio of $R$ under PIS's count input to $R$ under
PIS\_N602's own.  Full results: arXiv:2610.19229v2, Tables~S12--S15.}
\label{tab:tests}
\label{tab:e2}\\
\toprule
Test & Reading fixed in advance & Outcome \\
\midrule
\endfirsthead
\multicolumn{3}{@{}l}{\footnotesize\itshape Table~\ref{tab:tests} continued from the previous page.}\\
\toprule
Test & Reading fixed in advance & Outcome \\
\midrule
\endhead
\bottomrule
\endlastfoot
\textbf{E1}, matched cells
 & Five outcomes: relative difference $\le0.15$ at equal $\Nzero$;
   $\ge0.3$ at an $\Nzero$ ratio of $2$; binning by $Z'$ removes
   $\ge0.7$ between patients and $\ge0.5$ within one, or the rest is
   within noise; and across the ten baseline pairs the relative
   difference tracks $|\Delta\ln N_{\mathrm{base}}|$ (Spearman
   $\ge0.6$) more than $|\Delta\mathrm{WED}|$.
 & \textbf{All five met.}  Relative difference $0.00$ at equal
   $\Nzero$ (PIS and PIS\_NFLTD, one anatomy at two scales whose WEDs
   differ by $98$~mm) and $0.76$--$1.46$ at
   photon ratios of $2$ and above; removal $0.92$--$1.00$, the rest
   within noise in every pair; Spearman $0.83$ against $-0.17$ for
   WED (ten pairs of five volumes, indicative only).  The bias ratio
   was reported against its first-order $1/\Nzero$ guide, not tested:
   $0.32$--$0.46$ against $0.38$--$0.50$. \\
\addlinespace
\textbf{E2}, the same anatomy by the corpus rule
 & $R=1.69$ on PIS\_NFLTD under the baseline would fall to $\le1.0$ if
   the acquisition caused V8c's failure and stay $\ge1.3$ if the
   anatomy did.  Validity guard: V8c's error at $X_2<5$ must not grow,
   $\log_{10}N_{\mathrm{base}}=5.78$ lying above V8c's training range
   $5.25$--$5.72$.
 & \textbf{Acquisition.}  $R=0.85$ under the baseline, closing $83\%$
   of the log gap to the in-distribution geometric mean of $0.74$; the
   guard did not trigger (RMSE at $X_2<5$ fell from $0.095$ to
   $0.024$).  Under $\NoFC$ most of the deficit remains, $R=1.22$
   against $1.14$ on LIDC-0116, the closest in-distribution patient in
   size. \\
\addlinespace
\textbf{E3b} and the override test
 & E3b descriptive, read against a first-order count-level reference
   $R\approx1.02$.  Override test, cuts fixed: the transfer fraction
   $t\ge0.5$ means the count input carries most of E2's change,
   $t\le0.1$ none; $\xi_Q\le0.95$ would support a body-size reading of
   the input, $\xi_Q\ge1$ would not.
 & \textbf{Mixed; the mechanism is not resolved.}  E3b: $R=1.46$ on
   PIS\_N602 against $0.73$ on PIS and the reference $1.02$, so V8c's
   advantage is lost beyond what the photon level explains.  Override:
   $t=0.84$ (run~A) and $0.45$ (run~B); $\xi_Q=6.9$, so a body-size
   reading is not supported.  Both effects are present already at
   $5.72$ ($0.91$ and $1.02$ of the change), so neither is an
   extrapolation artifact. \\
\addlinespace
\textbf{E5}, the T2bp term
 & Five outcomes on bins~$7$--$9$: the own mapping accounts for $b_T$
   ($R^2$ with the level $\ge0.8$); shrinkage dominates (share
   $\ge0.6$) or the inherited term does ($\le0.4$); a corpus mapping
   either fails ($R^2$ of the differences $\le0.5$ and $\ge0.3$ below
   the own mapping's) or suffices ($\ge0.8$); the bin-level form holds
   ($R^2\ge0.6$) with the sign right ($\ge80\%$); and $s$ rises with
   the bin's share of the patient's bundles (Spearman $\ge0.5$, partial
   $\ge0.3$).
 & \textbf{Mixed; shrinkage dominates.}  Shrinkage takes $0.84$ and
   $0.87$ of the magnitudes in the two regimes and $0.93$ and $1.19$ of
   the differences between patients, while SNN2's bias alone gets those
   differences wrong ($R^2<0$) and the bin-level form reproduces them
   and their sign ($R^2$ $0.87$ and $0.85$; sign $0.84$ and $1.00$).
   The own mapping accounts for $b_T$ under the baseline ($R^2$ $0.97$)
   but only in part under $\NoFC$ ($0.87$ for the differences, $0.62$
   for the level).  Rarity holds in some bins only. \\
\end{longtable}
\endgroup

\paragraph{E1, matched cells.}\label{app:biasstruct:cells}
At equal $\Nzero$, PIS and PIS\_NFLTD, one anatomy at two scales whose
WEDs differ by $98$~mm, agree within noise
over the depths they share ($X_2<6$); a photon ratio of $2$ or $2.63$
gives a large difference that binning by $Z'$ removes, between patients
and within one.  The corpus-wide version of this test (each volume's
bias predicted from the others, and bundles of one volume compared at
the same $X$ but different $\Nzero$ under $\NoFC$) is E4, left to
Paper~3.

\paragraph{E2, the corpus acquisition rule.}\label{app:biasstruct:e2}
Because PIS\_NFLTDC shares PIS\_NFLTD's sinograms, the pair isolates
the acquisition at identical anatomy, and V8c's out-of-distribution
deficit largely closes under the baseline.  The same pair also tests
the shift law within one anatomy: at identical $X$ SNN2's bias differs
by a relative difference of $1.04$ with a bias ratio of $0.32$ against
$0.38$ for a $1/\Nzero$ scaling, and binning by $Z'$ removes $93\%$ of
the difference under the baseline and, within noise, all of it under
$\NoFC$.

\paragraph{E3b and V8c's count input.}\label{app:biasstruct:e3}
E2 changed the photon level and V8c's count input together; under the
baseline that input is the per-patient constant
$\log_{10}N_{\mathrm{base}}$, which the corpus rule ties to body size
(Spearman $0.97$ with WED).  E3b reverses the photon change at fixed
anatomy and the override test changes the input alone, on PIS\_NFLTD's
bundles from $5.36$ to the recalibrated $5.78$ (run~A) and on
PIS\_NFLTDC's from $5.78$ to $5.36$ (run~B), and the two
readings disagree: the input carries most of E2's change at the lower
counts and part of it at the higher, yet it is not acting as a body-size
proxy.  Where the input disagrees with the pairing of counts and
anatomy seen in training, V8c's error at $X_2<5$ rises
$10$--$15$-fold.

\paragraph{E5, the T2bp term.}\label{app:biasstruct:t2bp}
The shrinkage the posterior applies, not the bias inherited from SNN2,
carries T2bp's middle-path bias and the differences between patients.
The mapping from $(\Xhat_2,\Xhat_3)$ to that bias is patient-specific:
a corpus mapping reproduces the differences in part under the baseline
and not at all under $\NoFC$.  It depends on the patient through both
its prior and its photon level, which sets the width of the likelihood
at a given $\Xhat_2$, and the corpus comparison cannot separate the two
--- either one makes the mapping patient-specific, and the calibration
rule gives larger patients more photons.  Under $\NoFC$ a ray's
incident count follows its attenuation, so the likelihood may carry
information about the true depth that no mapping of
$(\Xhat_2,\Xhat_3)$ sees (\emph{hypothesis}); a post hoc refit of every
cell by a plane in $\Xhat_2$ and $\Xhat_3$ left the gap unchanged.

\section{Validation-pool test of the composition}
\label{app:valcomp}

\emph{Supports main Sections~\ref{sec:results:composition} and
\ref{sec:results:dif} and the held-out statement of main
Section~\ref{sec:discussion:limitations}: the four validation volumes,
the reading fixed in advance, and its outcome; the same reading of the
eight test cells; and, on those cells, T2bp on the box-constrained estimate.}

The composition comparison of main
Section~\ref{sec:results:composition} evaluates the four test volumes.
After the main results were fixed it was repeated, unchanged, on the four
volumes of the validation pool (Section~\ref{app:split}), at full volume
in both regimes: $319\,509\,504$ bundles per regime, against the
$\sim\!320$\,M that Table~\ref{tab:n0fc_val_x2_appendix} gives from an
independent count.

\paragraph{Their roles in development.}
The four volumes sit outside the test pool but not outside development.
V8c's checkpoint was selected on them (Section~\ref{app:v8c:train}).
Scored on the four test patients instead, the same epoch is the best of
the $\NoFC$ run's saved checkpoints (epochs $1$, $5$, $10$, $15$, $20$,
$25$ and $26$) on every one, and epoch~$26$ is worse by $6.5$--$22\%$;
whether the selection inflates V8c's numbers on these four volumes was
not tested.  LIDC-0766 is one of the four patients that set T2bp's
iteration count $K$, under the baseline acquisition
(Section~\ref{app:t2bp:deploy}).  The bow-tie $B^*$, which every $\NoFC$
cell uses, was fitted to reference slices that include LIDC-0628
(Section~\ref{app:n0fc:objective}).  The finding that motivated the
second training batch was read on the validation pool of that time,
which LIDC-0628 then joined from training (Section~\ref{app:split}).

Differences are paired within a cell and read by a block bootstrap:
contiguous blocks of $8$ slices, $2000$ resamples, every bundle of a
sampled block included.  A cell is called \emph{tied} when the $95\%$
interval does not exclude zero.  The eight test cells of main
Section~\ref{sec:results:composition} are read the same way at the end
of this section (Table~\ref{tab:tccomp}).

\begin{table}[htbp]
\centering
\caption{Validation-pool composition comparison, all eight cells at full
volume.  T2bp$\circ$V8c and T2bp$\circ$SNN2: pool $X_2$ RMSE.  $D$ is the
relative difference in pool $X_2$ RMSE,
$(\mathrm{T2bp}\circ\mathrm{V8c}-\mathrm{T2bp}\circ\mathrm{SNN2})$
divided by the latter; negative favors T2bp$\circ$V8c.  Intervals are
$95\%$, from a paired bootstrap over contiguous $8$-slice blocks with
$2000$ resamples; Blocks: the number of blocks per volume.}
\label{tab:valcomp}
\smallskip
\footnotesize
\begin{tabular}{@{}llrrrrl@{}}
\toprule
Volume & Regime & Blocks & T2bp$\circ$V8c & T2bp$\circ$SNN2 & $D$ (\%) & $95\%$ interval \\
\midrule
LIDC-0709 & $\NoFC$ & 13 & $0.026713$ & $0.030570$ & $-12.62$ & $[-13.15,-10.95]$ \\
LIDC-0052 & $\NoFC$ & 34 & $0.060502$ & $0.062032$ & $-2.47$  & $[-4.24,-0.96]$ \\
LIDC-0628 & $\NoFC$ & 36 & $0.066888$ & $0.069065$ & $-3.15$  & $[-4.84,-1.06]$ \\
LIDC-0766 & $\NoFC$ & 35 & $0.140253$ & $0.142456$ & $-1.55$  & $[-5.01,+1.69]$ \\
\addlinespace
LIDC-0709 & baseline & 13 & $0.095595$ & $0.109696$ & $-12.85$ & $[-14.67,-7.87]$ \\
LIDC-0052 & baseline & 34 & $0.200103$ & $0.203757$ & $-1.79$  & $[-7.70,+4.05]$ \\
LIDC-0628 & baseline & 36 & $0.192323$ & $0.162325$ & $+18.48$ & $[+11.62,+25.18]$ \\
LIDC-0766 & baseline & 35 & $0.394317$ & $0.322706$ & $+22.19$ & $[+17.55,+25.44]$ \\
\bottomrule
\end{tabular}
\end{table}

Under $\NoFC$ T2bp$\circ$V8c is ahead on three volumes and tied on the
fourth, and T2bp$\circ$SNN2 is ahead on none; the median difference across
the four is $-2.81\%$, resampling patients only.  Under the baseline
acquisition the picture splits: T2bp$\circ$V8c is far ahead on the
smallest volume, T2bp$\circ$SNN2 is ahead by $18$--$22\%$ on the two largest, and the
fourth is tied.

\paragraph{The reading fixed in advance.}
Before the comparison ran, a condition was recorded for reopening the
question of acquiring volumes from outside the corpus: under $\NoFC$,
T2bp$\circ$V8c ahead by more than $3\%$ with the interval excluding zero, on two
or more of the four.  LIDC-0709 ($-12.62\%$) and LIDC-0628 ($-3.15\%$, by
$0.15$ points) met it; LIDC-0052 fell short on the margin and LIDC-0766 on
the interval.  \textbf{The condition was met.}  A prediction recorded
alongside it, that it would not be, was wrong, and is recorded here as a miss.  The
question was nonetheless not reopened: the argument of main
Section~\ref{sec:bridge} does not rest on the composition ordering that
this condition tests, and the candidate volumes screened reach only
$288$~mm, while under $\NoFC$ the test cells favor T2bp$\circ$SNN2 by
more than $1\%$ only at $334.5$ and about $342$~mm (LIDC-0116 and
PIS\_NFLTD).

\paragraph{What these volumes do confirm.}
Their water-equivalent diameters --- $224.5$, $263.5$, $305.9$ and
$316.8$~mm --- fall between those of the test volumes LIDC-0951
($206.9$~mm), PIS ($244.4$~mm) and LIDC-0116 ($334.5$~mm).
Interpolating those cells' $\NoFC$ margins $D_{\mathrm{S}}$
(Table~\ref{tab:tccomp}) linearly in WED predicts $D=-1.2\%$ on
LIDC-0709 ($224.5$~mm) and $+2.8$, $+7.2$ and $+8.3\%$ on LIDC-0052
($263.5$~mm), LIDC-0766 ($305.9$~mm) and LIDC-0628 ($316.8$~mm).  Those
three come in on the other side, and LIDC-0709 favors T2bp$\circ$V8c by
about ten times the predicted margin.  The trend of main
Section~\ref{sec:results:v8c} is reproduced, however, for the final
network (an earlier network had shown it on validation and test
patients; Section~\ref{app:split}): under $\NoFC$, V8c-alone's advantage over
SNN2-alone falls from $22.1\%$ on LIDC-0709 to $6.9\%$, $2.8\%$ and
$5.3\%$ on LIDC-0052, LIDC-0766 and LIDC-0628, still positive
throughout, and negative on LIDC-0116.  Composing T2bp removes $40$--$64\%$ of that
advantage, helping the weaker base estimator more.

\paragraph{T2bp degrades V8c under the baseline acquisition.}
On three of the four baseline cells, composing T2bp raises V8c's
pool RMSE --- by $7.21\%$ on LIDC-0709, $5.41\%$ on LIDC-0766 and $4.16\%$
on LIDC-0052 --- and lowers it by $5.39\%$ on LIDC-0628.  On the same
volumes T2bp lowers SNN2's RMSE by $36$--$52\%$, and under
$\NoFC$ it lowers V8c's by $4.2$--$22.7\%$.  Main
Section~\ref{sec:results:composition} reports the single-cell version of
this, LIDC-0116 under the baseline at $+12\%$; across these volumes it is a
pattern, confined to the baseline regime, and T2bp degrades SNN2
on none of the sixteen cells now measured.  \emph{Hypothesis:} T2bp takes
its likelihood width from the CRB median, which assumes the base
estimator's error is
variance-dominated at roughly CRB scale.  Under the baseline the V8c error
is not: its standard deviation runs $0.097$, $0.047$ and $0.040$ of the
CRB in $X_2$ bins $7$, $8$ and $9$, with biases of $-0.20$, $-0.92$ and
$-1.63$, so the error is almost all bias with the variance far below the
bound, and T2bp models no bias.  Under $\NoFC$ the same bins run near
$1.0$--$1.4$ and the assumption holds.

\paragraph{The test cells.}
Table~\ref{tab:tccomp} applies the same reading to the eight test cells
of main Section~\ref{sec:results:composition}; the margins quoted there
are $D_{\mathrm{V}}$, relative to T2bp$\circ$V8c.  Of the four $\NoFC$
cells T2bp$\circ$SNN2 wins three, T2bp$\circ$V8c one, and none is tied;
of the four baseline cells T2bp$\circ$SNN2 wins one, T2bp$\circ$V8c one,
and two are tied.

\begin{table}[htbp]
\centering
\caption{Test-cell composition comparison: the eight cells of main
Table~\ref{tab:composition-10cell}, read as in Table~\ref{tab:valcomp}.
$D_{\mathrm{S}}$ is the difference in pool $X_2$ RMSE,
$(\mathrm{T2bp}\circ\mathrm{V8c}-\mathrm{T2bp}\circ\mathrm{SNN2})$,
divided by the latter (the $D$ of Table~\ref{tab:valcomp});
$D_{\mathrm{V}}$ divides it by T2bp$\circ$V8c.  Negative favors
T2bp$\circ$V8c.  Intervals are $95\%$, from a paired bootstrap over contiguous
$8$-slice blocks with $2000$ resamples: 25 blocks for PIS and
PIS\_NFLTD, 38 for LIDC-0951 and 17 for LIDC-0116.  Winner, as in main
Table~\ref{tab:composition-10cell}: $\star$ T2bp$\circ$SNN2 or
$\bullet$ T2bp$\circ$V8c when the interval excludes zero, otherwise
tied.}
\label{tab:tccomp}
\smallskip
\scriptsize
\setlength{\tabcolsep}{4pt}
\begin{tabular}{@{}llrrrlrll@{}}
\toprule
Volume & Regime & T2bp$\circ$V8c & T2bp$\circ$SNN2 & $D_{\mathrm{S}}$ (\%) & $95\%$ interval & $D_{\mathrm{V}}$ (\%) & $95\%$ interval & Winner \\
\midrule
PIS & $\NoFC$ & $0.029184$ & $0.028951$ & $+0.81$ & $[+0.11,+1.28]$ & $+0.80$ & $[+0.11,+1.26]$ & $\star$ \\
LIDC-0951 & $\NoFC$ & $0.023359$ & $0.024088$ & $-3.03$ & $[-3.96,-2.13]$ & $-3.12$ & $[-4.12,-2.18]$ & $\bullet$ \\
LIDC-0116 & $\NoFC$ & $0.140319$ & $0.127409$ & $+10.13$ & $[+5.44,+14.80]$ & $+9.20$ & $[+5.16,+12.89]$ & $\star$ \\
PIS\_NFLTD & $\NoFC$ & $0.113064$ & $0.094533$ & $+19.60$ & $[+13.05,+25.31]$ & $+16.39$ & $[+11.54,+20.20]$ & $\star$ \\
\addlinespace
PIS & baseline & $0.099170$ & $0.106464$ & $-6.85$ & $[-9.78,-2.40]$ & $-7.36$ & $[-10.84,-2.46]$ & $\bullet$ \\
LIDC-0951 & baseline & $0.101581$ & $0.106418$ & $-4.55$ & $[-12.89,+5.91]$ & $-4.76$ & $[-14.80,+5.58]$ & tied \\
LIDC-0116 & baseline & $0.363863$ & $0.302430$ & $+20.31$ & $[+17.21,+25.68]$ & $+16.88$ & $[+14.68,+20.44]$ & $\star$ \\
PIS\_NFLTD & baseline & $0.187977$ & $0.175198$ & $+7.29$ & $[-0.83,+13.16]$ & $+6.80$ & $[-0.83,+11.63]$ & tied \\
\bottomrule
\end{tabular}
\end{table}

\paragraph{T2bp on the box-constrained estimate.}
Table~\ref{tab:kktcomp} composes T2bp, unchanged, with the box-constrained
MLE of Section~\ref{app:snn2:mle} without the bias correction (main
Section~\ref{sec:results:dif}), on all slices of the eight test cells,
with the settings of main Table~\ref{tab:composition-10cell} and the
bootstrap of Table~\ref{tab:tccomp}.  Three readings were fixed before the
run.  The first, $G$, is the share of the box estimate's mean squared
error that T2bp removes: a median of at least $0.05$ over a regime's four
cells would mean that the per-patient prior adds on this base estimator,
one below
$0.02$ that the box estimate reproduces T2bp's gain.  The medians are
$0.194$ under $\NoFC$, where the gain is on the two larger volumes, and
$0.472$ under the baseline; every interval excludes zero.  The second,
$D_{\mathrm{SB}}$, compares the two classical base estimators; the choice
was to be
called immaterial in a regime if every cell were tied or within
$\pm2\%$.  It is immaterial under $\NoFC$, where every cell is within
$\pm2\%$ although three intervals exclude zero; under the baseline
acquisition, beyond $\pm2\%$, T2bp$\circ$Box is better on LIDC-0951 and
T2bp$\circ$SNN2 on LIDC-0116.  On LIDC-0116 the error of T2bp$\circ$Box
is lowest at iteration~$3$, where the two compositions tie ($0.2940$
against $0.2944$), and rises $5.9\%$ by $K=6$, against $3.1\%$ for
T2bp$\circ$SNN2 from its minimum at iteration~$4$
(\emph{post hoc}).  The third, $D_{\mathrm{V}}$ against T2bp$\circ$V8c,
was descriptive; its verdicts equal Table~\ref{tab:tccomp}'s in all
eight cells.

\begin{table}[htbp]
\centering
\caption{T2bp on the box-constrained estimate: the eight test cells of main
Table~\ref{tab:composition-10cell}.  Box: pool $X_2$ RMSE of the
box-constrained MLE without the bias correction; T2bp$\circ$Box: pool
$X_2$ RMSE after the fixed $K=6$ iterations.
$G=1-\mathrm{MSE}(\mathrm{T2bp}\circ\mathrm{Box})/\mathrm{MSE}(\mathrm{Box})$.
$D_{\mathrm{SB}}$ is the difference in pool $X_2$ RMSE,
$(\mathrm{T2bp}\circ\mathrm{Box}-\mathrm{T2bp}\circ\mathrm{SNN2})$,
divided by the latter; negative favors T2bp$\circ$Box.  $D_{\mathrm{V}}$ is
that of Table~\ref{tab:tccomp} with the box-constrained estimate in place
of SNN2,
$(\mathrm{T2bp}\circ\mathrm{V8c}-\mathrm{T2bp}\circ\mathrm{Box})$ divided
by T2bp$\circ$V8c; negative favors T2bp$\circ$V8c.  Intervals are $95\%$,
from the paired bootstrap of Table~\ref{tab:tccomp}.  Verdict (the
unlabeled column after each interval of $D_{\mathrm{SB}}$ and
$D_{\mathrm{V}}$): B, S or V (box, SNN2 or V8c base estimator better)
when the interval excludes zero,
otherwise t (tied).}
\label{tab:kktcomp}
\smallskip
\scriptsize
\setlength{\tabcolsep}{3pt}
\begin{tabular}{@{}lrrrlrlcrlc@{}}
\toprule
Volume & Box & T2bp$\circ$Box & $G$ & $95\%$ interval & $D_{\mathrm{SB}}$ (\%) & $95\%$ interval & & $D_{\mathrm{V}}$ (\%) & $95\%$ interval & \\
\midrule
\multicolumn{11}{@{}l}{\textit{$\NoFC$}} \\
PIS & $0.029208$ & $0.028910$ & $0.020$ & $[0.004,0.034]$ & $-0.14$ & $[-0.17,-0.10]$ & B & $+0.94$ & $[+0.23,+1.43]$ & B \\
LIDC-0951 & $0.024266$ & $0.024052$ & $0.018$ & $[0.007,0.026]$ & $-0.15$ & $[-0.18,-0.11]$ & B & $-2.97$ & $[-3.99,-2.00]$ & V \\
LIDC-0116 & $0.159126$ & $0.126551$ & $0.368$ & $[0.311,0.399]$ & $-0.67$ & $[-0.95,-0.27]$ & B & $+9.81$ & $[+5.63,+13.30]$ & B \\
PIS\_NFLTD & $0.121341$ & $0.094330$ & $0.396$ & $[0.281,0.457]$ & $-0.22$ & $[-0.47,+0.20]$ & t & $+16.57$ & $[+11.69,+20.38]$ & B \\
\addlinespace
\multicolumn{11}{@{}l}{\textit{Baseline}} \\
PIS & $0.143871$ & $0.105233$ & $0.465$ & $[0.196,0.574]$ & $-1.16$ & $[-1.40,-0.76]$ & B & $-6.11$ & $[-9.19,-1.77]$ & V \\
LIDC-0951 & $0.143351$ & $0.103498$ & $0.479$ & $[0.275,0.611]$ & $-2.74$ & $[-3.00,-2.41]$ & B & $-1.89$ & $[-12.10,+8.09]$ & t \\
LIDC-0116 & $0.415128$ & $0.311291$ & $0.438$ & $[0.376,0.511]$ & $+2.93$ & $[+1.42,+3.84]$ & S & $+14.45$ & $[+11.54,+19.20]$ & B \\
PIS\_NFLTD & $0.279204$ & $0.175126$ & $0.607$ & $[0.530,0.650]$ & $-0.04$ & $[-0.76,+0.67]$ & t & $+6.84$ & $[-0.06,+11.09]$ & t \\
\bottomrule
\end{tabular}
\end{table}

\section{The PIS\_NFLTD probe}
\label{app:oodprobe}

\subsection{V8c out of distribution, per bin}
\label{app:v8c:ood}

\emph{Supports main Section~\ref{sec:results:v8c-ood}: the values behind
Figure~\ref{fig:v8c-ood}, and why unseen attenuation alone does not
explain V8c's failure on this volume.}

PIS\_NFLTD is acquired at PIS's photon level, which under the baseline
detects $38\%$ of the corpus target count, so the volume sits outside
the training distribution in its exposure, though not in its anatomy
(main Section~\ref{sec:library:split}: the training WED reaches
$381.2$~mm).

\emph{Unseen attenuation does not explain the failure.}  The training
pool covers the high-attenuation range --- $0.26\%$ of its
$1.58\times10^{9}$ bundles lie at $X_2\ge8$, and $6.9\%$ of
LIDC-0127's do --- and PIS\_NFLTD is only about three times denser there
($0.78\%$).  V8c is already worse than SNN2 at moderate $X_2$ on this
volume: under the baseline its middle-path RMSE is $2.6$ times SNN2's at
$X_2\in[4,5)$, and under $\NoFC$ it is $1.2$--$1.5$ times SNN2's in
every bin holding at least $10^{4}$ bundles.  What dominates the pool
error is the negative bias of bins~$7$--$10$, despite their small
population: under $\NoFC$ the bin-$10$ bias is $-1.198$ and the bin-$9$
bias $-0.670$, while SNN2's bias stays within $\pm0.35$ throughout.
Bin~$10$ of PIS\_NFLTD holds $248$ bundles, which is why its markers are
plotted open in Figure~\ref{fig:v8c-ood}.

\emph{Recalibration.}  Regenerating the volume at the corpus rule's
$N_{\mathrm{base}}=602{,}461$ removes the deep bias under the baseline,
from $-1.06$ to $+0.06$ at $X_2\ge7$
(Section~\ref{app:biasstruct:e2}); under $\NoFC$ the bias structure is
unchanged.

\subsection{Paper~1's regime, and the two papers' dose conventions}
\label{app:paper1floor}

\emph{Supports main Section~\ref{subsec:paper1-floor}: the volume
Paper~1 evaluates on, why the inflation ratio is comparable across the
two papers' exposures, and the per-bin numbers behind the comparison.}

\paragraph{The volume.}  Paper~1 evaluates its network on single-patient
data, trained and tested on separate body regions of PIS\_NFLTD: the PIS
volume rescaled in-plane by $1.40$, from its native $1.24$\,mm spacing
to $1.736$\,mm.  The rescaling enlarges the patient
(WED~${\approx}342$~mm) and brings the line integrals near the support
ceiling \citep[Section~5.3]{besson2026pmb} --- here $\max X_2=9.16$ ---
beyond native PIS.  There the network improves on the constrained
maximum-likelihood estimate on the photon-starved middle path, for line
integrals of about $5$ to $8$, loses elsewhere, and nowhere falls below
the equal-dose floor $\sigfair$ \citep{besson2026pmb}.

\paragraph{Why the inflation ratio is a common yardstick.}  The
middle-path inflation ratio $\sigCRB_2/\sigfair$ is
$\sqrt{13/3}\approx2.08$ at equal attenuation and, in chest anatomy ---
where the endpoint paths are brighter than the heavily attenuated
center --- rises well above it, without bound as the dedicated Fisher
information collapses.  It is invariant to the incident exposure
$\Nzero$: the Fisher information matrix scales linearly in $\Nzero$, so
the ratio cancels.  That matters here because the two papers use
different exposures.  Paper~1's datasets use $\Nzero=10^5$ per ray and
reading \citep{besson2026pmb}; the PIS\_NFLTD bundles of this paper use
PIS's own $N_{\mathrm{base}}$ ($2.29\times10^5$), not recalibrated for
the rescaled volume (Section~\ref{app:n0fc:nbase}; main
Section~\ref{sec:results:v8c-ood} tests that recalibration).  The floor
ratio depends on neither exposure; an estimator's ratio can, through its
bias.

\begin{table}[ht]
\renewcommand{\arraystretch}{1.05}
\caption{Middle-path ($X_2$) error on the unbiased-floor scale
(RMSE ratios, $\sqrt{\DIF}$),
PIS\_NFLTD, baseline acquisition, pooled per bin as in Paper~1
(main Section~\ref{sec:model:dif}): the T2bp$\,\circ\,$SNN2 RMSE ratio
$\mathrm{RMSE}_b/\bar\sigma_{\mathrm{fair},b}$ ($=\sqrt{\DIF}$,
\eqref{eq:dif_mse}, bias charged), its bias, the realized floor
$\bar\sigma^{\mathrm{CRB}}_b/\bar\sigma_{\mathrm{fair},b}$, and the ratio of
the two.  Bins as in~\eqref{eq:x2bins}.  The equal-attenuation value
$\sqrt{13/3}\approx2.08$ is a reference, not a minimum.  T2bp$\,\circ\,$SNN2 rides the realized floor through $[6,7)$.}
\label{tab:paper1-floor}
\centering
\footnotesize
\setlength{\tabcolsep}{4pt}
\begin{tabular}{ccrrrr}
\toprule
Bin & Range & T2bp$\,\circ\,$SNN2 & Bias & Floor & Ratio \\
\midrule
3 & $[2,3)$ &  2.30 & $0.000$ &  2.31 & 1.00 \\
4 & $[3,4)$ &  2.67 & $+0.001$ &  2.66 & 1.00 \\
5 & $[4,5)$ &  3.78 & $+0.003$ &  3.71 & 1.02 \\
6 & $[5,6)$ &  6.08 & $+0.013$ &  5.69 & 1.07 \\
7 & $[6,7)$ &  8.11 & $+0.034$ &  8.24 & 0.98 \\
8 & $[7,8)$ &  7.10 & $-0.115$ & 14.46 & 0.49 \\
9 & $[8,9)$ &  9.26 & $-0.596$ & 20.36 & 0.45 \\
\bottomrule
\end{tabular}
\end{table}

\paragraph{The sparse high tail.} The last two rows of
Table~\ref{tab:paper1-floor} are where T2bp$\,\circ\,$SNN2 falls below
the realized floor, to $0.49$ at $[7,8)$ and $0.45$ at $[8,9)$.  These
bins hold under $9\%$ of the bundles; the rows through $[6,7)$, where it
rides the floor with negligible shrinkage bias, hold $91\%$.  The pooled
CRB of $X_2$ in those bins is $0.72$ and $1.49$, which approaches the
distance from the true $X_2$ to the box limit $X_{\max}=9.5$, so what
the ratio records there is the box truncating the error rather than an
estimator beating its bound.  The box truncates Paper~1's constrained
estimate at the same bins, and Paper~1's network goes below this floor
there as well, though not below $\sigfair$.

\paragraph{The same bins under $\NoFC$.} On this scale the middle-path
floor falls from $8.24$ to $3.26$ at $[6,7)$ and from $14.46$ to $4.51$
at $[7,8)$, and T2bp$\,\circ\,$SNN2 tracks it down, $8.11\to3.42$ and
$7.10\to3.50$, riding its realized floor within $0.99$--$1.05$ at every
bin through $[6,7)$.  Its RMSE ratio of $1.44$--$1.90$ across $[2,5)$, a
$\DIF$ of $2.1$--$3.6$, is below the equal-attenuation value
$\sqrt{13/3}\approx2.08$ without help from the prior: the reduction is
supplied by the acquisition redesign, which is why it carries across the
corpus (main Section~\ref{sec:results:floor}) without per-patient
training.


\begin{thebibliography}{}
\setlength{\itemsep}{2pt}
\setlength{\rightskip}{0pt plus 4em}  %

\bibitem[Armato et al.(2011)]{armato2011}
Armato S G III et al 2011 The Lung Image Database Consortium (LIDC) and
Image Database Resource Initiative (IDRI): a completed reference
database of lung nodules on CT scans \textit{Med. Phys.} \textbf{38}
915--931
(https://doi.org/10.1118/1.3528204)

\bibitem[Armato et al.(2015)]{armato2015data}
Armato S G III et al 2015 Data from LIDC-IDRI [Data set] \textit{The
Cancer Imaging Archive} (https://doi.org/10.7937/K9/TCIA.2015.LO9QL9SX)

\bibitem[Ba et al.(2016)]{ba2016S}
Ba J L, Kiros J R and Hinton G E 2016 Layer normalization
arXiv:1607.06450

\bibitem[Barrett et al.(1994)]{barrett1994}
Barrett H H, Wilson D W and Tsui B M W 1994 Noise properties of the EM
algorithm: I. Theory \textit{Phys. Med. Biol.} \textbf{39} 833--846
(https://doi.org/10.1088/0031-9155/39/5/004)

\bibitem[Bartolac et al.(2011)]{bartolac2011}
Bartolac S, Graham S, Siewerdsen J H and Jaffray D 2011 Fluence field
optimization for noise and dose objectives in CT \textit{Med. Phys.}
\textbf{38} S2
(https://doi.org/10.1118/1.3574885)

\bibitem[Besson(2015)]{besson2015medphys}
Besson G M 2015 New CT system architectures for high temporal
resolution with applications to improved geometric dose efficiency and
cardiac imaging \textit{Med. Phys.} \textbf{42} 2668--2678
(https://doi.org/10.1118/1.4918328)

\bibitem[Besson(2016)]{besson2016spie}
Besson G M 2016 Seventh-generation CT \textit{Proc. SPIE Med. Imaging}
\textbf{9783} 978350
(https://doi.org/10.1117/12.2214319)

\bibitem[Besson(2026)]{besson2026pmb}
Besson G M 2026 Per-bundle statistical limits and learned-prior
inversion in multiplexed x-ray imaging, with application to Temporal CT
\textit{Phys. Med. Biol.} \textbf{71} 225020
(https://doi.org/10.1088/1361-6560/aebfec)

\bibitem[Cox and Snell(1968)]{coxsnell1968S}
Cox D R and Snell E J 1968 A general definition of residuals
\textit{J. R. Stat. Soc. B} \textbf{30} 248--275
(https://doi.org/10.1111/j.2517-6161.1968.tb00724.x)

\bibitem[De Man et al.(2016)]{deman2016}
De Man B et al 2016 Multisource inverse-geometry CT. Part I. System
concept and development \textit{Med. Phys.} \textbf{43} 4607--4616
(https://doi.org/10.1118/1.4954846)

\bibitem[Efron(2011)]{efron2011}
Efron B 2011 Tweedie's formula and selection bias \textit{J. Am. Stat.
Assoc.} \textbf{106} 1602--1614
(https://doi.org/10.1198/jasa.2011.tm11181)

\bibitem[Engel et al.(2008)]{engel2008}
Engel K J, Herrmann C and Zeitler G 2008 X-ray scattering in single- and
dual-source CT \textit{Med. Phys.} \textbf{35} 318--332
(https://doi.org/10.1118/1.2820901)

\bibitem[Flohr et al.(2006)]{flohr2006}
Flohr T G et al 2006 First performance evaluation of a dual-source CT
(DSCT) system \textit{Eur. Radiol.} \textbf{16} 256--268
(https://doi.org/10.1007/s00330-005-2919-2)

\bibitem[Gang et al.(2017)]{gang2017}
Gang G J, Siewerdsen J H and Stayman J W 2017 Task-driven optimization
of fluence field and regularization for model-based iterative
reconstruction in computed tomography \textit{IEEE Trans. Med. Imaging}
\textbf{36} 2424--2435
(https://doi.org/10.1109/TMI.2017.2763538)

\bibitem[Gies et al.(1999)]{gies1999}
Gies M, Kalender W A, Wolf H and Suess C 1999 Dose reduction in CT by
anatomically adapted tube current modulation. I. Simulation studies
\textit{Med. Phys.} \textbf{26} 2235--2247
(https://doi.org/10.1118/1.598779)

\bibitem[Hassan et al.(2018)]{hassan2018}
Hassan A I, Skalej M, Schlattl H and Hoeschen C 2018 Determination and
verification of the x-ray spectrum of a CT scanner \textit{J. Med.
Imaging} \textbf{5} 013506
(https://doi.org/10.1117/1.JMI.5.1.013506)

\bibitem[Hero et al.(1996)]{hero1996}
Hero A O, Fessler J A and Usman M 1996 Exploring estimator
bias-variance tradeoffs using the uniform CR bound \textit{IEEE Trans.
Signal Process.} \textbf{44} 2026--2041
(https://doi.org/10.1109/78.533723)

\bibitem[Hsieh(2015)]{hsieh2015}
Hsieh J 2015 \textit{Computed Tomography: Principles, Design, Artifacts,
and Recent Advances} 3rd edn (Bellingham, WA: SPIE Press)
(https://doi.org/10.1117/3.2197756)

\bibitem[Hsieh and Pelc(2013)]{hsiehpelc2013}
Hsieh S S and Pelc N J 2013 The feasibility of a piecewise-linear
dynamic bowtie filter \textit{Med. Phys.} \textbf{40} 031910
(https://doi.org/10.1118/1.4789630)

\bibitem[Koenker and Mizera(2014)]{koenker2014S}
Koenker R and Mizera I 2014 Convex optimization, shape constraints,
compound decisions, and empirical Bayes rules \textit{J. Am. Stat.
Assoc.} \textbf{109} 674--685
(https://doi.org/10.1080/01621459.2013.869224)

\bibitem[La Rivi\`ere(2005)]{lariviere2005}
La Rivi\`ere P J 2005 Penalized-likelihood sinogram smoothing for low-dose
CT \textit{Med. Phys.} \textbf{32} 1676--1683
(https://doi.org/10.1118/1.1915015)

\bibitem[Loshchilov and Hutter(2017)]{loshchilov2017S}
Loshchilov I and Hutter F 2017 SGDR: stochastic gradient descent with
warm restarts \textit{Proc. Int. Conf. on Learning Representations
(ICLR)}

\bibitem[Loshchilov and Hutter(2019)]{loshchilov2019S}
Loshchilov I and Hutter F 2019 Decoupled weight decay regularization
\textit{Proc. Int. Conf. on Learning Representations (ICLR)}

\bibitem[Louis(1984)]{louis1984}
Louis T A 1984 Estimating a population of parameter values using Bayes
and empirical Bayes methods \textit{J. Am. Stat. Assoc.} \textbf{79}
393--398
(https://doi.org/10.1080/01621459.1984.10478062)

\bibitem[McCollough et al.(2014)]{mccollough2014}
McCollough C et al 2014 \textit{Use of Water Equivalent Diameter for
Calculating Patient Size and Size-Specific Dose Estimates (SSDE) in CT}
AAPM Report 220 (College Park, MD: American Association of Physicists in
Medicine)
(https://doi.org/10.37206/146)

\bibitem[Petersilka et al.(2010)]{petersilka2010}
Petersilka M, Stierstorfer K, Bruder H and Flohr T 2010 Strategies for
scatter correction in dual source CT \textit{Med. Phys.} \textbf{37}
5971--5992
(https://doi.org/10.1118/1.3504606)

\bibitem[Robbins(1956)]{robbins1956}
Robbins H 1956 An empirical Bayes approach to statistics
\textit{Proc. Third Berkeley Symp. on Mathematical Statistics and
Probability} vol~1 pp 157--163

\bibitem[Schmidt et al.(2004)]{schmidt2004}
Schmidt T G, Fahrig R, Pelc N J and Solomon E G 2004 An inverse-geometry
volumetric CT system with a large-area scanned source: a feasibility study
\textit{Med. Phys.} \textbf{31} 2623--2627
(https://doi.org/10.1118/1.1786171)

\bibitem[Szczykutowicz and Mistretta(2013)]{szczykutowicz2013}
Szczykutowicz T P and Mistretta C A 2013 Design of a digital beam
attenuation system for computed tomography: Part I. System design and
simulation framework \textit{Med. Phys.} \textbf{40} 021905
(https://doi.org/10.1118/1.4773879)

\bibitem[Szczykutowicz and Mistretta(2014)]{szczykutowicz2014}
Szczykutowicz T P and Mistretta C A 2014 Experimental realization of
fluence field modulated CT using digital beam attenuation
\textit{Phys. Med. Biol.} \textbf{59} 1305--1326
(https://doi.org/10.1088/0031-9155/59/5/1305)

\bibitem[van Aarle et al.(2016)]{vanaarle2016}
van Aarle W et al 2016 Fast and flexible X-ray tomography using the ASTRA
toolbox \textit{Opt. Express} \textbf{24} 25129--25147
(https://doi.org/10.1364/OE.24.025129)

\bibitem[Wang et al.(2006)]{wang2006}
Wang J, Li T, Lu H and Liang Z 2006 Penalized weighted least-squares
approach to sinogram noise reduction and image reconstruction for low-dose
X-ray computed tomography \textit{IEEE Trans. Med. Imaging} \textbf{25}
1272--1283
(https://doi.org/10.1109/TMI.2006.882141)

\bibitem[Wu et al.(2022)]{wu2022}
Wu M, FitzGerald P, Zhang J, Segars W P, Yu H, Xu Y and De Man B 2022
XCIST---an open access x-ray/CT simulation toolkit \textit{Phys. Med.
Biol.} \textbf{67} 194002
(https://doi.org/10.1088/1361-6560/ac9174)

\bibitem[Yu et al.(2000)]{yu2000}
Yu D F, Fessler J A and Ficaro E P 2000 Maximum-likelihood transmission
image reconstruction for overlapping transmission beams \textit{IEEE
Trans. Med. Imaging} \textbf{19} 1094--1105
(https://doi.org/10.1109/42.896785)

\end{thebibliography}
\end{document}